\documentclass[aps,prd,superscriptaddress,amsmath,groupedaddress,twocolumn,nofootinbib]{revtex4}

\usepackage{color,hangcaption,hhline,psfrag,rotating,amssymb}
\usepackage[hang,nooneline]{subfigure}
\usepackage{dcolumn}

\begin{document}
\title{Precision tests of the $J/\psi$ from full lattice QCD: mass, leptonic width and radiative decay rate to $\eta_c$ }
\author{G. C. Donald}
\affiliation{SUPA, School of Physics and Astronomy, University of Glasgow, Glasgow, G12 8QQ, UK}
\author{C. T. H. Davies}
\email[]{c.davies@physics.gla.ac.uk}
\affiliation{SUPA, School of Physics and Astronomy, University of Glasgow, Glasgow, G12 8QQ, UK}
\author{R. J. Dowdall}
\affiliation{SUPA, School of Physics and Astronomy, University of Glasgow, Glasgow, G12 8QQ, UK}
\author{E. Follana}
\affiliation{Departamento de F\'{\i}sica Te\'{o}rica, Universidad de Zaragoza, E-50009 Zaragoza, Spain}
\author{K. Hornbostel}
\affiliation{Southern Methodist University, Dallas, Texas 75275, USA}
\author{J. Koponen}
\affiliation{SUPA, School of Physics and Astronomy, University of Glasgow, Glasgow, G12 8QQ, UK}
\author{G. P. Lepage}
\affiliation{Laboratory of Elementary-Particle Physics, Cornell University, Ithaca, New York 14853, USA}
\author{C. McNeile}
\affiliation{Bergische Universit\"{a}t Wuppertal, Gaussstr.\,20, D-42119 Wuppertal, Germany}

\collaboration{HPQCD collaboration}
\homepage{http://www.physics.gla.ac.uk/HPQCD}
\noaffiliation

\date{\today}

\begin{abstract}
We calculate the $J/\psi$ mass, leptonic width and radiative decay 
rate to $\gamma \eta_c$ from lattice QCD including $u$, $d$ and 
$s$ quarks in the sea for the first time. 
We use the Highly Improved Staggered Quark formalism and nonperturbatively 
normalised vector currents for the leptonic and radiative decay rates. 
Our results are: $M_{J/\psi} -M_{\eta_c} = 116.5(3.2) \mathrm{MeV}$; 
$\Gamma(J/\psi \rightarrow e^+e^-) = 5.48(16) \mathrm{keV}$; 
$\Gamma(J/\psi \rightarrow \gamma \eta_c) = 2.49(19) \mathrm{keV}$. 
The first two are in good 
agreement with experiment, 
with $\Gamma(J/\psi \rightarrow e^+e^-)$ providing a test 
of a decay matrix element in QCD, independent 
of CKM uncertainties, to 2\%. 
At the same time results for the time moments of the correlation function 
can be compared to values from the charm contribution 
to $\sigma(e^+e^- \rightarrow \mathrm{hadrons})$,  
giving a 1.5\% test of QCD. 
Our results show that an 
improved experimental error would 
enable a similarly strong test from 
$\Gamma(J/\psi \rightarrow \gamma \eta_c)$.
\end{abstract}


\maketitle

\section{Introduction}
\label{sec:intro}

Precision tests of lattice QCD against experiment 
are critical to provide benchmarks against which 
to calibrate the reliability of predictions from 
lattice QCD~\cite{ourlatqcd}. Most tests to date have relied on 
the spectrum of gold-plated hadron masses - for 
example, the mass of the $D_s$ meson can be calculated 
in lattice QCD with an error of 3 MeV 
(having fixed the masses of 
the $c$ and $s$ masses from other mesons) 
and the result agrees with experiment~\cite{fdsupdate, rachelnew}. 
Here we give another such test by determining 
the mass of the $J/\psi$ to a precision of 3 MeV. 

Tests of decay matrix elements are harder 
to do very accurately. We need precision tests
of these because it is the
predictions of decay matrix elements from lattice QCD 
that enable, for example, 
progress with the flavor physics programme~\cite{cdlat11} of 
over-determining the CKM matrix to find signs of 
new physics~\cite{lunghilat11}.  
The leptonic decay rate of the 
$\pi$ via a $W$ boson provides one such test. 
The QCD input to this is the 
pion decay constant, which is determined to 1\% in lattice 
QCD~\cite{milc10}. If we take $V_{ud}$ from 
nuclear $\beta$ decay~\cite{pdg}, we have a 2\% determination 
of the leptonic decay rate to be compared to experiment. 
The leptonic decay 
rates of other charged pseudoscalars can also 
be determined to a few percent from lattice QCD~\cite{cdlat11} but 
then the comparison with experiment is generally needed
to determine the appropriate CKM element. 
Independent tests of matrix elements, without CKM 
uncertainties, come only from electromagnetic decays.
Here we provide two such tests through 
two different decay rates of the $J/\psi$: annihilation 
to $e^+e^-$ via a photon and radiative decay to 
the $\eta_c$. We give the first results from full 
lattice QCD including $u$, $d$ and $s$ quarks in 
the sea, although earlier calculations have been 
done in quenched QCD~\cite{dudekcharm} and including 
$u$ and $d$ sea quarks~\cite{clqcdcharm, etmccharm}.  

We are able to determine these 
matrix elements to a few percent  because of our 
development of an accurate and fully relativistic 
approach to $c$ quarks (as well as $u$, $d$ and 
$s$) in lattice QCD 
called the Highly Improved Staggered Quark (HISQ) 
formalism~\cite{hisqdef}. In this formalism we are able to 
normalise the vector current which mediates the 
electromagnetic decay accurately and nonperturbatively 
and we show how to do that here. 

The layout of the paper is as follows: section~\ref{sec:latt} 
describes the lattice calculation and then section~\ref{sec:results} gives 
results for the $J/\psi$ mass, leptonic width (along 
with time moments of the $J/\psi$ correlator) and 
radiative decay rate in turn. We compare 
our results to experiment and 
to previous lattice QCD calculations in section~\ref{sec:discussion}. 
Section~\ref{sec:conclusions} gives our conclusions.
The Appendices discuss the more technical issues of 
discretisation errors and our two different 
methods for current renormalisation. 

\section{Lattice calculation}
\label{sec:latt}

We use 6 ensembles of lattice gluon 
configurations at 4 different, widely separated, values of 
the lattice spacing, provided by the MILC 
collaboration~\cite{milcreview}. 
The configurations include the effect of $u$, $d$ and $s$
quarks in the sea with the improved staggered (asqtad) formalism. 
The $u$ and $d$ masses are taken to be the 
same with $m_{u/d}/m_s$ approximately 0.2 on most 
of the ensembles.
Based on our experience of other gold-plated mesons~\cite{fdsupdate}
we expect sea quark mass effects to be small 
for the $J/\psi$ because it has no valence light
quarks.  We can test this by comparison of results on 
sets 1 and 2 where the sea value of $m_{u,d}$
changes by a factor of two and with set 3 where 
the sea value of $m_s$ changes by 70\%. 
Table~\ref{tab:params} lists the 
parameters of the ensembles. 

The lattice spacing is determined on an ensemble-by-ensemble basis 
using a parameter $r_1$ that comes from fits to the static quark 
potential calculated on the lattice~\cite{milcreview}. This 
parameter has small statistical/fitting errors but  
its physical value is not accessible to experiment and 
so must be determined using other quantities, calculated on the 
lattice, that are. We have determined $r_1$ = 0.3133(23) fm
using four different quantities ranging from the (2S-1S) splitting 
in the $\Upsilon$ system to the decay constant of the $\eta_s$ (fixing 
$f_K$ and $f_{\pi}$ from experiment)~\cite{oldr1paper}. 
Using our value for $r_1$ and 
the MILC values for $r_1/a$ given in Table~\ref{tab:params} we can 
determine $a$ in fm on each ensemble or, equivalently, $a^{-1}$ in 
GeV needed to convert lattice quantities to physical units. 

\begin{table}
\begin{tabular}{lllllllll}
\hline
\hline
Set &  $r_1/a$ & $au_0m_{l}^{asq}$ & $au_0m_{s}^{asq}$ & $L_s/a$ & $L_t/a$ & $\delta x_l$ & $\delta x_s$\\
\hline
1 &  2.647(3) & 0.005 & 0.05 & 24 & 64 & 0.11 & 0.43 \\
2 &  2.618(3) & 0.01 & 0.05 & 20 & 64 & 0.25 & 0.43  \\
3 &  2.658(3) & 0.01 & 0.03 & 20 & 64 & 0.25 & -0.14  \\
\hline
4 & 3.699(3) & 0.0062 & 0.031 & 28 & 96  & 0.20 & 0.19 \\
\hline 
5 &  5.296(7) & 0.0036 & 0.018 & 48 & 144 & 0.16 & -0.03 \\
\hline
6 & 7.115(20) & 0.0028 & 0.014 & 64 & 192 & 0.17 & 0.04 \\
\hline
\hline
\end{tabular}
\caption{Ensembles (sets) of MILC configurations used for this analysis. 
The sea 
asqtad quark masses $m_l^{asq}$ ($l = u/d$) and $m_s^{asq}$ 
are given in the MILC convention where $u_0$ is the plaquette 
tadpole parameter. 
The lattice spacing values in units of $r_1$ after `smoothing'
are given in the second column~\cite{milcreview}. 
Here sets 1, 2 and 3 are `coarse'; set 4, 
`fine'; set 5 `superfine' and set 6 `ultrafine'.  The size of 
the lattices is given by $L_s^3 \times L_t$. The final two columns give the 
difference between the sea quark mass and its physical value in 
units of the $s$ quark mass~\cite{fdsupdate}. 
}
\label{tab:params}
\end{table}

On these ensembles we calculate $c$ quark propagators 
using the HISQ action and combine them 
into meson correlation functions. The quark 
propagators are made from a `random wall' source - 
a color 3-vector 
of U(1) random numbers -  
on a given timeslice
to reduce the statistical noise. 
An added reduction 
comes from the use of a random starting point for 
the equally spaced time-sources we use on the coarse and fine 
ensembles. We include only connected correlation functions here - 
disconnected contributions for the $J/\psi$ are related to its
hadronic width which is in keV and therefore negligible here. 

The $c$ quark 
mass is tuned from the $\eta_c$ meson mass~\cite{fdsupdate}. 
The appropriate `experimental' mass for the $\eta_c$ 
for our calculations is 2.986(3) GeV, differing from 
the experimental result of 2.981(1) GeV~\cite{pdg} 
because of missing electromagnetic, $\eta_c$ 
annihilation and $c$-in-the-sea effects 
that we estimate perturbatively~\cite{gregory}.
The HISQ lattice $c$ quark masses for the ensembles we are
using were determined in~\cite{fdsupdate}. 

Meson masses and decay constants are determined from 
simple `2-point' meson correlation functions made from 
combining quark propagators with appropriate spin 
matrices at source and sink to project onto the correct 
$J^{PC}$. For staggered quarks, where the spin degree 
of freedom has disappeared, the spin projection matrices 
are replaced with space-time-dependent phases of $\pm 1$. 
Because of fermion-doubling, there are in fact 16 `tastes' 
of every meson made by combining a point-splitting 
of the quark and antiquark source and sink along with 
the appropriate $\pm 1$ phases. 
The most accurate meson correlation functions come 
from either local or 1-link separated sources and sinks 
and we will restrict ourselves to these here. 
Because the taste-splittings are discretisation effects 
we are free to use whichever taste is the most convenient 
for a given calculation. 

For the pseudoscalar mesons the mass differences between 
the different tastes have a simple picture with 
the mass increasing as the amount of point-splitting 
in the source/sink operator increases. 
The lightest mass particle is the Goldstone  
meson whose correlator is simply the modulus squared 
of the propagator and whose squared mass vanishes linearly 
with the quark mass. This is the one that is used 
to tune the quark mass. The other taste pseudoscalar mesons 
have a mass for which the difference of mass-squared with the 
Goldstone meson is a constant with quark 
mass which vanishes as $\alpha_s^2a^2$. 
These taste-splitting discretisation 
errors are particularly small with the HISQ action~\cite{hisqdef}. 
They also become smaller, in proportion to the meson mass, 
as the meson mass increases and so are very small for 
mesons made of $c$ quarks~\cite{hisqdef}. 
The mass difference between the Goldstone meson 
and the next heaviest pseudoscalar meson is visible, 
however. Both masses can be determined very 
accurately in lattice QCD because they both correspond 
to local operators. The Goldstone meson corresponds 
to the local $\gamma_5$ operator and the local non-Goldstone 
to the local $\gamma_0 \gamma_5$ operator. 
We will use both of these mesons in our 
calculation of the radiative decay rate of 
the $J/\psi$. 

Vector meson taste-splittings are significantly 
smaller than for pseudoscalars and typically not 
visible for light mesons above the statistical errors. 
For the charmonium vectors we use the local $\gamma_i$ 
operator to determine the leptonic decay rate and two 
different
1-link split operators for the radiative decay. 
We discuss mass differences from taste-splittings further 
in Appendix~\ref{appendix:taste}. 

\section{Results }
\label{sec:results}

\subsection{$M_{J/\psi}$}
\label{sec:psimass}

The determination of the mass of the $J/\psi$ 
is most accurately done through the determination 
of the charmonium hyperfine splitting, i.e. 
the mass difference with the pseudoscalar $\eta_c$ 
meson. For the $\eta_c$ we use the Goldstone 
meson, as discussed in section~\ref{sec:latt}, 
because this is the most accurately determined 
in lattice QCD and is the meson we use to 
fix the $c$ quark mass. We studied this meson in 
detail in~\cite{fdsupdate}. 
For the $J/\psi$ we use the local $\gamma_i$ operator 
to create and destroy the vector meson. The $J/\psi$ 
correlators are then obtained by combining 
quark propagators from the default random wall 
with antiquark propagators from a source using 
the same random wall but patterned with phases, 
for example 
$(-1)^x$ for the vector polarised in the $x$ direction. 
$(-1)^x$ is also inserted at the sink where the 
propagators are tied together. 

\begin{table*}
\begin{tabular}{lllllllll}
\hline
\hline
Set & $N_{\mathrm{cfg}}\times N_t$ & $m_ca$ & $\epsilon$ & $aM_{\eta_c}$ & $aM_{J/\psi}$ & $a\Delta M_{hyp}$  & $af_{J/\psi}/Z$ & $Z_{cc}$ \\
\hline
1 & $2099 \times 8$ & 0.622 & -0.221 & 1.79118(4) & 1.85934(8) & 0.06817(6) & 0.2810(2) & 0.979(12)\\
\hline
2 & $2259 \times 4$ & 0.63 & -0.226 & 1.80851(5) & 1.87797(10) & 0.06946(8) & 0.2855(2) & 0.979(12)\\
2 & $2259 \times 8$ & 0.66 & -0.244 & 1.86667(4) & 1.93430(9) & 0.06763(7) & 0.2925(2) & 0.974(12)\\
\hline
3 & $323 \times 8$ & 0.617 & -0.218 & 1.78212(12) & 1.85081(23) & 0.06869(17) & 0.2804(5) & 0.979(12) \\
\hline
4 & $566 \times 4$ & 0.413 & -0.107 & 1.28052(7) & 1.32901(12) & 0.04849(10) & 0.1829(2) & 0.983(12)\\
\hline 
5 & $200 \times 2$ & 0.273 &  -0.0487 & 0.89948(8)  & 0.93369(13) & 0.03421(11)  & 0.1244(3) & 0.986(12)\\
\hline
6 & $208 \times 1$ & 0.193 & -0.0247 & 0.66649(6) & 0.69217(11) & 0.02568(10) & 0.0925(3) & 0.990(12)\\
\hline
\hline
\end{tabular}
\caption{ Results in lattice units for the masses of $\eta_c$ and $J/\psi$ and their 
difference on each ensemble along with the 
raw (unrenormalised) decay constant and $Z$ factor for the $J/\psi$. 
Columns 3 and 4 give the bare HISQ charm quark mass, tuned from the $\eta_c$
and the corresponding coefficient $\epsilon$ used in the 
Naik discretization improvement term of the HISQ action~\cite{fdsupdate}. 
All of the charm quark masses are very well tuned except for the 
lower result on set 2 ($m_ca=0.66$), which was deliberately mistuned to 
assess the sensitivity of quantities to the tuning. 
Of the remaining masses the least well-tuned is on superfine set 
5 where $M_{\eta_c}$ is 0.5\% too high. 
Column 2 gives the number of configurations used and the number 
of time sources for propagators on each configuration. 
Results are binned on time sources and binned over neighbouring configurations 
for sets 5 and 6. The $J/\psi$ correlators are averaged over polarisations 
except on sets 2 and 3 where only one polarisation was calculated. 
The results for the $\eta_c$ masses 
are also given in~\cite{fdsupdate}. They differ slightly from these in some 
cases because 
of fitting simultaneously with $J/\psi$ correlators.  
The $Z$ factors are taken from moment 4 of the nonperturbative (on the lattice) 
current-current correlator 
method described in Appendix~\ref{appendix:currcurr}.
 }
\label{tab:massres}
\end{table*}

The $J/\psi$ and $\eta_c$ correlators at zero spatial 
momentum are fit simultaneously 
so that correlations between them are taken into account. 
The fit form for the average $J/\psi$ correlator as a function of 
time separation between source and sink, $t$, is: 
\begin{equation}
\overline{C}_{2pt}(t) = \sum_{i_n,i_o} a^2_{i_n} \mathrm{fn}(M_{i_n},t) - \tilde{a}^2_{i_o}\mathrm{fo}(\tilde{M}_{i_o},t)
\label{eq:fit1}
\end{equation}
with 
\begin{eqnarray}
\mathrm{fn}(M,t) &=& e^{-Mt} + e^{-M(L_t-t)} \nonumber \\
\mathrm{fo}(M,t) &=& (-1)^{t/a} \mathrm{fn}(M,t)
\label{eq:fnfo}
\end{eqnarray}
and $L_t$ the time extent of the lattice. 
$i_n=0$ is the ground state and larger $i_n$ values 
denote radial or other excitations with the same $J^{PC}$ quantum 
numbers. The $M_{i_n}$ are the masses 
of the corresponding particles. 
There are `oscillating' terms coming 
from opposite parity states, denoted $i_o$. 
The Goldstone $\eta_c$ meson has the same fit form 
except that there are no 
oscillating contributions (when the $\eta_c$ is at rest). 
Note that we do not use 
any `smearing' functions for the propagators at either 
source or sink.  

\begin{figure}
\begin{center}
\includegraphics[width=0.9\hsize]{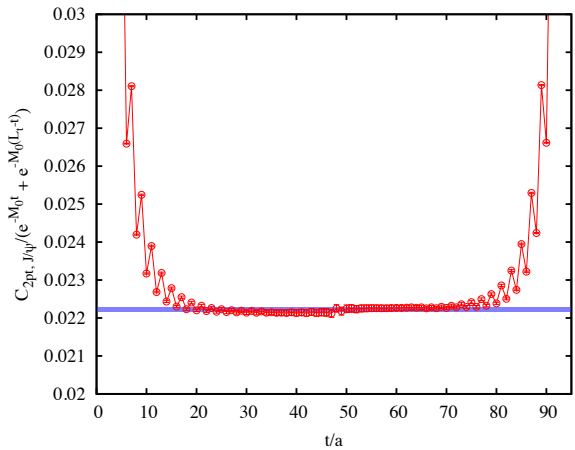}
\end{center}
\caption{Our average $J/\psi$ correlator divided 
by the ground state exponential ($\mathrm{fn}(M_0,t)$ 
from eq.~(\ref{eq:fnfo})) as a function of lattice time. 
Lines are drawn to join the points (which include 
statistical errors) for clarity. The 
fitted result
for the ground state amplitude, $a_0^2$, 
is given by the blue band. 
The fit includes 6 normal exponentials and 6 
oscillating ones, which are responsible for the oscillating 
behaviour clearly seen in the results. 
}
\label{fig:corr}
\end{figure}

To fit we use a number of exponentials $i_n$, 
and where appropriate $i_o$, 
in the range 2--6, loosely constraining the higher order exponentials 
by the use of Bayesian priors~\cite{gplbayes}. 
As the number of exponentials increases, 
we see the $\chi^2$ value fall below 1 and the results
for the fitted values 
and errors for the parameters for the ground state $i=0$ stabilise. 
This allows us to determine the ground state parameters $a_0$ and $M_0$ as accurately 
as possible whilst including the full systematic error from the presence 
of higher excitations in the correlation function. 
We take the fit parameters 
to be the logarithm of the ground state masses $M_0$ and $\tilde{M}_{0}$ 
and the logarithms of 
the differences in mass between successive radial excitations (which 
are then forced to be positive). 
The Bayesian prior value for $M_0$ for the $\eta_c$ is obtained from a simple `effective 
mass' in the correlator and the prior width on the value is taken as 0.3.  
The prior value on $M_0$ for the $J/\psi$ is taken to be $100 \pm 50$ 
MeV above the $\eta_c$. 
The prior value for mass splittings to and between 
excitations is taken as 600(300) MeV. 
The amplitudes $a_{i_n}$ and $a_{i_o}$ are given prior 
widths of 1.0. 
We apply a cut on the range of eigenvalues from the 
correlation matrix that are used in the fit of $10^{-4}$. 
We also cut out small $t/a$ (and $(L_t-t)/a$) values below 6 from our 
fit to reduce the effect of higher excitations. 

Figure~\ref{fig:corr} shows the quality of our results 
with a plot of the $J/\psi$ correlation function. It is 
divided by the ground-state exponential function so 
that it shows a plateau in the centre of value $a_0^2$. 
The results for the ground-state masses in lattice units of the $J/\psi$
and $\eta_c$ and the difference between them, $a\Delta M_{hyp}$, are given 
in Table~\ref{tab:massres}. The difference is typically 
more accurate than that obtained by simply 
subtracting the masses because of the correlation between 
the correlators. 

\begin{figure}
\begin{center}
\includegraphics[width=0.95\hsize]{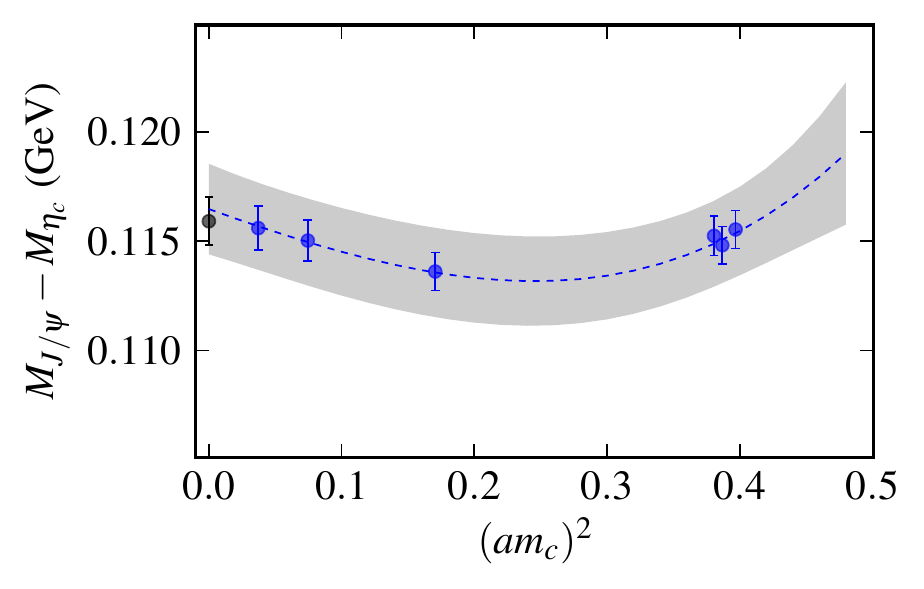}
\end{center}
\caption{Results for the charmonium hyperfine splitting 
plotted as a function of lattice spacing. 
For the $x$-axis we use $(m_ca)^2$ to allow  the $a$-dependence 
of our fit 
function (eq.~(\ref{eq:fithyp})) (blue dashed line 
with grey error band) to be displayed simply. 
The data points have been corrected for $c$ quark mass 
mistuning and sea quark mass effects, but the corrections 
are smaller than the error bars. We do not include 
on the plot the deliberately mistuned $c$ mass but 
it is included in the fit to constrain the $c$ mass 
dependence. 
The errors shown 
include (and are dominated by) uncertainties from the 
determination of the lattice spacing $a$ (from the 
physical value of the parameter $r_1$) that are
correlated between the points. 
The experimental average is plotted as the black point 
at the origin, offset slightly from the $y$-axis for clarity. 
}
\label{fig:hyp}
\end{figure}

The hyperfine splitting is converted to physical units 
using the values for $a$ on each ensemble as 
discussed in section~\ref{sec:latt}. 
The results are shown in Figure~\ref{fig:hyp}. 
Figure~\ref{fig:hyp} includes the error from the 
determination of the lattice spacing on each point. 
This dominates the error but is correlated between the points
and that should be borne in mind in looking at the figure. 
It is important to realise that 
the naive lattice spacing error is magnified by a 
factor of approximately two in the hyperfine splitting because 
of the inverse relationship between hyperfine splitting 
and quark mass. 
For example, a shift by uncertainty $\delta$ upwards 
in the inverse lattice spacing causes a shift upwards in the 
meson mass by the same proportion. To determine the total 
effect of this on the hyperfine splitting we must include 
the effect of retuning the $c$ quark mass to make 
the meson mass correct again. This means in this case
retuning the quark mass down by fraction $\delta$ which 
shifts the hyperfine splitting upward by a further factor 
of $\delta$ to that coming simply from the lattice spacing 
change. Thus the change in the hyperfine splitting, representing 
its uncertainty, is approximately $2\delta$~\cite{daviesoldups}\footnote{This point
has frequently been overlooked in lattice QCD calculations.}.
Thus lattice spacing uncertainties are typically much 
more important in the determination of hyperfine 
splittings than statistical errors.

We fit the hyperfine splitting as a function of lattice 
spacing and sea quark masses to the form:
\begin{eqnarray}
\label{eq:fithyp}
f(a, \delta x_l, \delta x_s) &=& f_0 \times \\
&& \sum_{ijkl} c_{ijkl} (am_c)^{2i} (\frac{\delta x_1}{10})^{j} (\frac{\delta x_2}{10})^{k} (\frac{\delta x_3}{10})^{l} \nonumber \\
&+& (d_0 + d_1(am_c)^2)(M_{\eta_c, \mathrm{latt}}-M_{\eta_c,\mathrm{expt}}). \nonumber 
\end{eqnarray}
Here $f_0$ is the physical result, the sum over $ijkl$ 
allows for discretisation errors and sea quark effects and 
the final term allows for mistuning of the $c$ quark mass.  
We allow the discretisation errors, which are evident 
in our results, to have a scale set 
by the $c$ quark mass. These appear only as even powers 
of $a$ for staggered quarks. 
$\delta x_l$ and $\delta x_s$ 
are the mistuning of the sea quark masses: 
\begin{equation}
\delta x_q = \frac{m_{q,\mathrm{sea}}-m_{q,\mathrm{phys}}}{m_{s,\mathrm{phys}}} .
\end{equation}
$\delta x_l$ and $\delta x_s$ values are given for each 
ensemble in Table~\ref{tab:params} and are taken 
from Appendix A of~\cite{fdsupdate}. 
Eq.~(\ref{eq:fithyp}) includes a term for each sea quark 
($u/d$ appearing twice, and $s$), with the coefficients 
constrained to be the same so that the fit function is
symmetric with respect to interchange of any two. 
The division by 10 is because the scale for dependence 
on light quark masses from chiral perturbation theory 
is $4\pi f_{\pi} \approx 10 m_s$. 
We see no significant sea quark mass dependence
in the hyperfine splitting. A fairly strong dependence 
was seen in the twisted mass calculations~\cite{etmccharm}. 
However, at least some of that dependence could be 
attributed to the sea quark mass dependence of the 
lattice spacing, since that is determined only in the 
chiral limit. Here we determine the lattice spacing 
for each ensemble and hence separate lattice spacing 
dependence from physical sea quark mass effects. 
The sum over $ijkl$ in eq.~(\ref{eq:fithyp}) allows for the possibility of 
lattice spacing dependent sea quark mass effects. 

We take a Bayesian prior~\cite{gplbayes} on $f_0$ of 0.1(1) 
and then fix $c_{0000}$ to 1. The other $c_{ijkl}$ are 
given priors of $0.0\pm 1.0$ except for the 
$c_{0jkl}$ which determine the $a$-independent
sea quark mass dependence. These are taken to have 
priors $0.0\pm 0.33$ because we expect sea quark 
mass effects to be typically a factor of 3 smaller than 
valence quark mass effects which would have chiral perturbation 
theory coefficients of $\mathcal{O}(1)$. 
We include 5 terms in the $a$-dependence and 
3 in the $\delta x$ dependence. Including additional 
terms makes no difference to the value for $f_0$ or its error. 
The priors for $d_0$ and $d_1$ are taken as 0.00(5), informed 
by the expectation that the hyperfine splitting should 
be inversely proportional to the mass, and 
by the effect of our mistuned $c$ mass on set 2 which 
agrees roughly with that expectation. 

The fit gives $f_0=116.5(2.1) \mathrm{MeV}$, 
as the result for the hyperfine splitting 
in the absence of electromagnetism, $c$-in-the-sea and 
$\overline{c}c$ annihilation. The first two affect the $\eta_c$ 
and $J/\psi$ equally and so have no effect on 
the hyperfine splitting. The third affects the $\eta_c$ 
more than the $J/\psi$, which has negligible width. 
A perturbative estimate of the shift of the $\eta_c$ 
mass resulting from its annihilation to two gluons~\cite{hisqdef} 
related this to the total $\eta_c$ width and obtained 
a shift downwards of the $\eta_c$ mass of 2.4 MeV\footnote{This would 
now amount to 2.9 MeV given that the average experimental 
width of the $\eta_c$ has increased to 30 MeV~\cite{pdg}.}. 
Using this, we have since applied a shift of 2.4(1.2) MeV for this 
effect to determine the $\eta_c$ mass to which to tune our 
$c$ quark mass, as in section~\ref{sec:latt}. 
For this purpose the impact of the shift is completely negligible, 
amounting to less than 0.1\% of the $\eta_c$ mass. 
For the hyperfine splitting, however, this shift could 
be a relatively large effect. 
Nonperturbative calculations of the contribution of 
`disconnected diagrams' to the $\eta_c$ mass have 
agreed on a small value of a few MeV for the shift 
from $\eta_c$ annihilation 
but obtained the opposite sign~\cite{levkova}. The argument 
is that the perturbative result may be modified significantly 
by the $gg$ intermediate state forming a resonance such 
as a glueball which is lighter in mass than the $\eta_c$, 
or a lighter hadron state. 
To allow for this possibility  and be consistent with the 
nonperturbative calculations 
we do not apply a shift to the hyperfine splitting obtained 
from our fit above, but instead take an additional 
systematic error of 2.4 MeV, corresponding 
to our original shift, to allow for the effect.  

Our final result for the hyperfine splitting is then:
\begin{equation}
\Delta M_{\mathrm{hyp}} = 116.5(2.1)(2.4) \, \mathrm{MeV} 
\label{eq:hypgive}
\end{equation}
where the errors are in turn from statistics/fitting 
and $\eta_c$ annihilation. The uncertainty from $\eta_c$ annihilation 
dominates the error. A complete error budget is given 
in Table~\ref{tab:errorbudget}. 

\begin{table}
\begin{tabular}{lccc}
\hline
\hline
 &  \hspace{-1em}$M_{J/\psi}-M_{\eta_c}$ & $f_{J/\psi}$ & $V_{J/\psi \rightarrow \eta_c\gamma}(0)$ \\
\hline
$(am_c)^2$ extrapolation & 0.45 & 0.45 & 3.5 \\
statistics & 0.50 & 0.41 & 0.74\\
lattice spacing & 1.6 & 0.42 & 0.0 \\
sea quark extrapolation & 0.29 & 0.26 & 1.3 \\
$M_{\eta_c}$ tuning & 0.11 & 0.09 & 0.0 \\
Z & - & 1.23 & 0.14\\
$M_{\eta_c}$ annihilation & 2.1 & 0.0 & 0.0\\
electromagnetism & 0.0 & 0.5 & 0.5 \\ 
\hline 
Total (\%) & 2.7 & 1.5 &  3.8 \\
\hline
\hline
\end{tabular}
\caption{Complete error budget for hyperfine splitting, leptonic 
width and vector form factor as a percentage of the final 
answer. 
}
\label{tab:errorbudget}
\end{table}

This is to be compared to the difference of the experimental 
averages of the two masses of 115.9(1.1) MeV~\cite{pdg}. 
Quite a spread of results make up the average.
Recent values tend to be at the lower end of the hyperfine splitting 
range. For example, the 2011 Belle result for the $\eta_c$ 
mass gives a hyperfine splitting of $111.5( {+2.5\atop -1.6})$ MeV~\cite{bellehyp}, 
and a recent result from BESIII gives 112.6(0.9) MeV~\cite{bes3etac}. 

\subsection{$\Gamma(J/\psi \rightarrow e^+e^-)$ and $R_{e^+e^-}$}
\label{sec:leptwidth}

The amplitude, $a_0$, from the fit in 
equation~(\ref{eq:fit1}) to our $J/\psi$ correlators
is directly related to the matrix element for the local 
vector operator to create or destroy the ground-state vector meson 
from the vacuum. The vector meson decay constant, $f_v$, for 
meson $v$ is 
defined by:
\begin{equation}
\langle 0 | \overline{\psi}\gamma^i \psi | v \rangle = f_vm_v \epsilon^i
\label{eq:fvdef}
\end{equation}
where $\epsilon^i$ is the meson polarization. 
$f_v$ for the $J/\psi$ is then determined from 
our lattice QCD correlators, in terms of the 
ground-state parameters from our fit (eq.~(\ref{eq:fit1})) by: 
\begin{equation}
\frac{f_v}{Z} =  a_0\sqrt{\frac{2}{M_0}},
\label{eq:atof}
\end{equation}
where $Z$ is the renormalisation constant 
required to match the local vector current in lattice QCD used here to 
that of continuum QCD at each value of the lattice spacing. 

$f_v$ is clearly a measure of the internal structure 
of a meson and in turn is related 
to the experimentally measurable leptonic branching fraction:
\begin{equation}
\Gamma(v_h \rightarrow e^+e^-) = \frac{4\pi}{3}\alpha_{QED}^2 e_h^2 \frac{f_v^2}{m_v}
\label{eq:vdecay}
\end{equation} 
where $e_h$ is the electric charge of the heavy quark 
in units of $e$ (2/3 for $c$). 
The experimental average, 
$\Gamma(J/\psi \rightarrow e^+e^-) = 5.55(14) \mathrm{keV}$~\cite{pdg} 
gives $f_{J/\psi}$ = 407(5) MeV, remembering 
that the electromagnetic coupling constant runs with 
scale and using $1/\alpha_{QED}(m_c) = 134$~\cite{alpha-em}. 
This can then provide a test of QCD at the 1\% level. 
Electromagnetic corrections are small since the $J/\psi$ must 
decay to an odd number of photons~\cite{cleo3gamma}. 

Our results for $f_{J/\psi}/Z$ are given in Table~\ref{tab:massres}. 
The final column of that table gives the values of
$Z$ determined from current-current correlators as 
described in Appendix~\ref{appendix:z}. 
This method uses continuum perturbation theory through 
$\mathcal{O}(\alpha_s^3)$ to normalise the lattice QCD correlators 
at small times. $Z$ then results from a combination of non-perturbative lattice 
QCD calculations with continuum perturbation 
theory in a similar approach to that of the RI-MOM scheme\footnote{This method is often called `nonperturbative' in the lattice QCD literature.} used to renormalise 
the currents for the same calculation using twisted 
mass quarks in~\cite{etmccharm}. The current-current 
correlator method has the advantage that 
we can use the same correlators from which we also extract, 
at large times, the nonperturbative information on the ground-state 
mass and decay constant. Indeed this allows some 
cancellation of discretisation errors apparent in 
the unrenormalized decay constant. 

\begin{figure}
\begin{center}
\includegraphics[width=0.95\hsize]{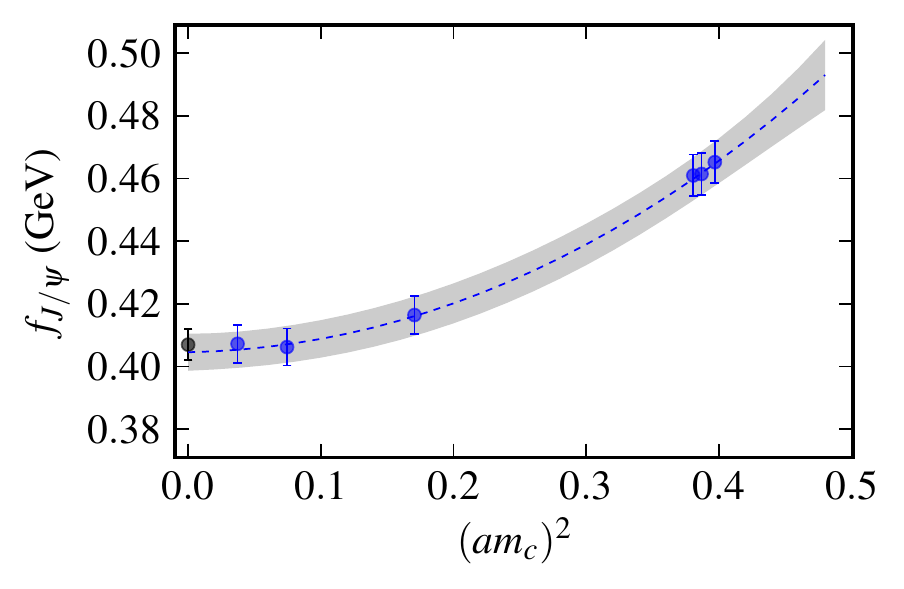}
\end{center}
\caption{Results for the charmonium vector decay constant
plotted as a function of lattice spacing. 
For the $x$-axis we use $(m_ca)^2$ to allow  the $a$-dependence 
of our fit 
function (eq.~(\ref{eq:fithyp})) (blue dashed line with 
grey error band) to be displayed simply. 
The data points have been corrected for $c$ quark mass 
mistuning and sea quark mass effects, but the corrections 
are smaller than the error bars. 
We do not include 
on the plot the deliberately mistuned $c$ mass but 
it is included in the fit to constrain the $c$ mass 
dependence. 
The errors shown 
include (and are dominated by) uncertainties from the 
determination of the current renormalization 
factor, $Z$, that are 
correlated between the points. 
The experimental average is plotted as the black point 
at the origin, offset slightly from the $y$-axis for clarity. 
}
\label{fig:fpsi}
\end{figure}

Multiplying $f_{J/\psi}/Z$ by $Z$ and then by $a^{-1}$ in GeV 
gives the physical results for the decay constant plotted 
in Figure~\ref{fig:fpsi}. We fit these to the same 
function of lattice spacing and sea quark mass used 
for the hyperfine splitting, eq.~(\ref{eq:fithyp}).  
The only differences are that the prior on $f_0$ is 
taken as 0.5(5) in this case and the priors on the slope of 
the variation of $f_{J/\psi}$ with $M_{\eta_c}$ are
taken as: $d_0$, 0.065(5) and $d_1$, 0.00(25). 
These are informed by the variation we see for the 
deliberately mistuned $c$ mass on set 2 and also by 
our extensive study of the behaviour of $f_{\eta_c}$ 
with $M_{\eta_c}$ in~\cite{fdsupdate}. There we find 
a strong $a$-dependence in the slope of the decay 
constant with mass and so we allow for that here. 

The physical result that we obtain in the continuum 
limit is:
\begin{equation}
f_{J/\psi} = 405(6)(2) \mathrm{MeV}.
\label{eq:fvres}
\end{equation}
The first error is from the fit and is dominated 
by the error from the $Z$ factor. The second
error is an estimate of systematic effects from 
missing electromagnetism in our lattice QCD 
calculation~\cite{fdsupdate}. The effect of missing 
$c$-in-the-sea is negligible in this case. 
A complete error budget is given in Table~\ref{tab:errorbudget}. 

The leptonic width is determined by the amplitude of the 
ground-state that dominates the correlator at large times. 
We can also determine the charm contribution to $R_{e^+e^-}$ through 
the time moments of the $J/\psi$ correlator which depend on 
the behaviour at short times. The moments are defined by:
\begin{equation}
G_n^{V} = Z^2C_n^{V} = Z^2\sum_{\tilde{t}} \tilde{t}^n \overline{C}_{J/\psi}(\tilde{t})
\label{eq:timemomv1}
\end{equation}
where $\tilde{t}$ is lattice time 
symmetrised around the centre of the lattice (see Appendix~\ref{appendix:z}). 
Results for $(G^V_n/Z^2)^{1/(n-2)}$ in lattice units 
on each of our ensembles are given in Table~\ref{tab:moments} 
for $n=4, 6, 8$ and 10. The power $1/(n-2)$ is taken to 
reduce all the moments to the same dimension. 
We take the $Z$ factor for the vector current to be the same one used 
for the leptonic width above, determined in Appendix~\ref{appendix:z}. 
Figure~\ref{fig:rn} then shows the physical results for
the moments as a function of lattice spacing.
The gray bands show our fits which use the same 
function of lattice spacing 
and sea quark masses as given in eq.~(\ref{eq:fithyp}). 
We reduce the prior width on the lattice spacing 
dependent terms by a factor of 4 because the moments 
are not as sensitive to short distances as the leptonic 
width or hyperfine splitting. 
 
The physical results that we obtain for each moment 
in the continuum limit are given by: 
\begin{eqnarray}
(G^V_4)^{1/2} &=& 0.3152(41)(9) \, \mathrm{GeV}^{-1} \nonumber \\
(G^V_6)^{1/4} &=& 0.6695(57)(13) \, \mathrm{GeV}^{-1} \nonumber \\
(G^V_8)^{1/6} &=& 0.9967(65)(10) \, \mathrm{GeV}^{-1} \nonumber \\
(G^V_{10})^{1/8} &=& 1.3050(65)(6) \, \mathrm{GeV}^{-1} .
\end{eqnarray}
The first error comes from the fit and the 
second allows for electromagnetism (e.g. photons in 
the final state) missing from our calculation but 
present in experiment. The error is estimated by 
substituting $\alpha_{QED}$ for $\alpha_s$ in the 
perturbative QCD analysis of the moments~\cite{kuhnmc07}. 
A complete error budget for our results is given in Table~\ref{tab:momerrs}. 

The results agree well with the values extracted for 
the $q^2$ derivative moments, $\mathcal{M}_k$ ($n=2k+2$), 
of the charm quark vacuum 
polarization using experimental values for 
$R_{e^+e^-} = \sigma(e^+e^- \rightarrow \mathrm{hadrons})/\sigma_{pt}$~\cite{kuhnmc07, hoangmc11}. 
The values, extracted from experiment by~\cite{kuhnmc07} and 
appropriately normalised for the comparison to ours, are:
\begin{eqnarray}
(M^{\mathrm{exp}}_1 4!/(12\pi^2 e_c^2))^{1/2} &=& 0.3142(22) \, \mathrm{GeV}^{-1} \nonumber \\
(M^{\mathrm{exp}}_2 6!/(12\pi^2e_c^2))^{1/4} &=& 0.6727(30) \, \mathrm{GeV}^{-1} \nonumber \\
(M^{\mathrm{exp}}_3 8!/(12\pi^2e_c^2))^{1/6} &=& 1.0008(34) \, \mathrm{GeV}^{-1} \nonumber \\
(M^{\mathrm{exp}}_{4} 10!/(12\pi^2e_c^2))^{1/8} &=& 1.3088(35) \, \mathrm{GeV}^{-1}. 
\label{eq:rnexp}
\end{eqnarray}
Our results from lattice QCD have approximately double 
the error of the experimental values but together these results 
provide a further test of QCD to better than 1.5\%. 

\begin{table}
\begin{tabular}{lllllllll}
\hline
\hline
Set & $m_ca$ & $\left(\frac{G^V_4}{Z^2a^2}\right)^{1/2}$ & $\left(\frac{G^V_6}{Z^2a^4}\right)^{1/4}$  & $\left(\frac{G^V_8}{Z^2a^6}\right)^{1/6}$ & $\left(\frac{G^V_{10}}{Z^2a^8}\right)^{1/8}$ \\
\hline
1 & 0.622 & 0.5399(1) & 1.2162(1) & 1.7732(1) & 2.2780(1) \\
\hline
2 & 0.63 & 0.5339(1) & 1.2054(1) & 1.7581(1) & 2.2584(1) \\
2 & 0.66 & 0.5135(1) & 1.1692(1) & 1.7081(1) & 2.1941(1) \\
\hline
3 & 0.617 & 0.5434(1) & 1.2223(1) & 1.7817(1) & 2.2888(1) \\
\hline
4 & 0.413 & 0.7586(1) & 1.6351(1) & 2.3887(2) & 3.0952(2) \\
\hline 
5 & 0.273 & 1.0681(1)  & 2.2705(2) & 3.3454(3)  & 4.3601(4) \\
\hline
6 & 0.193 & 1.4323(3) & 3.0397(5) & 4.4990(7) & 5.8738(8) \\
\hline
\hline
\end{tabular}
\caption{ Results in lattice units for time moments of the 
$J/\psi$ correlator as defined in eq.~(\ref{eq:timemomv1}).  We give 
results for $n$=4, 6, 8 and 10. 
 }
\label{tab:moments}
\end{table}

\begin{figure}
\begin{center}
\includegraphics[width=0.95\hsize]{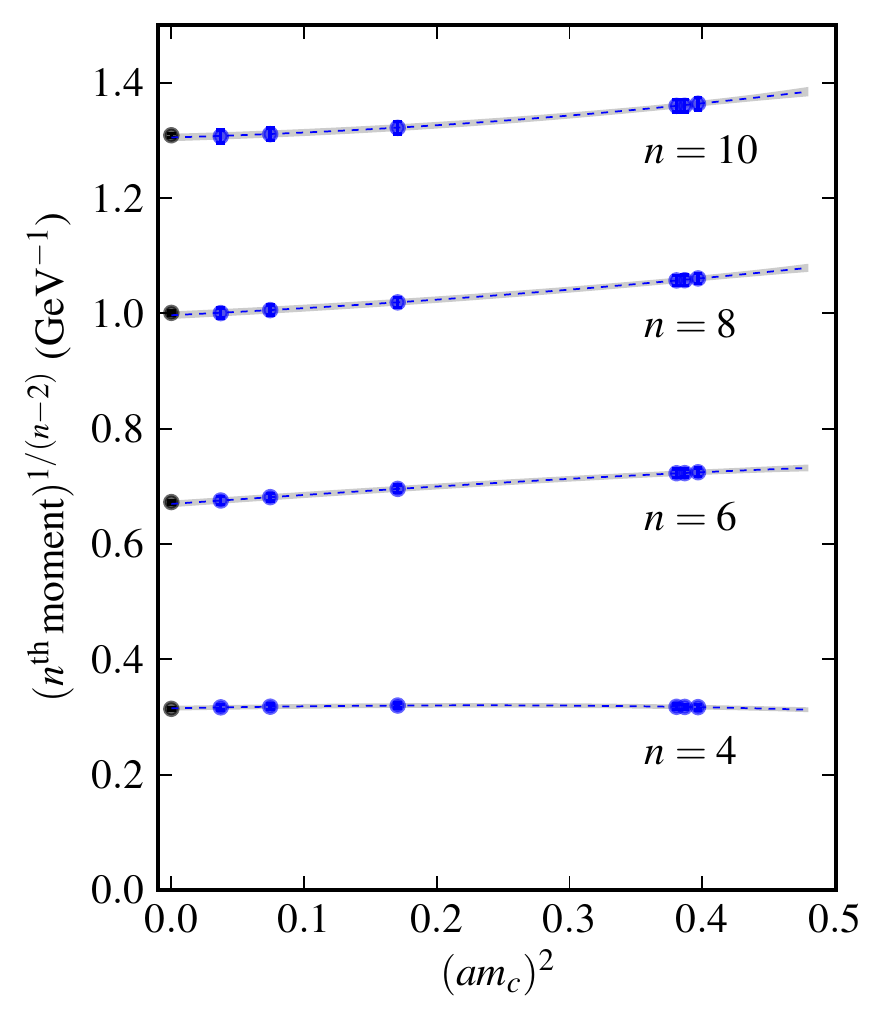}
\end{center}
\caption{Results for the 4th, 6th, 8th and 10th 
time moments of the charmonium vector correlator
shown as blue points and 
plotted as a function of lattice spacing. 
The errors shown (the same size or smaller than the points) 
include (and are dominated by) uncertainties from the 
determination of the current renormalization 
factor, $Z$, that are 
correlated between the points. 
The data points have been corrected for $c$ quark mass 
mistuning and sea quark mass effects, but the corrections 
are smaller than the error bars (the value for 
the deliberately mistuned $c$ mass on set 2 is not shown). 
The blue dashed line with 
grey error band displays our continuum/chiral fit. 
Experimental results determined from $R_{e^+e^-}$ (eq.~(\ref{eq:rnexp}))
are plotted as the black points 
at the origin 
offset slightly from the $y$-axis for clarity. 
}
\label{fig:rn}
\end{figure}

\begin{table}
\begin{tabular}{lcccc}
\hline
\hline
 &  $(G^V_4)^{1/2}$ & $(G^V_6)^{1/4}$ & $(G^V_8)^{1/6}$ & $(G^V_{10})^{1/8}$\\
\hline
$(am_c)^2$ extrapolation & 0.18 & 0.18 & 0.16 & 0.16\\
statistics & 0.05 & 0.04 & 0.03 & 0.03\\
lattice spacing & 0.32 & 0.51 & 0.43 & 0.30 \\
sea quark extrapolation & 0.14 & 0.13 & 0.12 & 0.12\\
$M_{\eta_c}$ tuning & 0.15 & 0.18 & 0.17 & 0.16 \\
Z & 1.23 & 0.61 & 0.41 & 0.31\\
electromagnetism & 0.3 & 0.2 & 0.1 & 0.05 \\
\hline 
Total (\%) & 1.3 & 0.9 &  0.7 & 0.5\\
\hline
\hline
\end{tabular}
\caption{Complete error budget for the time moments of 
the $J/\psi$ correlator 
as a percentage of the final 
answer. 
}
\label{tab:momerrs}
\end{table}

\subsection{$\Gamma(J/\psi \rightarrow \gamma \eta_c)$}
\label{sec:raddecay}

The radiative decay of the $J/\psi$ meson to the 
$\eta_c$ requires the emission of a photon from 
either the charm quark or antiquark and a spin-flip, 
so it is an M1 transition. 
Because it is sensitive to relativistic corrections 
this rate is hard to predict in nonrelativistic 
effective theories and potential models (see, for 
example,~\cite{eichtenrosner, qwg10})
Here we use a fully relativistic method in lattice 
QCD with a nonperturbatively determined current 
renormalisation and so none of these issues apply. 
In addition, of course, the lattice QCD result 
is free from model-dependence. 

The quantity that parameterises the nonperturbative 
QCD information (akin to the decay constant of the 
previous section) is the vector 
form factor, $V(q^2)$, where $q^2$ is the square 
of the 4-momentum transfer from $J/\psi$ to $\eta_c$. 
The form factor is related to the matrix element 
of the vector current between the two mesons by: 
\begin{equation}
\langle \eta_c(p^{\prime}) | \overline{c}\gamma^{\mu} c | J/\psi(p) \rangle = 
\frac{2V(q^2)}{(M_{J/\psi}+M_{\eta_c})} \varepsilon^{\mu \alpha \beta \gamma} p^{\prime}_{\alpha}p_{\beta} \epsilon_{J/\psi, \gamma}
\label{eq:ffdef}
\end{equation}
Note that the right-hand-side vanishes unless all the vectors are in 
different directions. 
Here we use a normalisation for $V(q^2)$ appropriate to a 
lattice QCD calculation in which the vector current is inserted 
in one $c$ quark line only and the quark electric charge ($2e/3$) 
is taken as a separate factor. 
The decay rate is then given by~\cite{dudekcharm}:
\begin{equation}
\Gamma(J/\psi \rightarrow \eta_c \gamma) = \alpha_{QED}\frac{64|\vec{q}|^3}{27(M_{\eta_c}+M_{J/\psi})^2} |V(0)|^2,
\label{eq:gammarad}
\end{equation}
where it is the form factor at $q^2=0$ that contributes because 
the real photon is massless. $|\vec{q}|$ is the corresponding momentum of the 
$\eta_c$ in the $J/\psi$ rest-frame.  

The most recent experimental result from CLEO-c~\cite{cleopsieta} of 
1.98(31)\% for the branching fraction, combined with the total 
width of the $J/\psi$ of 92.9(2.8) keV~\cite{pdg} gives  
\begin{equation}
V(0)_{\mathrm{expt}} = 1.63(14), 
\end{equation}
where we have used $\alpha_{QED} = 1/137$ and 
$|\vec{q}| = (M_{J/\psi}-M_{\eta_c})(M_{J/\psi}+M_{\eta_c})/(2M_{J/\psi})$.  
The value of $|\vec{q}|$ from experiment is 0.1137(11) GeV 
where the error comes from the uncertainty in the 
$\eta_c$ mass. 
$V(0)$ is then the quantity that can 
be calculated in lattice QCD and compared to experiment. 

\begin{table*}
\begin{tabular}{cccccccccccc}
\hline
\hline
Set & $N_{\mathrm{cfg}}\times N_t$ & $T$ values & $m_ca$ & $aM_{J/\psi}$ & $aM_{\eta_c}$ & $2\pi\theta$ & $aE^{\theta}_{\eta_c}$ & $V^{nn}_{00}$ & $V(0)/Z$ & $Z_{ff}$ & $a^2q^2$ \\
& & & & $\gamma_0\gamma_i \otimes \gamma_0\gamma_i\gamma_j$ & $\gamma_5 \otimes \gamma_5$ & & & & & \\
\hline
1 & $2088 \times 4$ & 15,18,21 & 0.622 & 1.86084(10) & 1.79116(4) & 1.6410 & 1.79243(4) & 0.0362(2) & 1.900(11) & 0.9896(11) & $1(4)\times 10^{-5}$ \\
\hline
2 & $2259 \times 4$ & 15,18,21 & 0.63 & 1.87972(12) & 1.80842(7) & 1.4007 & 1.81023(5) & 0.0368(2) & 1.897(12) & 0.9894(8) & $-7(1)\times 10^{-5}$ \\
 &  & 15,18,21 & 0.63 & 1.87962(14) & 1.80839(8) & 1.3880 & 1.81019(4) & 0.0362(4) & 1.883(20) & 0.9894(8) & $1(5)\times 10^{-5}$ \\
\hline
4 & $1911 \times 4$ & 20,23,26,29 & 0.413 & 1.32905(9) & 1.28046(3) & 1.3327 & 1.28133(3) & 0.0348(2) & 1.876(8) & 1.0049(10) & $6(4)\times 10^{-5}$ \\

\hline
\hline
& & & & $\gamma_i \otimes \gamma_i\gamma_j$ & $\gamma_0\gamma_5 \otimes \gamma_0\gamma_5$ & & & & & \\
\hline
1 & $2088 \times 4$ & 15,18,21 & 0.622 & 1.86035(15) & 1.79621(4) & 1.5120 & 1.79725(4) & 0.0338(6) & 1.925(35) & 0.9896(11) & $3(5)\times 10^{-5}$ \\
\hline
2 & $2259 \times 4$ & 15,18,21 & 0.63 & 1.87887(13) & 1.81369(6) & 1.2814 & 1.81480(5) & 0.0334(8)  & 1.896(45) & 0.9894(8) & $0(4)\times 10^{-5}$ \\
 &  & 15,18,21 & 0.66 & 1.93604(15) & 1.87254(6) & 1.2490 & 1.87355(6) & 0.0322(8) & 1.934(42) & 0.9863(17) & $1(5)\times 10^{-5}$ \\
\hline
4 & $1911 \times 4$ & 19,20,23,26 & 0.413 & 1.32904(11) & 1.28160(4) & 1.3116 & 1.28243(4) & 0.0342(4) & 1.872(21) & 1.0049(10) & $-2(1)\times 10^{-5}$ \\
\hline 
\hline
\end{tabular}
\caption{ Results from simultaneous fits for 3-point and 2-point correlators 
for $J/\psi \rightarrow \gamma \eta_c$ decay. The upper table gives results from our 
preferred jpsigamma0 method; the lower table from etacgamma0. See the 
text for a definition of the two methods. Column 2 gives the number of 
configurations and time sources for $0$ on each configuration. Column 3 gives 
the different values of the end-point of the 3-point function, $T$, included 
in the fit. The lattice $c$ quark mass and $\epsilon$ parameter are 
the same as those used in section~\ref{sec:psimass} (the lower table 
includes the deliberately mistuned mass on set 2 for comparison). $aM_{J/\psi}$ and 
$aM_{\eta_c}$ are the zero-momentum meson masses for the tastes of $J/\psi$ and 
$\eta_c$ mesons used here. $2\pi\theta$ indicates the value of the phase 
at the boundary used to achieve the kinematics of $q^2=0$ in 
the $J/\psi \rightarrow \eta_c$ decay. The $a^2q^2$ values actually obtained 
with those kinematics are given in the final column (rows 2 and 3 of the 
upper table compare two different values of $a^2q^2$ close to zero). 
$aE_{\eta_c}$ gives 
the energy of the $\eta_c$ at the value of the spatial momentum corresponding 
to $\theta$. $V_{nn}^{00}$ from the 3-point fit of eq~(\ref{eq:3ptfit}) is 
given in column 9 and 
this is converted to a value of $V(0)/Z$ in column 10 using 
eq.~(\ref{eq:getff}). Column 
11 gives the values of the renormalisation parameter, $Z$, obtained from 
the vector form factor method of Appendix~\ref{appendix:vecff}.    
}
\label{tab:3ptres}
\end{table*}

\begin{figure}
\begin{center}
\includegraphics[width=0.9\hsize]{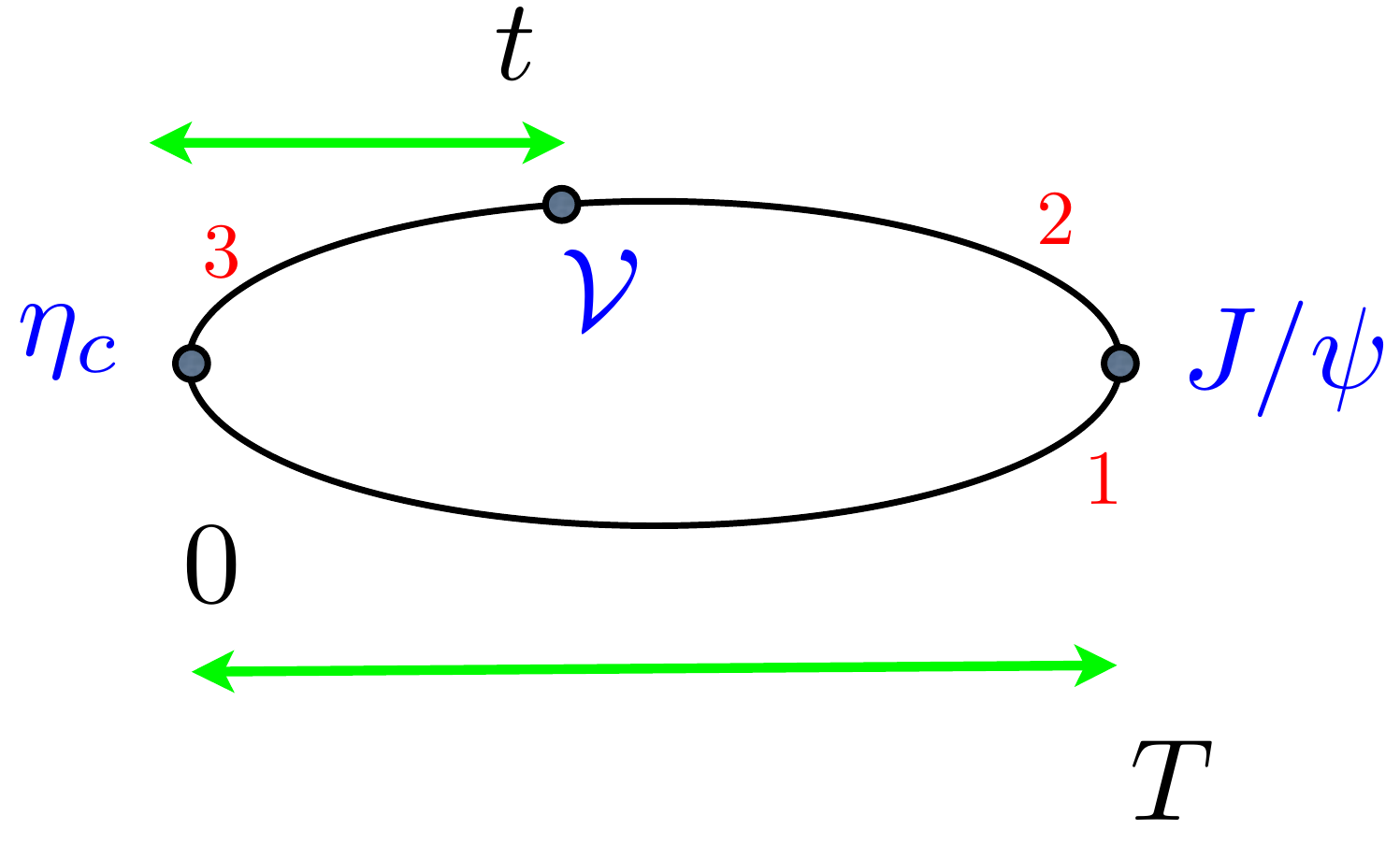}
\end{center}
\caption{ A schematic diagram of the connected `3-point' 
function in lattice QCD for $J/\psi$ to $\eta_c$ 
radiative decay. The lines all represent $c$ quark 
propagators in this case. The propagator labelled 
1 is the spectator quark; 2 and 3 are the initial 
and final active quarks respectively. 
$0$ and $T$ label the position in 
time of the $\eta_c$ and $J/\psi$ operators. 
The vector current is inserted at time $t$  which 
takes all values between $0$ and $T$. 
}
\label{fig:3ptpic}
\end{figure}

The radiative decay of the $J/\psi$ to $\eta_c$ meson 
needs the calculation of a `3-point' function in 
lattice QCD.  The 3 points (in lattice time) 
correspond to: the position 
of the $\eta_c$ operator, which we take as the origin; 
the position of the $J/\psi$ operator which we denote 
$T$ and the position of the insertion 
of a vector operator, 
$\mathcal{V}=\overline{c}\gamma_{\mu}c$, which couples to the photon 
at time $t$. $t$ varies from $0$ to $T$. Sums over 
spatial points are implied at each time. 
The `connected' correlator that we calculate 
is illustrated in Figure~\ref{fig:3ptpic}. 
Disconnected correlators are expected to be 
negligible here based on perturbative and 
phenomenological arguments~\cite{dudekcharm} and we 
do not include them.

The 3-point function is calculated in lattice QCD 
by combining 3 quark propagators together with 
appropriate spin projection matrices. As discussed 
in section~\ref{sec:latt} for staggered quarks 
these $\gamma$ matrices become $\pm 1$ phases. 
Tastes must be combined in a staggered quark correlator 
so that the overall correlation function is 
`tasteless'. What this means for a 3-point function is 
that only certain taste combinations of $J/\psi$, 
$\eta_c$ and $\mathcal{V}$ operators are allowed.  To optimise 
statistical errors we need to keep to a minimum the 
amount of point-splitting in the operators. It is also convenient, 
for renormalisation purposes, to have a vector current, 
$\mathcal{V}$, which corresponds to a local operator (and this 
is also what we used for the decay constant in 
section~\ref{sec:leptwidth}). 

We therefore choose the $\eta_c$ operator to be the 
local $\gamma_5$ operator (so that the $\eta_c$ 
is the Goldstone pseudoscalar with spin-taste 
$\gamma_5 \otimes \gamma_5$) and the 
$J/\psi$ operator to be a one-link separated 
$\gamma_0\gamma_i$ operator in which the polarisation of 
the $J/\psi$ and the one-link separation are both 
in an orthogonal spatial direction to the polarisation 
of the vector current, $\mathcal{V}=\overline{c}\gamma_k c$ 
(this $J/\psi$ has spin-taste structure 
$\gamma_0\gamma_i \otimes \gamma_0\gamma_i\gamma_j$). 

To implement this configuration is simple. 
The spectator quark propagator 
(number 1 in Figure~\ref{fig:3ptpic}) 
is generated from the default random wall 
at time $0$. 
Active propagator 2 is then generated from a 
source which is made from a symmetric point-splitting 
of propagator 1 at time $T$ patterned by a phase. 
For a $J/\psi$ with polarisation $x$ we 
take a point-splitting in the $y$ direction and 
phase $(-1)^{x+z}$. 
Active propagator 3 is made from the same default 
random wall as 1. 
Finally 2 and 3 are combined together at $t$ by 
summing over space with a patterning of $(-1)^z$ 
to implement a local vector current in the $z$ direction. 

To achieve the configuration corresponding 
to $q^2=0$ we keep the $J/\psi$ at rest in the 
frame of the lattice and give the $\eta_c$ an 
appropriate spatial momentum. 
The $\eta_c$ momentum 
is implemented by calculating propagator 3 with 
a `twisted boundary condition'~\cite{firsttwist,etmctwist}. 
If propagator 3 is calculated with boundary condition: 
\begin{equation}
\chi(x + \hat{e}_j L) = e^{2\pi i \theta_j} \chi(x),
\label{eq:twist}
\end{equation}
then the momentum of the $\eta_c$ meson made 
by combining propagators 1 and 3 with our random wall 
sources and summing over spatial sites at the sink is:
\begin{equation}
p_j = \frac{2\pi}{L_s} \theta_j.  
\label{eq:momtwist}
\end{equation}
The boundary condition in eq.~(\ref{eq:twist}) is 
actually implemented by multiplying the gluon links 
in the $j$ direction by phase $\exp(2\pi i \theta_j/L_s)$. 
We take $j$ to be the $y$ direction here so that 
the momentum is in an orthogonal direction to the 
polarisation of both the $J/\psi$ and $\mathcal{V}$. 

The 3-point function is then given by: 
\begin{eqnarray}
&& C_{3pt}(0,t,T) = \sum_{s_T,s_t}\frac{1}{4}(-1)^{x_T+z_T}(-1)^{z_t} \times \\
&&\mathrm{Tr}\left\{g(t,T)[g(T+1_y,0)+g(T-1_y,0)]g^{\dag}_{\theta}(t,0)\right\} \nonumber
\label{eq:3pt}
\end{eqnarray}
where $g$ represent staggered $c$ quark propagators, 
with $g_{\theta}$ computed with a phase on the gluon 
field,
the trace is over color and sums are done over spatial 
sites $s_t$ and $s_T$ at $t$ and $T$.
The $1/4$ is the taste factor for the normalisation of a 
staggered quark loop. 
The corresponding 2-point function for
the $\eta_c$ meson is 
\begin{equation}
C_{\eta_c,2pt}(0,t) = \sum_{s_t} \frac{1}{4} \times 
\mathrm{Tr}\left[g(t,0)g^{\dag}_{\theta}(t,0)\right].
\label{eq:2pt}
\end{equation}
The 2-point function for the $J/\psi$ is given by
\begin{eqnarray}
&&C_{J/\psi,2pt}(0,t) = \sum_{s_t} \frac{1}{4}(-1)^{y_0+t_0}(-1)^{y_t+t_t} \times  \\
&&\mathrm{Tr}\left[g(t,0)(g^{\dag}(t+1_y,1_y)+g^{\dag}(t-1_y,1_y)+ \{1 \leftrightarrow -1 \})\right]. \nonumber
\label{eq:2ptpsi}
\end{eqnarray}

As an alternative configuration we can take 
the $\eta_c$ operator to be the local 
$\gamma_0\gamma_5$ operator (so that the $\eta_c$ is 
the local non-Goldstone meson with spin-taste 
structure $\gamma_0\gamma_5 \otimes \gamma_0\gamma_5$) 
and the $J/\psi$ 
operator to be a one-link separated $\gamma_i$ operator in 
which the polarisation of the $J/\psi$ and the 
one-link separation are both in an orthogonal spatial 
direction to the polarisation of the vector current, $\mathcal{V}$ 
(this has spin-taste structure $\gamma_i \otimes \gamma_i\gamma_j$). 
The 3-point function is then given by: 
\begin{eqnarray}
&&C_{3pt}(0,t,T) = \sum_{s_T,s_t}\frac{1}{4}(-1)^{x_0+y_0+z_0}(-1)^{y_T}(-1)^{z_t} \times \nonumber \\
&&\mathrm{Tr}\left[g(t,T)(g(T+1_y,0)+g(T-1_y,0))g^{\dag}_{\theta}(t,0)\right] 
\label{eq:3ptalt}
\end{eqnarray}
and the corresponding 2-point functions are:
\begin{eqnarray}
C_{\eta_c,2pt}(0,t) &=& \sum_{s_t} \frac{1}{4}(-1)^{x_0+y_0+z_0}(-1)^{x_t+y_t+z_t} \times \nonumber \\
&&\mathrm{Tr}\left[g(t,0)g^{\dag}_{\theta}(t,0)\right].
\label{eq:2ptalt}
\end{eqnarray}
and
\begin{eqnarray}
&& C_{J/\psi,2pt}(0,t) = \sum_{s_t} \frac{1}{4}(-1)^{x_0+z_0+t_0}(-1)^{x_t+z_t+t_t} \times  \\
&&\mathrm{Tr}\left[g(t,0)(g^{\dag}(t+1_y,1_y)+g^{\dag}(t-1_y,1_y)+ \{1 \leftrightarrow -1 \})\right]. \nonumber
\label{eq:2ptpsialt}
\end{eqnarray}
We call this configuration the `etacgamma0' configuration and 
the original configuration of eq.~(\ref{eq:3pt}) the 
`jpsigamma0' configuration. In fact, as we shall see, the 
jpsigamma0 configuration is to be preferred on the basis 
of statistical errors but the results agree between the two.  

\begin{figure}
\begin{center}
\includegraphics[width=0.9\hsize]{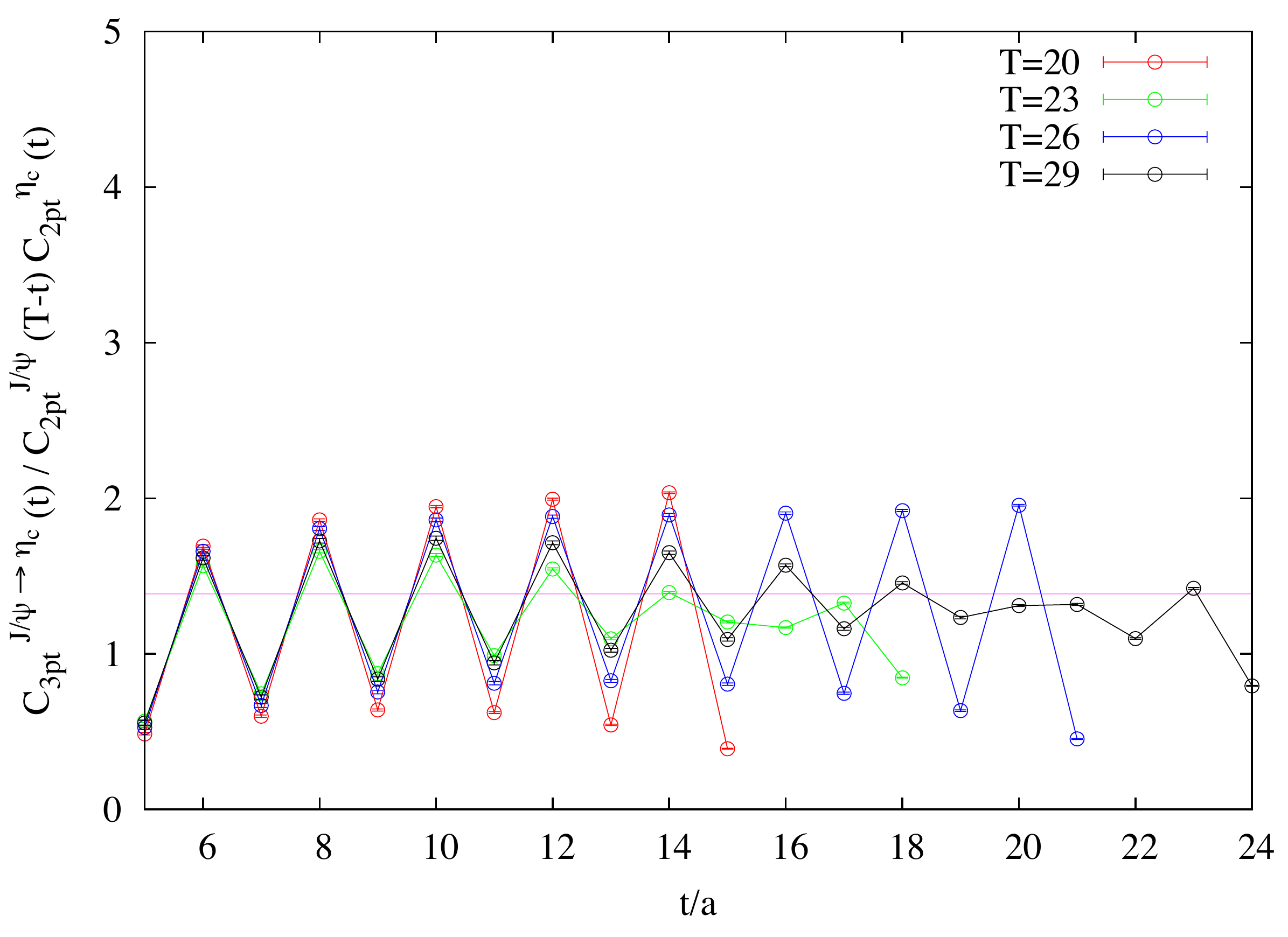}
\includegraphics[width=0.9\hsize]{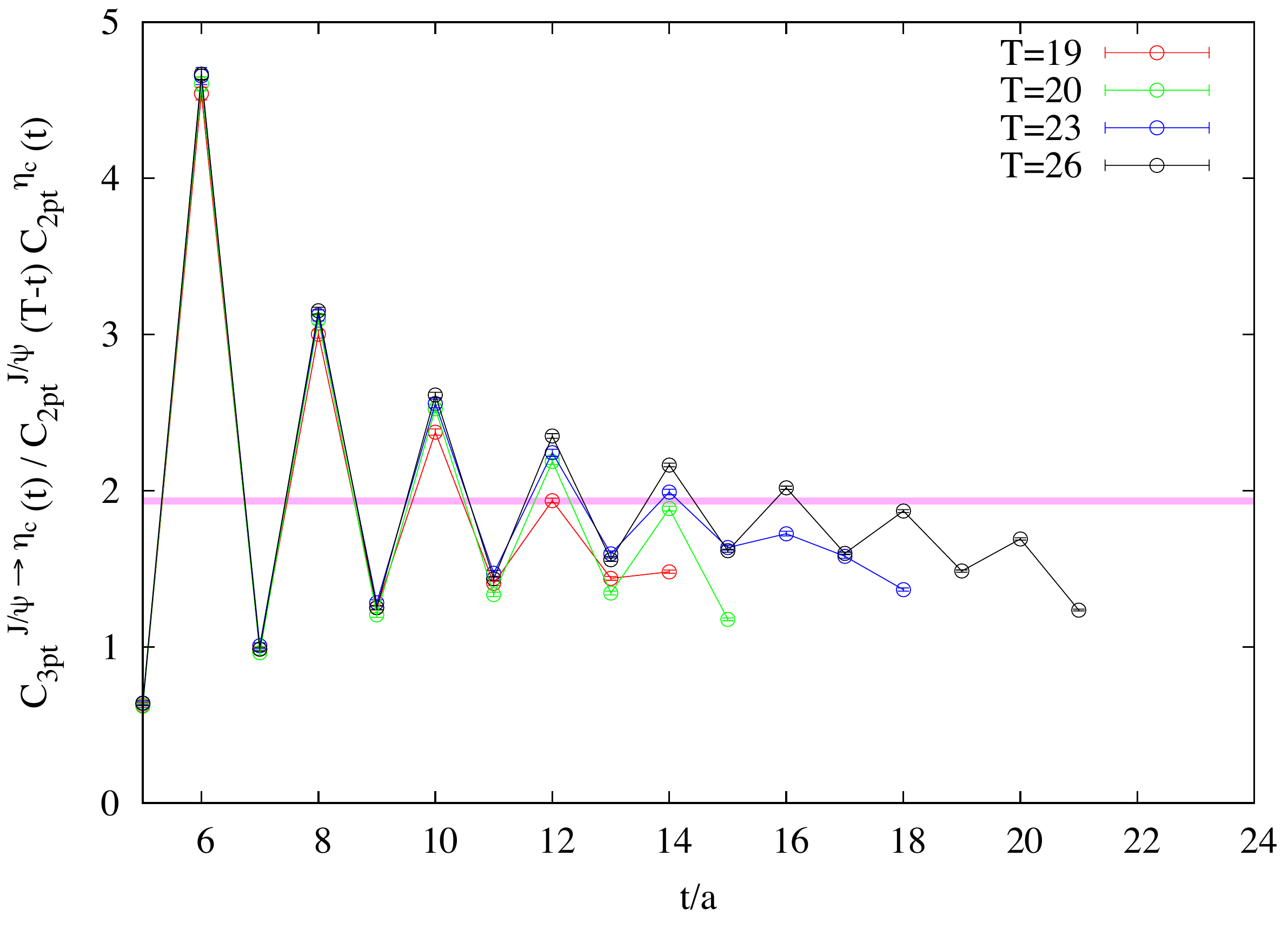}
\end{center}
\caption{ A plot of the 
ratio $C_{3pt}(t,T) / (C_{2pt, \eta_c}(t) C_{2pt, J/\psi}(T-t))$ 
for the three values of $T$ on the fine ensemble, set 4 
and for our two different methods.
Lines join the points (which have statistical errors on 
them) for clarity. We only include points in the central 
region of $t$ i.e. 
$t \geq 5$ or $t \leq T-5$. 
The pink shaded band shows the ratio of fit 
parameters ${V^{nn}_{00}}/{a_0 b_0}$ which is the 
ground-state contribution to this ratio. 
These come from a fit that included 6 normal exponentials 
and 5 oscillating ones (which produce the oscillations 
evident in the figure). 
The top plot shows the results for the case where 
$\gamma_0$ is included in the $J/\psi$ operator (jpsigamma0 method) and 
the lower plot shows the results for the 
case where $\gamma_0$ is included in the $\eta_c$
operator (etacgamma0 method). 
}
\label{fig:3pt2ptratio}
\end{figure}

The 3-point function in both cases is calculated along with the 
2-point functions for the $\eta_c$ and $J/\psi$ mesons
that appear in it. We use multiple time-sources 
for point $0$ on each configuration and also multiple 
values for $T$. 
Figure~\ref{fig:3pt2ptratio} shows results for the 3-point 
function on fine set 4, normalising it to the product of the relevant 
2-point functions. The two plots compare results for 
the jpsigamma0 method and the etacgamma0 method.
The two differ in the amount of oscillation that is seen at 
the two ends of the plot. Not surprisingly the jpsigamma0 
method shows more oscillation on the $J/\psi$ end ($t$ near $T$) 
since the $\eta_c$ in this case  would not oscillate at rest. 
The etacgamma0 method has relatively large oscillations for 
the $\eta_c$ side but smaller oscillations on the 
$J/\psi$ side. In both cases statistical errors are very small 
enabling us to fit both normal and oscillating terms.  
The figure also shows how having multiple $T$
values improves our determination of the ground-state 
transition amplitude. 

We fit the 3-point function 
and 2-point functions simultaneously to a multi-exponential 
that determines the ground-state amplitudes accurately 
because it includes excited state contributions. 
The fit form for the 2-point functions was already given 
in eq.~(\ref{eq:fit1}).  Here both the $J/\psi$ and 
$\eta_c$ correlators have oscillating contributions and, 
in the $\eta_c$ case, the exponent gives the energy of 
the meson at momentum $p_j$ (eq.~(\ref{eq:momtwist}))
rather than the mass. 
The fit form for the 3-point function is then: 
\begin{eqnarray}
&&C_{3pt}(t,T) = \\
&& \sum_{i_n,j_n} a_{i_n} \mathrm{fn}(E_{a,i_n},t) V^{nn}_{i_n,j_n} b_{j_n} \mathrm{fn}(E_{b,j_n},T-t) \nonumber \\
&& -\sum_{i_n,j_o} a_{i_n} \mathrm{fn}(E_{a,i_n},t) V^{no}_{i_n,j_o} \tilde{b}_{j_o} \mathrm{fo}(\tilde{E}_{b,j_o},T-t) \nonumber \\
&& -\sum_{i_o,j_n} \tilde{a}_{i_o} \mathrm{fo}(\tilde{E}_{a,i_o},t) V^{on}_{i_o,j_n} b_{j_n} \mathrm{fn}(E_{b,j_n},T-t) \nonumber \\
&& +\sum_{i_o,j_o} \tilde{a}_{i_o} \mathrm{fo}(\tilde{E}_{a,i_o},t) V^{oo}_{i_o,j_o} \tilde{b}_{j_o} \mathrm{fo}(\tilde{E}_{b,j_o},T-t) \nonumber 
\label{eq:3ptfit}
\end{eqnarray}
and, again: 
\begin{eqnarray}
\mathrm{fn}(E,t) &=& e^{-Et} + e^{-E(L_t-t)} \nonumber \\
\mathrm{fo}(E,t) &=& (-1)^{t/a} \mathrm{fn}(E,t) 
\end{eqnarray}
with $L_t$ again the time extent of the lattice. 
Here $n$ denotes the normal contributions and 
$o$ the contributions from oscillating states. 
The ground-state energies/masses that we need 
are $E_{\eta_c,0}$ and $E_{J/\psi,0}=M_{J/\psi}$ 
and the matrix element between them that is 
proportional to $V^{nn}_{0,0}$. By matching to 
a continuum correlator with a relativistic 
normalisation of states and allowing for a 
renormalisation of the lattice vector current we see that:
\begin{equation}
\langle \eta_c | V | J/\psi \rangle = 2 Z \sqrt{M_{J/\psi}E_{\eta_c}} V^{nn}_{0,0}.
\label{eq:fixv}
\end{equation}
The vector form factor that we need, $V(0)$, is then, from eq.~(\ref{eq:ffdef}), given by: 
\begin{equation}
\frac{V(0)}{Z} = \frac{M_{J/\psi}+M_{\eta_c}}{2M_{J/\psi}p_j} 2 \sqrt{M_{J/\psi}E_{\eta_c}} V^{nn}_{0,0} ,
\label{eq:getff}
\end{equation}
with $p_j$ from eq.~(\ref{eq:momtwist}). The determination of 
$Z$ will be discussed below. 

To perform the joint fit to the 3pt correlators 
using eq.~(\ref{eq:3ptfit}) and the 2pt correlators 
using eq.(\ref{eq:fit1}) we use the same approach 
as outlined in section~\ref{sec:psimass}.  
For both the $\eta_c$ and $J/\psi$, the prior for 
the ground-state mass comes from the effective mass of the correlator.
We use priors of 600 MeV with a width of 300 MeV for 
the difference in mass between the ground state and the lowest oscillating 
mass and between all radial excitations, both normal and oscillating. 
The 2-point amplitudes, $a_i$ and $b_i$, have prior widths of 0.5 
and the 3-point amplitudes, $V_{ij}$, have widths of 0.25.
We omit $t$ values below a certain $t_{\mathrm{min}}$ 
to reduce the effect of excited states. 
$t_{\mathrm{min}} = 4(5)$ for the coarse(fine) lattices 
for the etacgamm0 method and 6 
for the jpsigamm0 method. 

Table~\ref{tab:3ptres} gives our results from fits that 
include 6 normal exponentials and 5 oscillating. We work 
on ensembles 1, 2 and 4 of Table~\ref{tab:params} 
but using more configurations than in section~\ref{sec:psimass} 
to reduce statistical errors. Table~\ref{tab:3ptres} 
gives the number of configurations and time sources 
as well as the values of $T$ used in the 3-point functions. 
It is important to use both even and odd values of $T$ 
to separate clearly the normal and oscillating contributions. 
Having determined the mass of the local non-Goldstone $\eta_c$ 
and 1-link vector 
from separate 2-point function fits we then determine 
the value of $\theta$ needed to achieve $q^2=0$.  
The final fits are done as a simultaneous fit to the 
3-point function and 2-point functions for zero 
momentum and finite momentum $\eta_c$ and zero 
momentum $J/\psi$. 

The key parameters to be determined 
from the fit, as discussed above, are the ground-state masses of the 
$\eta_c$ and $J/\psi$, the ground state energy 
at non-zero momentum of the $\eta_c$ and the 
ground-state to ground-state amplitude of the 3-point function. 
Our fit returns excited state to ground-state and 
oscillating to ground state amplitudes also. Most 
of these do not have a significant signal. Indeed 
the excited state to ground state amplitudes 
are very small, as expected since they correspond 
to a hindered M1 transition. A non-zero result is 
seen for the transition between the oscillating 
partner of the $\eta_c$ (in the etacgamma0 method) 
and the $J/\psi$. This 
corresponds to the E1 $\chi_{c0}$ to $J/\psi$ decay, 
but not at the correct kinematics for that decay.  
Likewise a signal is seen for E1 $h_c \rightarrow \eta_c$ 
decay in jpsigamma0 method.  
We will discuss these transitions further elsewhere. 

\begin{figure}
\begin{center}
\includegraphics[width=0.9\hsize]{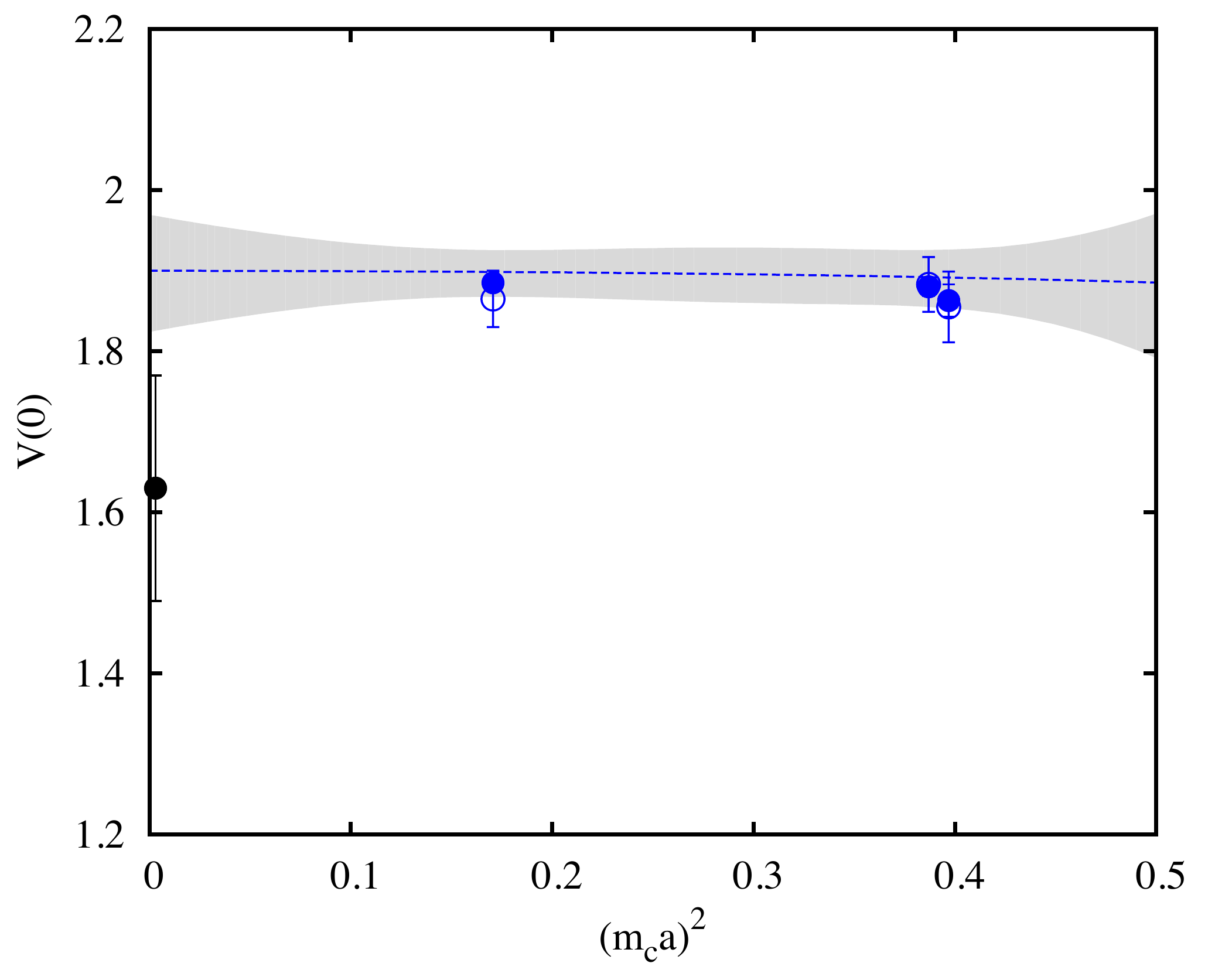}
\end{center}
\caption{Results for the vector form 
factor at $q^2=0$ for $J/\psi \rightarrow \eta_c$ 
decay 
plotted as a function of lattice spacing. 
The filled blue circles are from our preferred 
jpsigamma0 method; the open blue circles are 
from the etacgamma0 method.  
For the $x$-axis we use $(m_ca)^2$ to allow  the $a$-dependence 
of our fit 
function to be displayed simply 
(blue dashed line and grey band). 
The fit is to results from the jpsigamma0 method. 
The errors shown 
include statistical errors and errors from the 
$Z$ factor. 
The experimental result extracted from the branching 
fraction for $J/\psi \rightarrow \gamma \eta_c$ is plotted as the black point 
offset slightly from the origin for clarity. 
}
\label{fig:v0}
\end{figure}

From eq.~(\ref{eq:getff}) we can determine 
$V(0)$ given a value for $V^{nn}_{00}$ and 
a renormalisation factor, $Z$. For $Z$ we use 
the fully nonperturbative vector form factor 
method described in Appendix~\ref{appendix:vecff}
which normalises the local charm-charm vector current 
that we are using here by demanding that 
its form factor is 1 between identical 
mesons at $q^2=0$. This requires a 
non-staggered spectator quark and we 
use NRQCD for this. The determination of $Z_{ff}$ 
then needs the calculation of the form 
factor of the temporal component of the 
vector current between two $B_c$-like mesons (the 
mesons do not have to be real $B_c$ mesons) at rest. 
$Z_{ff}$ can be determined with a statistical uncertainty 
of 0.1\% this way. Details are given in Appendix~\ref{appendix:vecff}.   

The values for $Z$ are given in Table~\ref{tab:zvtable} 
of Appendix~\ref{appendix:vecff}
and the values we use here are reproduced in Table~\ref{tab:3ptres}
along with our results for $V^{nn}_{00}$, $V(0)/Z$ and 
the $\eta_c$ and $J/\psi$ masses and energies. 
The table is divided into two with the upper results from 
the jpsigamma0 method and the lower results from the 
etacgamma0 method. The two methods give results for $V(0)/Z$ 
in good agreement, but the jpsigamma0 results are statistically 
more accurate. This is then our preferred method and the one 
that we 
will use for our final result. 
The agreement between the two methods to within the 2\% 
statistical errors is a strong test of the control 
of discretisation errors in the HISQ formalism. 

Table~\ref{tab:zvtable} also gives results that allow us 
to test to what extent $V(0)$ depends on $m_c$ and the 
precise tuning of $q^2$ to zero. 
On set 2 we have deliberately mistuned the $c$ quark 
mass by 5\% and see that it makes no significant difference to 
$V(0)$ within our 2\% statistical errors. 
$q^2$ is tuned to zero typically within our statistical 
errors of $(10\mathrm{MeV})^2$. 
On set 2 comparison between two different values of 
$q^2$ shows no effect within our 1\% statistical errors. 
We use the value closest to $q^2=0$ in our fits below. 
These are both good tests of the robustness of our 
results to the tuning of parameters. 
 
Figure~\ref{fig:v0} shows our results for $V(0)$ 
plotted as a function of the lattice spacing. 
To determine the physical value 
we use a fit similar to that for the hyperfine splitting 
and leptonic decay constant 
given in eq.~\ref{eq:fithyp}. We simplify the fit slightly 
in dropping the tuning for the physical $c$ mass since 
our results in Table~\ref{tab:3ptres} show negligible
dependence on the $c$ quark mass. 
We take the prior on the physical value to be 2.0(0.5) 
and allow for terms in $(m_ca)^{2i}$ up to $i=5$. We take the prior 
on the leading $(m_ca)^2$ term to be 0.0(3) since tree-level 
$a^2$ errors are removed in the HISQ action. We take 
linear and quadratic terms in $2\delta x_l+\delta x_s$ 
and allow $a^2$ dependence multiplying the linear term. 

The physical value for $V(0)$ from the fit is 1.90(7) from 
the jpsigamma0 method. The etacgamma0 method gives a result 
in good agreement with a very similar error.  
The error is dominated by that from the extrapolation in 
the lattice spacing. In fact there is no visible lattice 
spacing dependence in our results and it could be argued 
that, in a 
transition from $J/\psi$ to $\eta_c$ that probes relatively 
low momenta, the 
relevant scale for discretisation errors is well 
below $m_c$. However, to be conservative, we 
allow discretisation errors 
to depend on $(m_ca)^2$ and allow for multiple powers 
to appear.

We have also tested extrapolations of $V(0)$ to 
the physical point using alternative definitions 
of the renormalisation of the current. 
We get the same answer using $Z_{ff}$ values taken 
from $B_c \rightarrow B_c$ form factors with a 
heavier $b$ quark mass, as given in Appendix~\ref{appendix:vecff}. 
We also get a result in good agreement if 
we use values for $Z$ from $Z_{cc}$ given 
in Appendix~\ref{appendix:currcurr}. 

Our physical result for $V(0)$ is for a world 
that does not include electromagnetism, 
$c$-in-the-sea or allow for $\eta_c$ annihilation. The effect 
of missing electromagnetism is similar to that for the 
decay constant and so we allow the same additional systematic error 
of 0.5\%. We expect $c$-in the sea effects to be negligible, 
as for the decay constant. 
$\eta_c$ annihilation 
affects the mass difference between the $J/\psi$ and $\eta_c$ 
(as discussed in section~\ref{sec:psimass})
and therefore affects the momentum of the $\eta_c$ that 
corresponds to $q^2=0$ for this decay. Equivalently it means  
that the real $q^2=0$ point corresponds to a non-zero 
$q^2$ in our calculation. Since we allow an uncertainty in 
the $\eta_c$ mass of 2.4 MeV (Table~\ref{tab:errorbudget}) 
this corresponds to an uncertainty around $q^2=0$ 
of $6 \times 10^{-6} \mathrm{GeV}^2$, keeping the spatial 
momentum fixed. From Table~\ref{tab:3ptres} we see that this 
would produce a negligible change in $V(0)$, not visible 
beneath our statistical errors. In addition we can use 
information from~\cite{dudekcharm} which used results 
at different $q^2$ values to extrapolate to $q^2=0$ albeit
in the quenched approximation. The $q^2$ dependence gave 
a change in $V(q^2)$ from $V(0)$ of 20\% when 
$q^2$ was $1\mathrm{GeV}^2$. From this it is clear that 
effects from a slight mistuning because of $\eta_c$ 
annihilation effects should be completely negligible.   
We take as our final result then: 
\begin{equation}
V(0) = 1.90(7)(1). 
\label{eq:v0give}
\end{equation}
The complete error budget is given in Table~\ref{tab:errorbudget}. 

\section{Discussion}
\label{sec:discussion}

\begin{figure}
\begin{center}
\includegraphics[width=0.9\hsize]{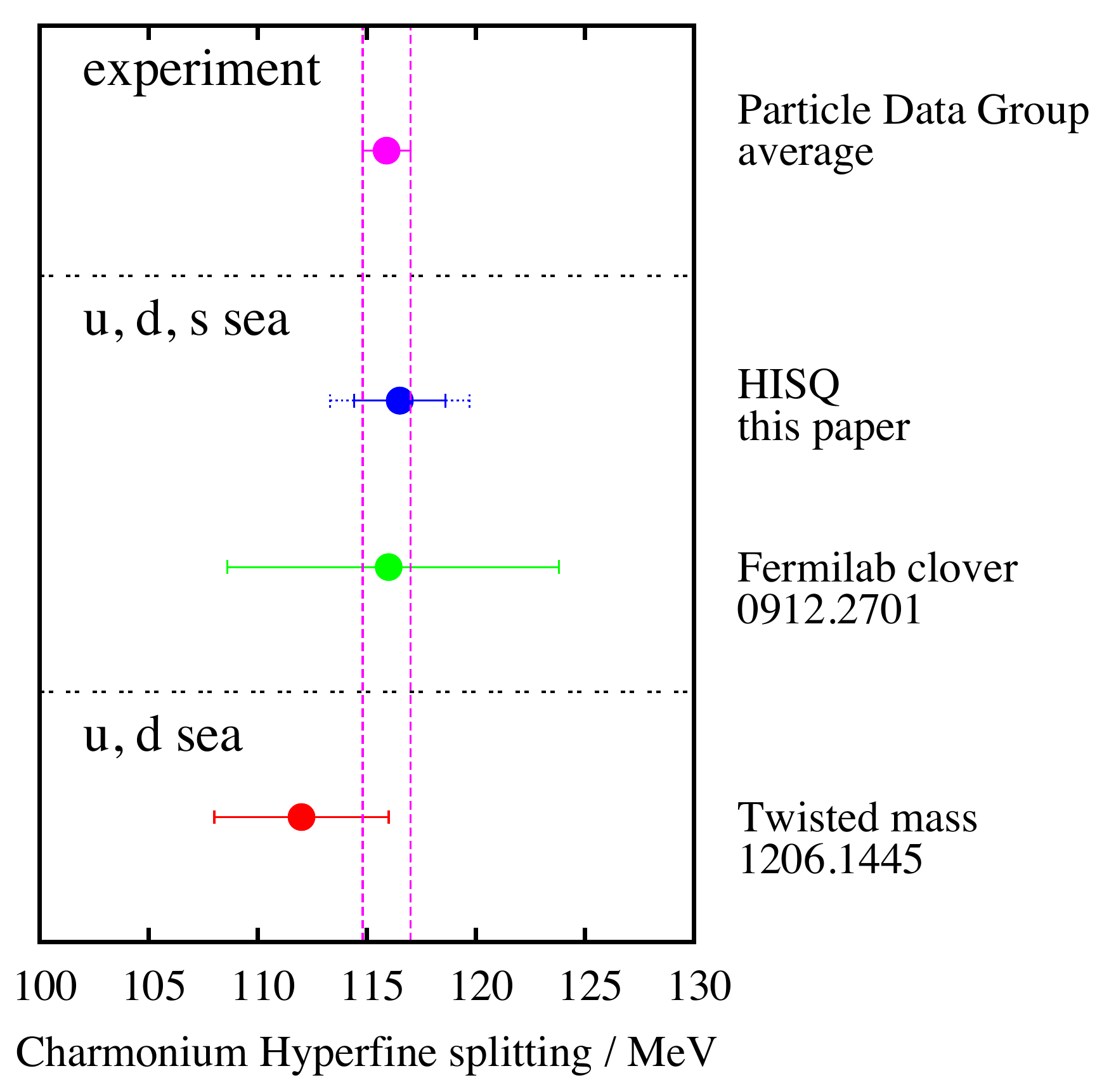}
\end{center}
\caption{ A comparison of results 
for the charmonium hyperfine splitting from 
lattice QCD with experiment.
We show only results that include sea quarks 
and make use of multiple 
lattice spacings to derive a continuum value. 
The experimental average~\cite{pdg} is given 
at the top, followed by the result for HISQ 
quarks from this paper. The Fermilab clover~\cite{fnalcharm} 
and twisted mass~\cite{etmccharm} results follow. 
Neither of 
these lower two results include an error for 
missing $\eta_c$ annihilation effects. 
This error is the 
dominant error for our calculation. 
Here we show our error bar excluding this effect 
as a solid line and the total error including 
this effect as a dotted line. 
}
\label{fig:hypall}
\end{figure}

Figure~\ref{fig:hypall} compares our result for 
the charmonium hyperfine splitting to experiment and 
to that from other lattice QCD calculations. 
We only show results that have been obtained including 
sea quark effects and making use of multiple lattice 
spacing values to derive a physical continuum result. 
Values are also given for different forms of the 
clover action in~\cite{ryancharm, saracharm, pacscharm, mohlercharm} 
but either at only one value of the lattice spacing 
or without giving a value from continuum extrapolation. 
Some of these latter calculations obtain values well below experiment 
because of the large discretisation errors, particularly 
for the hyperfine interaction, in the clover formalism. 

Our result agrees well with experiment and is more accurate 
than earlier values, especially since earlier values do 
not generally include any error for missing $\eta_c$ 
annihilation effects. 

\begin{figure}
\begin{center}
\includegraphics[width=0.9\hsize]{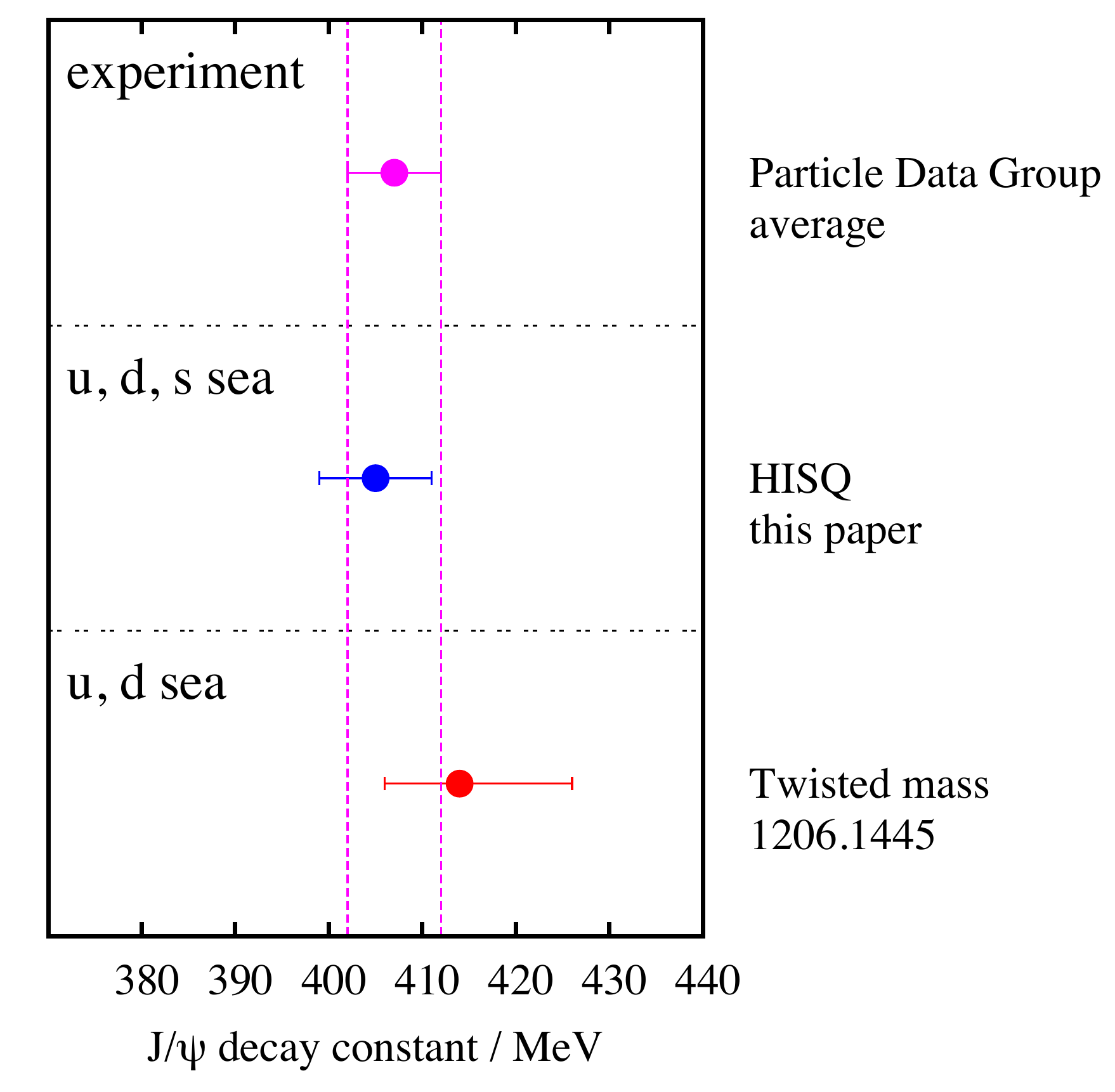}
\end{center}
\caption{ A comparison of results 
for the decay constant of the $J/\psi$ from 
lattice QCD with experiment.
We include only results that include sea quarks 
and make use of multiple 
lattice spacings to derive a continuum value. 
The experimental average~\cite{pdg} is given 
at the top, followed by the result for HISQ 
quarks from this paper. The 
twisted mass~\cite{etmccharm} results follow. 
}
\label{fig:fpsiall}
\end{figure}

\begin{figure}
\begin{center}
\includegraphics[width=0.8\hsize]{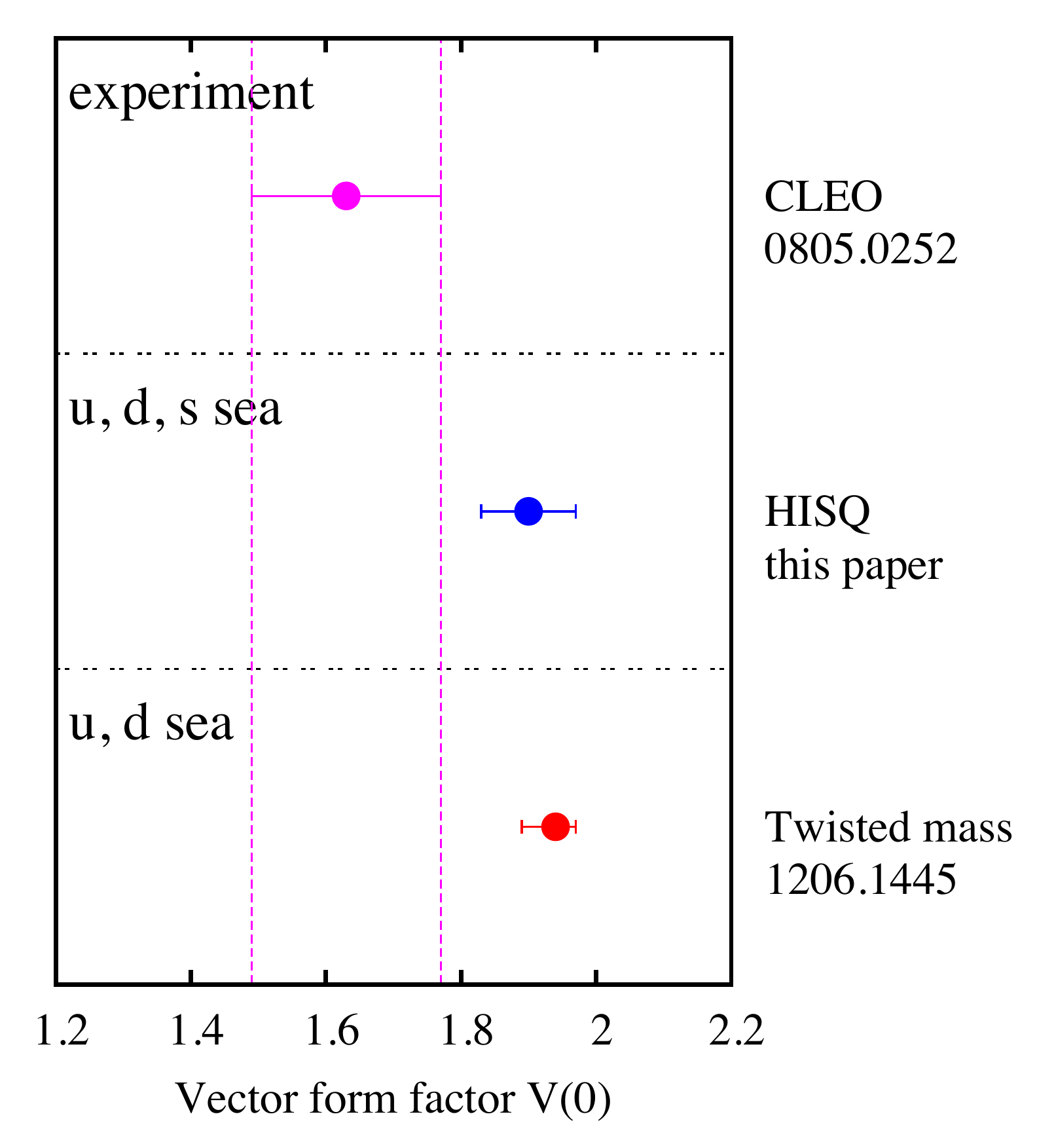}
\end{center}
\caption{ A comparison of results 
for the vector form factor, $V(0)$ for 
$J/\psi \rightarrow \eta_c \gamma$ from 
lattice QCD with experiment.
We include only results that include sea quarks 
and make use of multiple 
lattice spacings to derive a continuum value. 
The experimental result~\cite{cleopsieta} is given 
at the top, followed by the result for HISQ 
quarks from this paper. The 
twisted mass~\cite{etmccharm} results follow. 
}
\label{fig:v0all}
\end{figure}

Figure~\ref{fig:fpsiall} similarly compares our 
result for $f_{J/\psi}$ to that from twisted mass quarks including 
only $u$ and $d$ quarks in the sea~\cite{etmccharm} 
and to experiment (from eq.~(\ref{eq:vdecay})).  
Both lattice results agree well with experiment at 
the ~2\% level of accuracy achieved. 
Our value for $f_{J/\psi}$ gives a value for 
$\Gamma(J/\psi \rightarrow e^+e^-)$ of 5.48(16) keV
using eq.~(\ref{eq:vdecay}). 

Figure~\ref{fig:v0all} shows the same comparison 
for the vector form factor at $q^2=0$, $V(0)$, 
for $J/\psi \rightarrow \eta_c \gamma$ decay. 
Our result here using HISQ quarks and including 
$u$, $d$ and $s$ quarks in the sea agrees well, 
at the 4\% level of accuracy achieved, 
with the result using twisted mass quarks and 
including only $u$ and $d$ sea quarks. 

The value of $V(0)$ extracted from the 
experimental branching fraction~\cite{cleopsieta} 
is $1.7\sigma$ lower than the lattice numbers 
where $\sigma$ is dominated by the 8\% uncertainty 
from experiment. This situation is an improvement 
over that before the CLEO 
measurement~\cite{dudekcharm}. However, it is clear 
that a more stringent test of QCD would be 
possible with a smaller experimental error for 
the $J/\psi \rightarrow \eta_c \gamma$ branching 
fraction and this may become possible with BES III 
although it is a challenging mode~\cite{besbook,yuannote}. 

Our value for $V(0)$ corresponds to a width 
for $J/\psi \rightarrow \gamma \eta_c$ of 2.49(18)(7) keV
using eq.~(\ref{eq:gammarad}). The first error is from 
our result and the second from the experimental 
error in $|\vec{q}|$. Note that in using 
eq.~(\ref{eq:gammarad}) we put in the experimental masses 
for the $J/\psi$ and $\eta_c$. This is appropriate 
because these factors are kinematic ones and therefore 
should be taken to match the experiment. What we 
calculate in lattice QCD is $V(0)$. In fact, as discussed 
above, we have good agreement between our results and 
experiment for $M_{J/\psi}-M_{\eta_c}$ and so the 
kinematic factors would also be correct from 
lattice QCD. However, extra uncertainty would be introduced 
by using the lattice QCD results and that is not 
necessary or appropriate.  
Our result for the decay width corresponds to 
a branching fraction for $J/\psi \rightarrow \eta_c$ 
of 2.68(19)(11)\%, where the first error is from our 
calculation and the second from experiment, including 
the experimental width of the $J/\psi$. 

Figures~\ref{fig:hyp},~\ref{fig:fpsi} and~\ref{fig:v0}, 
which show our results as a function of lattice spacing, 
confirm that discretisation errors are small (although visible) 
for the HISQ formalism and that the approach to the 
continuum limit is well-controlled. 
This is discussed further in Appendix~\ref{appendix:discretisation}
where we compare the dependence on lattice spacing to 
that for twisted mass quarks~\cite{etmccharm}. 

\section{Conclusions}
\label{sec:conclusions}

We have given results for 3 key quantities associated 
with the $J/\psi$ meson from lattice QCD, for the first
time including the effect of all three $u$, $d$ and $s$ quarks 
in the sea. The quantities are the mass difference 
with its pseudoscalar partner, the $\eta_c$ meson, 
the decay constant  and the vector form factor at 
$q^2=0$ for $J/\psi \rightarrow \eta_c$ decay.  

Our first key result is for the $J/\psi$ decay constant. We obtain:
\begin{equation}
f_{J/\psi} = 405(6) \, \mathrm{MeV},
\label{eq:finalf}
\end{equation}
leading to 
$\Gamma(J/\psi \rightarrow e^+e^-)$ = 5.48(16) keV.
This is to be compared to the experimental result 
of $\Gamma(J/\psi \rightarrow e^+e^-)$ = 5.55(14) keV~\cite{pdg}. 
We have therefore achieved a 4\% test of lattice 
QCD from an electromagnetic decay rate (a 2\% test 
from the decay constant), that does not suffer 
from CKM uncertainties. This is itself a stringent 
test of QCD  and one for which lattice QCD is absolutely 
necessary; $f_{J/\psi}$ could not be calculated this 
accurately with any other method. 
At the same time we are able to verify that the 
time-moments of the $J/\psi$ correlator agree 
as they should with results for the charm 
contribution to $\sigma(e^+e^- \rightarrow \mathrm{hadrons})$
extracted from experiment. This is a test of 
QCD to better than 1.5\%. 

Our $f_{J/\psi}$ result is a critically important test 
for our calculations that determine the decay constants 
of the $D_s$~\cite{fdsorig, fdsupdate} and 
the $D$~\cite{fdsorig, nafd} to a similar level of 
precision. In particular it tests the HISQ formalism 
for $c$ quarks~\cite{hisqdef} even more stringently than in the 
$D$ and $D_s$ cases because the $J/\psi$ contains 
two $c$ quarks and is a smaller meson, more sensitive 
to discretisation effects on the lattice.  
Combined with our earlier work on using the 
HISQ formalism for light quarks in $f_{\pi}$ and 
$f_K$~\cite{fdsorig, oldr1paper, dowdallr1}, our result 
for $f_{J/\psi}$ provides compelling evidence that 
we have the systematic errors in $f_{D_s}$ and $f_D$ 
under control.  

We can improve our result for $f_{J/\psi}$ further 
in future by using the vector form factor method
of renormalisation rather than the current-current 
correlator method. This will only be useful if improved 
experimental results become available. This is 
expected from BESIII~\cite{yuannote}. 

A further test of QCD/Lattice QCD comes from the $J/\psi$ mass. We find: 
\begin{equation}
M_{J/\psi}-M_{\eta_c} = 116.5 \pm 3.2 \, \mathrm{MeV} 
\label{eq:finalhyp}
\end{equation}
giving $M_{J/\psi} = 3.0975(32)(11) \,\mathrm{GeV}$ 
where the second error comes from the experimental 
average for $M_{\eta_c}$~\cite{pdg}. Experiment 
gives $M_{J/\psi} = 3.0969 \,\mathrm{GeV}$. This is another 
strong test of lattice QCD, and indeed QCD, against 
experiment to be compared to that of the determination 
of $M_{D_s}$~\cite{fdsupdate} and $M_D$~\cite{rachelnew}. 
The hyperfine splitting is a relatively small relativistic 
correction in the broader context of charmonium 
meson masses and the fact that we can do this well (with 
no free parameters) is 
because the HISQ formalism is such a highly improved 
relativistic formalism. This is underlined by a study 
of the meson dispersion relation (and associated `speed of 
light') in Appendix~\ref{appendix:discretisation}. 
In fact our error on $M_{J/\psi}$ is dominated by uncertainties 
from the effect of annihilation of the $\eta_c$ meson to 
gluons, and it is important to pin these down more accurately. 

Our third result for the $J/\psi$ is that for its M1 radiative 
decay mode to the $\eta_c$. 
We find: 
\begin{equation}
\Gamma(J/\psi\rightarrow \gamma \eta_c) =  2.49 \pm 0.19 \, \mathrm{keV} 
\label{eq:finalrad}
\end{equation}
to be compared to the current experimental 
value of 1.84(30) keV~\cite{cleopsieta}. 
The agreement is reasonably good, but the experimental error 
is large and the lattice QCD result would allow a much stronger 
test of QCD 
if this were reduced. This should be possible at BESIII~\cite{besbook}. 
Since the error in our lattice QCD result is dominated 
by the continuum extrapolation it will be improved in 
calculations on superfine and ultrafine lattices as we 
have done for the decay constant, and 2\% errors should 
also be achievable here. Again, this is only possible 
in lattice QCD. 

The $J/\psi \rightarrow \gamma \eta_c$ decay rate is another 
test of QCD, along with our leptonic decay constant, that
is free of CKM uncertainties. 
It provides a validation of semileptonic decay rate
calculations for $D$ and $D_s$ 
mesons~\cite{na1, na2, jonnalat11, gordonlat11, usinprep}, 
that also use HISQ quarks as well as a test of 
our techniques for nonperturbative current renormalisation 
that we are using for a range of semileptonic and 
radiative decays~\cite{jonnalat11, gordonlat11, usinprep}. 

{\bf{Acknowledgements}} We are grateful to the MILC collaboration 
for the use of 
their configurations and to C. DeTar, F. Sanfilippo, 
J. Simone and M. Steinhauser for useful 
discussions.  
Computing was done on the Darwin supercomputer at the University 
of Cambridge as part of the DiRAC facility jointly funded by 
STFC, BIS and the Universities of Cambridge and Glasgow 
and at the Argonne 
Leadership Computing Facility at Argonne National Laboratory, supported
by the Office of Science of the U.S. Department of Energy under 
Contract DOE-AC02-06CH11357. 
We acknowledge the use of Chroma~\cite{chroma} for part 
of our analysis. 
Funding for this work came from MICINN (grants FPA2009-09638 
and FPA2008-10732 and the Ramon y Cajal program), DGIID-DGA (grant 
2007-E24/2), the DoE (DE-FG02-04ER41299), the EU (ITN-STRONGnet, PITN-GA-2009-238353), the NSF, 
the Royal Society, the Wolfson Foundation, the Scottish Universities 
Physics Alliance and STFC. 

\appendix

\section{Taste effects in staggered mesons}
\label{appendix:taste} 

\begin{figure}
\begin{center}
\includegraphics[width=0.9\hsize]{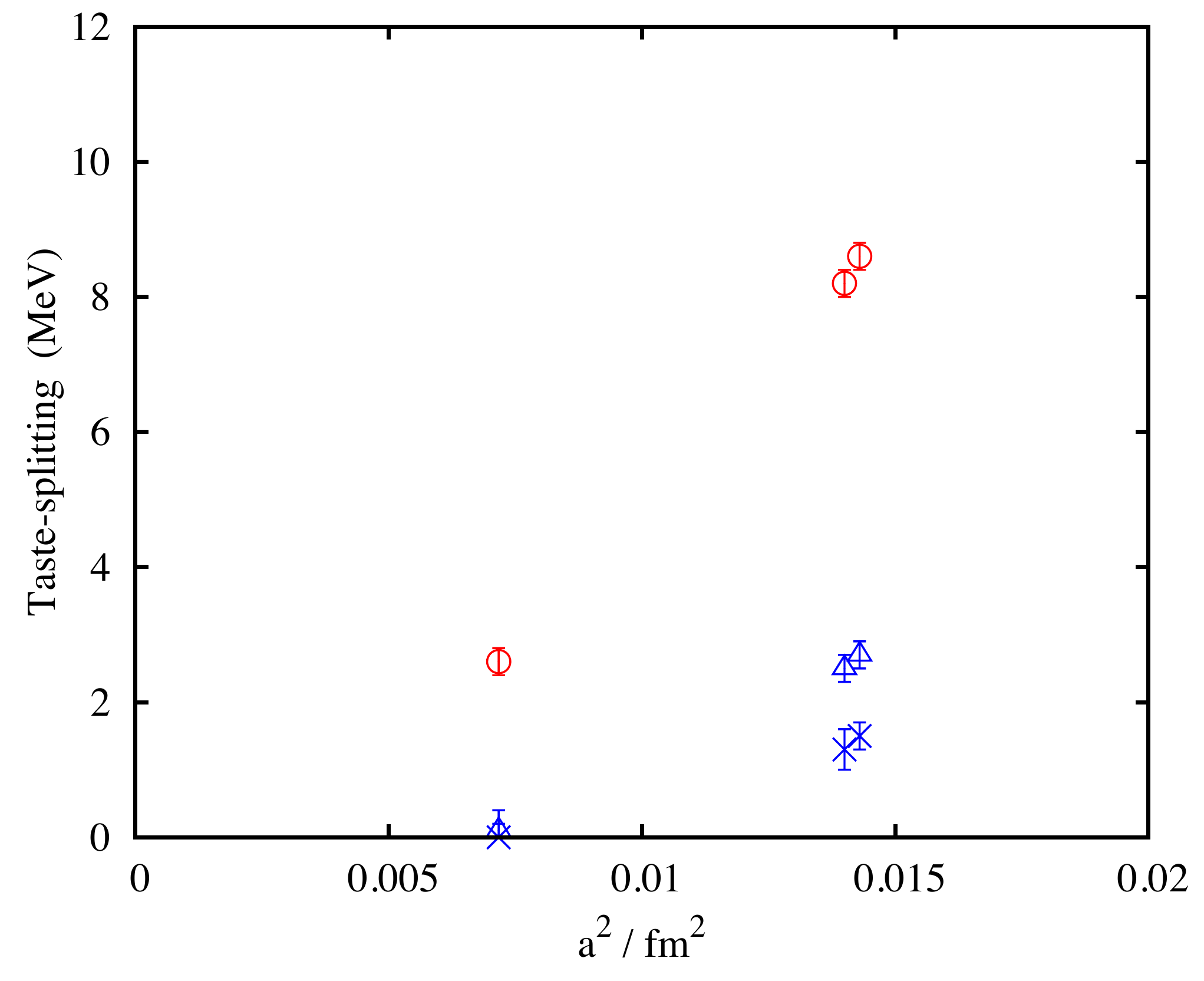}
\end{center}
\caption{ Difference in mass in MeV for different 
meson `tastes' for the $\eta_c$ and $J/\psi$ used here, 
plotted against the square of the lattice spacing. 
Red open circles show the mass difference between 
the local non-Goldstone and goldstone $\eta_c$ mesons. 
For the vector we have the mass difference between 
the local $J/\psi$ meson and the 
1-link $\gamma_i \otimes \gamma_i \gamma_j$ vector (blue crosses) 
and the 1-link $\gamma_0 \gamma_i \otimes \gamma_0 \gamma_i \gamma_j$ 
vector (blue open triangles).  The local $J/\psi$ meson is 
the lighter in both cases. 
Results are for sets 1, 2 and 4 from Tables~\ref{tab:massres} 
and~\ref{tab:3ptres}. Errors are statistical. 
}
\label{fig:tastesplit}
\end{figure}

Each staggered meson comes in 16 different tastes, 
most easily seen in terms of naive quark operators 
made with different point splittings:
\begin{equation}
J_n^{(s)} = \overline{\psi}(x) \gamma_n \psi(x+s). 
\label{eq:mesops}
\end{equation}
Here $s$ is a 4-dimensional vector with 0 or 1 in 
each component. 
The different $J_n^{(s)}$ operators are orthogonal 
to each other. 
To work out the corresponding staggered quark 
correlators we need the staggering matrix, $\Omega(x)$. 
In our convention this is: 
\begin{equation}
\Omega(x) = \prod_{\mu=1}^4 (\gamma_{\mu})^{x_{\mu}},
\label{eq:omega}
\end{equation}
with $\gamma_4 \equiv \gamma_0$. 
Then the connection between naive quark propagators, 
which carry a spin index, and staggered quark propagators, 
which do not, is: 
\begin{equation} 
S_F(x,y) \equiv \langle \psi(x) \overline{\psi}(y) \rangle_{\psi} = g(x,y)\Omega(x)\Omega^{\dag}(y). 
\label{eq:ksnaive}
\end{equation}
To work out the phases that appear in the correlator of a particular 
taste we then simply have to calculate spin traces over 
products of $\Omega$ and $\gamma_n$ factors, see, 
for example,~\cite{hisqdef}.  

Here we use two different tastes for the $\eta_c$ and three for 
the $J/\psi$ and they all have slightly different masses. 
The mass differences between the different tastes 
of a given meson, however, vanish as $\alpha_s^2a^2$. 
For the $\eta_c$ we use the Goldstone meson (in spin-taste
notation this is the $\gamma_5 \otimes \gamma_5$ meson) 
and the local non-Goldstone meson 
(the $\gamma_0 \gamma_5 \otimes \gamma_0 \gamma_5$ meson), 
which are the first two mesons on the ladder of pseudoscalar 
tastes. We show the mass difference between them in 
Figure~\ref{fig:tastesplit} for coarse and fine lattices. 
The mass difference amounts on the coarse lattices 
to a little less than 10 MeV 
(for a 3 GeV particle) and clearly falls 
with $a^2$ as expected. 
For the $J/\psi$ we use the local vector 
($\gamma_i \otimes \gamma_i$) and two 1-link operators 
which have a point-splitting in an orthogonal direction 
to the polarisation ($\gamma_i \otimes \gamma_i\gamma_j$ 
and $\gamma_0 \gamma_i \otimes \gamma_i \gamma_j$). 
Note that these are not taste-singlet vectors. 
The mass difference between tastes for mesons of 
other $J^{PC}$ is typically much 
smaller than for pseudoscalars and that is clear here. 
Figure~\ref{fig:tastesplit} shows that the mass 
difference for the vectors is 1-2 MeV 
on the coarse lattices and not resolvable on the fine
lattices. 

In Section~\ref{sec:raddecay} we showed results for 
$J/\psi \rightarrow \gamma \eta_c$ using different 
tastes of $J/\psi$ and $\eta_c$ at the two ends of 
the 3-point function. No 
difference was seen in the vector form factor at 
$q^2=0$ in the two cases, either on the coarse or fine 
lattices (within our statistical errors of 2\%). 
This is another demonstration that taste 
effects are very small with HISQ quarks.  

\section{Determining nonperturbative $Z$ factors for local 
vector currents}
\label{appendix:z}

\subsection{The current-current renormalization method}
\label{appendix:currcurr}

\begin{table}
\begin{tabular}{llllll}
\hline
\hline
 &  $c_n^{(0)}$ & $c_n^{(1)}$ & $c_n^{(2)}$ & $c_n^{(3)}$ & $c_n^{(4)} \ldots$ \\
\hline
$n$=4 &  1.0 & 0.235 & 0.354 & -0.187 & 0.0(5) $\ldots$ \\
$n$=6 &  1.0 & 0.246 & 0.460 & 0.198 & 0.0(5) $\ldots$  \\
$n$=8 & 1.0 & 0.253 & 0.563 & 0.511 & 0.0(5) $\ldots$ \\
\hline 
\hline
\end{tabular}
\caption{ The perturbative series~\cite{qcdpt1, qcdpt2, qcdpt3, qcdpt4, qcdpt5} for the ratio 
$r_{n+2}^P/r_n^V$ for different moments, $n$, in continuum QCD perturbation theory. 
$c_n^{(i)}$ is the coefficient of $\alpha^i_{\overline{MS}}(\mu=m)$; 
$c_n^{(0)}$ is 1.0 for all cases, by definition. 
$c_n^{(i)}$ for $i>3$ were included in the determination of 
$Z$, allowing for a coefficient of $0.0\pm 0.5$. 
}
\label{tab:pth}
\end{table}

\begin{table}
\begin{tabular}{lllll}
\hline
\hline
Set & $m_ca$ &  $Z(4)=Z_{cc}$ & $Z(6)$ & $Z(8)$ \\
\hline
1 & 0.622 &  0.979(12) & 0.945(14) & 0.927(17) \\
2 & 0.63 & 0.979(12) & 0.945(14) & 0.926(17) \\
2 & 0.66 & 0.974(12) & 0.941(14) & 0.921(17) \\
4 & 0.413 & 0.983(12) & 0.953(14) & 0.953(17) \\
5 & 0.273 & 0.986(12) & 0.970(14) & 0.975(18) \\
6 & 0.193 & 0.990(12) & 0.982(14) & 0.986(18) \\
\hline 
\hline
\end{tabular}
\caption{ Renormalisation constants determined from 
the current-current correlator method on each 
configuration set used for the determination of $f_{J/\psi}$. 
The $Z$ value we use is that from moment 4. 
The errors include an estimate of effects from a 
gluon condensate contribution, and unknown fourth 
order and higher terms in continuum perturbation theory. 
The errors are highly correlated between configuration 
sets (to better than 1\% of the error). 
For set 2 we include both the tuned value of $am_c$ (0.63) 
and the heavier, detuned, value (0.66). Very little 
difference is seen between them. 
}
\label{tab:zres}
\end{table}

Time-moments of lattice QCD 
correlators for zero-momentum heavyonium mesons 
can be compared very accurately~\cite{firstcurrcurr} to continuum 
QCD perturbation theory~\cite{qcdpt1, qcdpt2, qcdpt3, qcdpt4, qcdpt5} 
developed for 
the analysis of the $e^+e^-$ annihilation cross-section. 
This has been used with pseudoscalar meson 
correlators made with HISQ quarks to extract $c$ and $b$ masses and 
$\alpha_s$ to better than 1\%~\cite{bcmasses}.  
These results used the goldstone pseudoscalar 
correlator, which is absolutely normalised 
because of the HISQ PCAC relation. 
Here we apply the same techniques to vector 
meson correlators but use it to determine the 
renormalisation factor, $Z$, required for 
the lattice vector current
to match the continuum current. 

The time moments of our lattice QCD correlators 
are defined as: 
\begin{equation}
C_n^{V} = \sum_{\tilde{t}} \tilde{t}^n \overline{C}_{J/\psi}(\tilde{t})
\label{eq:timemomv}
\end{equation}
and
\begin{equation}
C_n^{P} = \sum_{\tilde{t}} \tilde{t}^n (am_c)^2\overline{C}_{\eta_c}(\tilde{t})
\label{eq:timemomp}
\end{equation}
where $\tilde{t}$ is a symmetrised version of $t$ 
around the centre of the lattice, i.e. going forward 
in time, $\tilde{t}$ runs from $0$ 
to $L_t/2$ and then from $-L_t/2+1$ to 
$-1$. 
The extra factor of $(am_c)^2$ in the pseudoscalar 
case is to make a correlator moment which is finite 
as $a \rightarrow 0$. 
For both correlators we expect 
the small $n$ moments to behave perturbatively, 
since they probe small times. 
Then our match to continuum 
perturbation theory is: 
\begin{equation}
C_n^{P} = \frac{g_n^P(\alpha_{\overline{MS}}(\mu), \mu/m_c)}{(am_c(\mu))^{n-4}} + \mathcal{O}((am_c)^m).
\label{eq:pmom}
\end{equation}
where $g_n$ is the continuum QCD perturbation theory 
series in the $\overline{MS}$ scheme~\cite{qcdpt1, qcdpt2, qcdpt3, qcdpt4, qcdpt5}.
For the vector correlator a $Z$ factor is needed to multiply 
the lattice current and so: 
\begin{equation}
C_n^{V} = \frac{1}{Z^2}\frac{g_n^V(\alpha_{\overline{MS}}(\mu), \mu/m_c)}{(am_c(\mu))^{n-2}} + \mathcal{O}((am_c)^m).
\label{eq:vmom}
\end{equation}
$Z$ is then a function of the bare lattice strong coupling 
constant at each lattice spacing 
and so this match must be performed separately on each ensemble. 
By taking the ratio of pseudoscalar and vector moments we 
can cancel the factors of the quark mass.  
In fact we also divide each moment by its tree-level value 
(calculated with the gluon fields set to 1) to reduce 
discretisation errors, i.e. instead of $C_n$ we use 
$R_n$ where 
\begin{equation}
R_n = \frac{C_n}{C_n^{(0)}} 
\label{eq:Rn}
\end{equation}
and also take the same ratio, calling it $r_n$, 
in the continuum perturbation theory. 
Finally $Z$ is given by: 
\begin{equation}
Z(n) = \sqrt{\frac{R_{n+2}^{P}/R_n^V}{r_{n+2}^{P}/r_n^V}}.
\label{eq:zcalc}
\end{equation}

Table~\ref{tab:pth} gives the perturbative coefficients 
for the series in $\alpha_{\overline{MS}}(m)$ for 
$r_{n+2}^P/r_n^V$. This is known to 4-loops ($\alpha_s^3$) 
and we include the possibility of unknown higher order 
terms (to 20th order) with prior values 
for the coefficients of $0\pm 0.5$. 
We take $\mu = m$ but including the possibility of higher 
order corrections means that the results are almost 
completely insensitive 
to $\mu$. 
We also allow for gluon condensate contributions taking 
$\alpha_sG^2/\pi = 0 \pm 0.012 \,\mathrm{GeV}^4$. This 
increases the errors on the determination of the $Z$ values 
as $n$ increases. 

Table~\ref{tab:zres} gives the results for $Z$ on each 
ensemble and for moments 4, 6 and 8. The differences 
between the $Z$ values on a given ensemble 
arise from discretisation errors. 
We take our final 
result for use in section~\ref{sec:leptwidth} from 
moment 4 (and so these numbers are reproduced 
in Table~\ref{tab:massres}) since, even though
discretisation errors fall 
as the moment number increases, the errors from the 
gluon condensate rise more steeply. 
The error in $Z(4)$ is around 1\%. It is dominated by the 
uncertainty in higher orders in perturbation theory and 
so strongly correlated from one lattice spacing to 
the next. We have checked that we obtain the same final 
result for $f_{\psi}$ using Z(6) (but with an error of 1.6\% 
rather than 1.4\%). 

\subsection{The vector form factor method}
\label{appendix:vecff}

Vector currents can be normalised completely 
nonperturbatively by requiring that the vector 
form factor at $q^2=0$ between two identical mesons be 1 since 
this would be true for a conserved vector current. 
We use this to normalise the staggered taste-singlet 1-link vector current 
between two mesons made of staggered quarks in~\cite{jonnalat11}. 
Here, however, we want to normalise the taste-nonsinglet local 
staggered vector current, and we cannot do this using a 3-point 
function made entirely of staggered quarks.  
We have to include a non-staggered (and non-doubling) spectator 
quark in order to remove the requirement for overall taste-cancellation
in the 3-point function~\cite{mattfirsthl}. 
The Fermilab Lattice/MILC collaboration use 
this method~\cite{fermilabdecay11} with 
a clover spectator quark to normalise a light staggered 
vector current for the nonperturbative part of their 
mixed perturbative/nonperturbative approach to normalising 
the clover-staggered operators for heavy-light meson decay 
constants.  
Here we use an NRQCD spectator quark 
to normalise the local charm staggered vector current.
In fact we can use any NRQCD-like action for the spectator 
since it does not need to correspond to a physical quark. 
However, for simplicity we do use the NRQCD action developed 
for our NRQCD-light spectrum calculations~\cite{rachelnew}. 

\begin{figure}
\begin{center}
\includegraphics[width=0.9\hsize]{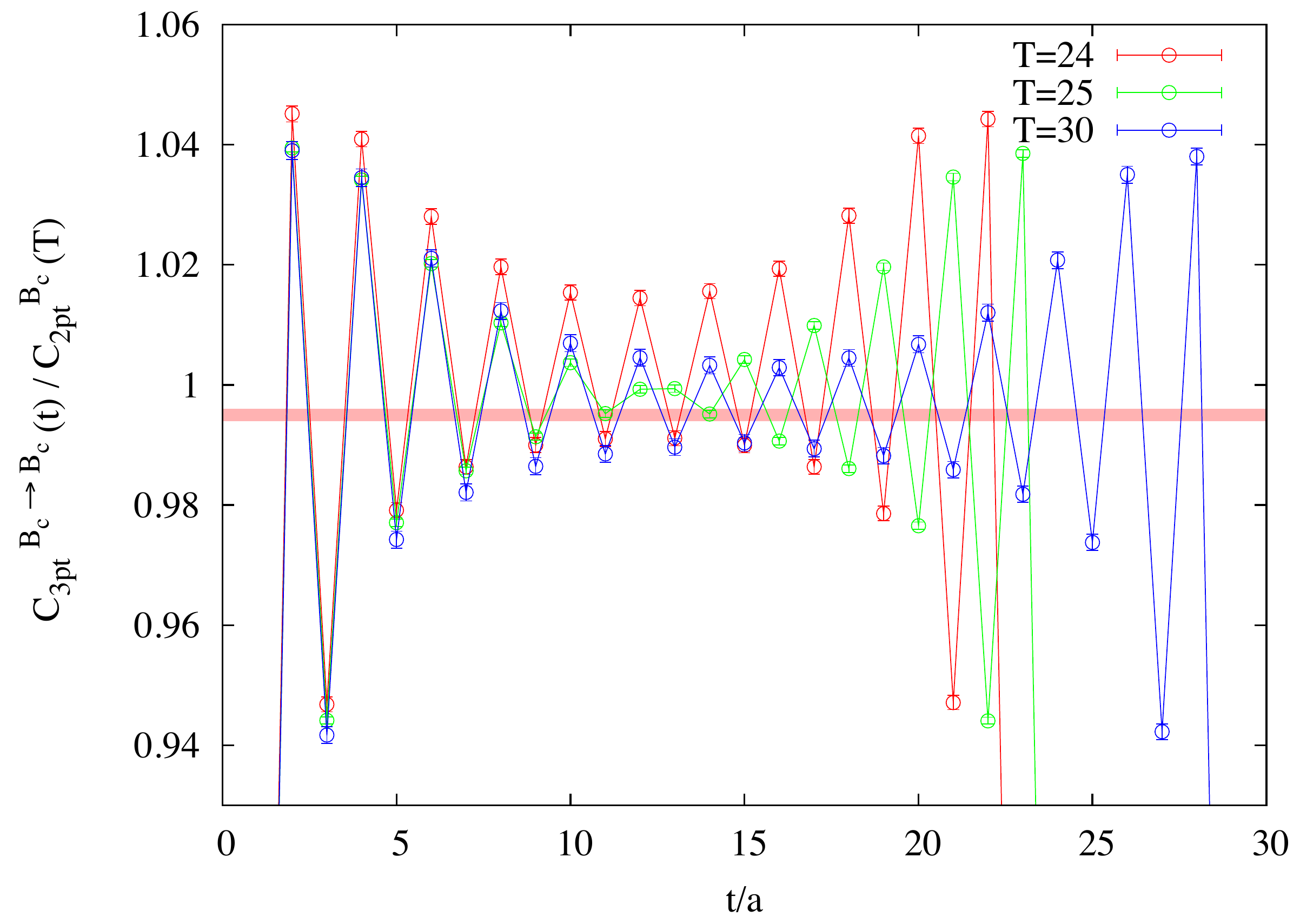}
\includegraphics[width=0.9\hsize]{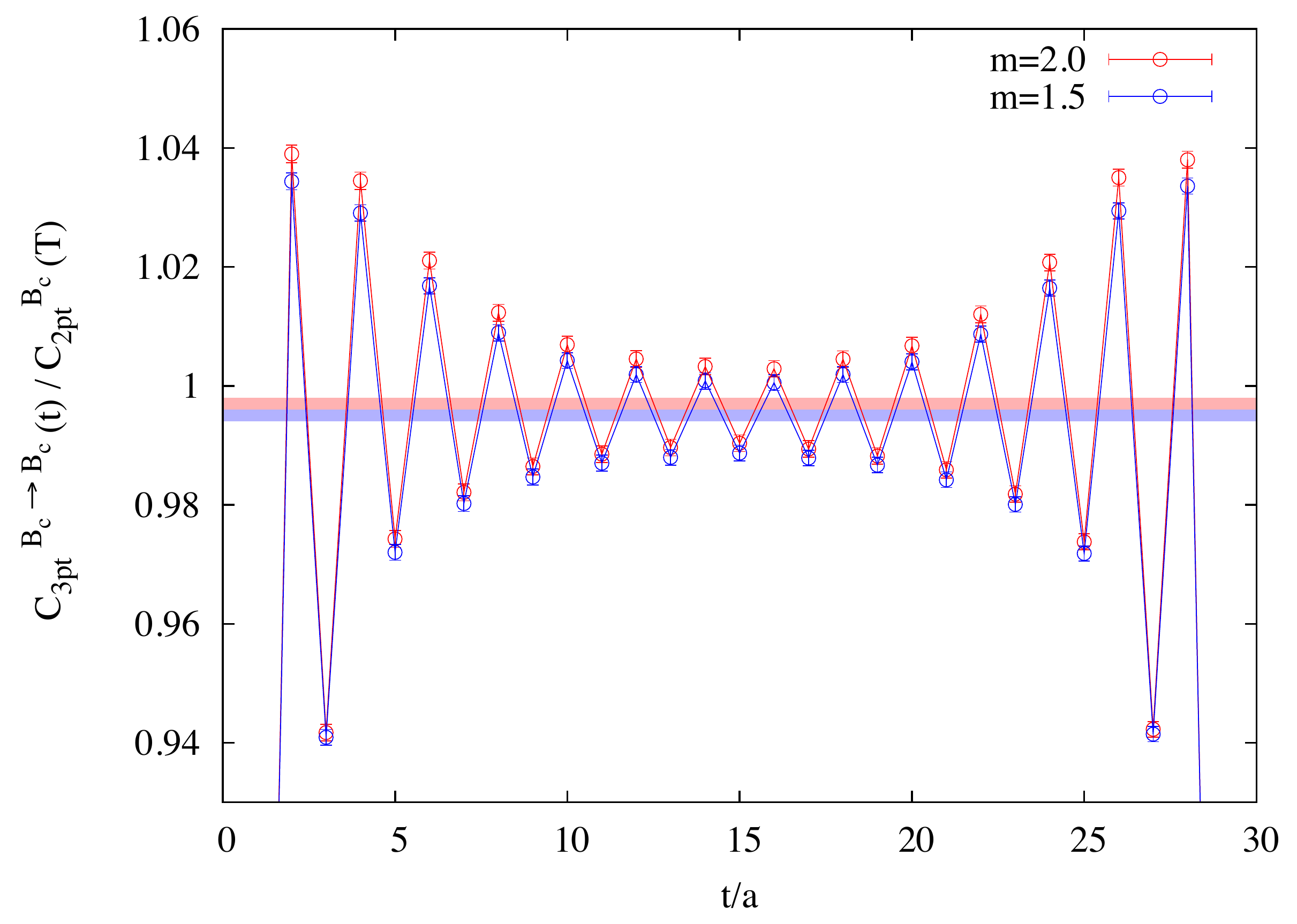}
\end{center}
\caption{ The ratio of the 3-point 
function for the `$B_c$' to `$B_c$' vector 
form factor to the `$B_c$' 2-point function
against the source-current separation, $t/a$. 
The top plot shows results for $m_ba=2.0$ on 
fine lattices, set 4, for various values of $T$. 
Lines join the points for clarity. 
The shaded red band gives our fit result for 
the ground state matrix element. 
The lower plot shows results, also on set 4, 
for two different values of $m_ba$. 
The shaded red and blue bands show the fit 
results for the ground state matrix elements. 
}
\label{fig:bc3pt}
\end{figure}

To combine an NRQCD (or other non-staggered) quark 
with a staggered quark we convert 
it to a naive quark~\cite{mattfirsthl}, re-instating 
the spin degree of freedom  
as in eq.~(\ref{eq:ksnaive}). 
A pseudoscalar meson correlator which, with two idential 
quarks, would simply be the sum over spins and colors 
of the squared modulus of the propagator, becomes in 
this case: 
\begin{equation}
C_{Qq}(t) = \sum_{\vec{x},\vec{y}}\mathrm{Tr}\left\{G(x,y)\Omega(y)g^{\dag}(x,y)\Omega^{\dag}(x)\right\} .
\label{eq:hlcorr1}
\end{equation}
The trace is over spin and color but separates since the 
staggered propagator, $g(x,y)$, has no spin and $\Omega$ 
no color. Then:  
\begin{equation}
C_{Qq}(t) = \sum_{\vec{x},\vec{y}}\mathrm{Tr}_c\left\{\mathrm{Tr}_s[\Omega^{\dag}(x)G(x,y)\Omega(y)]g^{\dag}(x,y)\right\}.
\label{eq:hlcorr2}
\end{equation}
This makes clear that we can transfer the $\Omega$ matrices to 
the non-staggered quark propagator, $G(x,y)$, and this allows us, as 
shown in~\cite{gregory}, to combine staggered and non-staggered 
quarks from random wall sources. We simply make the source of 
the non-staggered quark the product of $\Omega(y)$ (where $y$ 
runs over a time slice at the source) with the 
random number at each $y$ value in the random wall, 
convoluted with smearing functions if required. 
At the sink we multiply by $\Omega^{\dag}$ before tracing over 
spins to combine with the spinless staggered quark propagator. 
Here we develop this method further for 3-point functions.   

For the matrix element of the temporal charm-charm vector current 
between identical charmed pseudoscalar mesons, at rest, we have:
\begin{equation}
\langle P_c | V_t | P_c \rangle = 2 m_{P_c} f_+(0) 
\label{eq:vectorff}
\end{equation}
and we can demand that $f_+ (0) = 1$, i.e. we can multiply 
the left-hand-side by $Z$ to make this true.
The temporal component of the vector current is the 
easiest to use for this purpose although 
the spatial component of the vector current is 
the one that we use for $J/\psi \rightarrow \eta_c$ decay. 
For a relativistic action such as 
HISQ the renormalisation of the spatial and temporal 
components will be the same up to discretisation errors.

For the 3-point function needed to evaluate the 
matrix element above we use an NRQCD heavy quark 
propagating from $0$ to $T$ and 
HISQ charm quarks from $0$ to $t$ and $T$ to $t$ (see Fig.~\ref{fig:3ptpic}).
The local staggered temporal vector current ($\gamma_0 \otimes \gamma_0$ in 
spin-taste notation) is inserted at $t$  
and the restriction for staggered 3-point functions of an overall 
tasteless correlator is avoided by the spin content of the NRQCD propagator.

The 3-point function is given by:
\begin{eqnarray}
&&C_{3pt}(0,t,T) = \sum_{s_T,s_t}(-1)^{x_T+y_T+z_T+t_T}(-1)^{t_t}  \\
&&\times \mathrm{Tr_c}\left\{g(t,T)\mathrm{Tr_s}[\gamma_0\Omega^{\dag}(T)G(T,0)\Omega(0)]g^{\dag}(t,0)\right\} \nonumber
\label{eq:nrqcd3pt}
\end{eqnarray}
The $g$ are HISQ $c$ propagators and $G$ is an NRQCD heavy quark propagator. 
The $\gamma_0$ factor comes from the local temporal vector current.  
The 3-point function is therefore calculated in a very similar way 
to the 2-point function. $\Omega(x)$ multiplied by the random wall 
at the source time slice, 0, is used as the source for 
the NRQCD propagator. At time T this is multiplied by 
$\Omega^{\dag}$ and $\gamma_0$, source and sink spins are set equal 
and summed over. This is then the source for the HISQ propagator 
from $T$ to $t$ which is finally combined with the HISQ propagator 
generated from the random wall at time slice $0$.  

We calculate the 2-point and 3-point functions described above for 
several different NRQCD masses, $m_ha$, and $c$ quark masses, $m_ca$, on the configuration 
sets 1, 2 and 4. We also use several different values of $T$ so 
that our fit benefits from both $T$ and $t$ dependence to improve 
the extraction of the ground-state masses, amplitudes and matrix elements. 
The 3-point and 2-point correlators are fit simultaneously 
to the forms given in eqs~(\ref{eq:fit1}) and~(\ref{eq:3ptfit}).
We use the same priors as in section~\ref{sec:raddecay}. 
Note that for the 3-point fit we can now impose symmetry 
under interchange of the mesons at 0 and $T$ since they 
are the same. This means that the amplitudes $V^{nn}$ and $V^{oo}$ are 
square symmetric matrices and $V^{no}=V^{on}$. 

The key quantity that we extract from the fit is 
the ground state matrix element $V^{nn}_{00}$. 
This is proportional to the vector matrix element 
on the left-hand-side of eq.~(\ref{eq:vectorff}). 
We can work out the constant of proportionality by 
matching our fit equations, eq~(\ref{eq:3ptfit}) 
and~(\ref{eq:fit1}) to the form expected by 
inserting a complete set of states in a continuum 
3-point function. 
In this case factors of the mass of the meson\footnote{For 
a meson containing an NRQCD quark the energy obtained 
from the 2-point and 3-point fits is not its mass. However that 
is irrelevant here since it cancels.} 
cancel and we find $V^{nn}_{00} = f_+(0)$.  
Then the renormalisation factor we need is given by: 
\begin{equation}
Z = \frac{1}{V^{nn}_{00}}.
\label{eq:znrqcd}
\end{equation}

The results for $Z$ are taken from fits with 5 normal and 4 
oscillating exponentials and given in Table \ref{tab:zvtable}.
We obtain very precise results for $Z$, with errors of 0.1\%,
without even using the full statistics available for each ensemble. 
The values are much more precise, for example, than 
for the implementation given in~\cite{fermilabdecay11} (although their 
values are for light quarks rather than charm). 

Figure~\ref{fig:bc3pt} shows the quality of our results through 
plots of the ratio of the 3-point to the 2-point function. 
Note that we fit the 3-point and 2-point functions simultaneously 
and {\it not} just the ratio. It is convenient to plot the 
ratio, however, because the ground state contribution 
to this is simply $V^{nn}_{00}$. Our fit allows us to include the 
effect of excited states, both normal states and oscillating 
states. The presence of oscillating states is evident in 
the plots. The upper plot of Figure~\ref{fig:bc3pt} compares 
results at different $T$ values. All are included in the 
simultaneous fit. The lower plot shows results at different 
values of $m_ha$ for a given $m_ca$. 

\begin{table*}
\begin{tabular}{lllllllll}
\hline
\hline
Set & $N_{\mathrm{cfg}}\times N_t$ & $T$ & $am_h$ & $am_c$ & $\epsilon$ & $aE_{P_c}$ & $V^{nn}_{00}$ & $Z=Z_{ff}$ \\
\hline
1 & $450 \times 4$ & 20,21 & 2.0 & 0.622 & -0.221 & 0.9630(2) & 1.0104(12) & 0.9896(11)\\
\hline
2 & $408 \times 4$ & 20,21,24 & 2.8 & 0.63 & -0.226 & 1.0239(2) & 1.0220(10) & 0.9784(9)\\
 & & 20,21,24 & 2.0 & 0.63 & -0.226 & 0.9719(2) & 1.0106(8) & 0.9894(8)\\
 &  & 20,21 & 1.5 & 0.63 & -0.226 & 0.9311(3) & 1.0026(14) & 0.9974(14)\\
 &  & 20,21 & 2.0 & 0.66 & -0.244 & 0.9994(3) & 1.0138(18) & 0.9863(17)\\
\hline
4 & $322 \times 4$ & 24,25,30 & 2.0 & 0.413 & -0.107 & 0.6454(2) & 0.9966(14) & 1.0033(14)\\
 & & 24,25,30 & 1.5 & 0.413 & -0.107 & 0.5939(2) & 0.9950(10) & 1.0049(10)\\
\hline 
\hline
\end{tabular}
\caption{ 
The $Z$ factors (column 9) obtained from the vector form factor method on different 
configurations sets (column 1) and for different NRQCD quark masses, $m_ha$ (column 4), 
and $c$ quark masses, $m_ca$ (column 5) with $\epsilon$ factor (column 6). 
The $m_ca$ values are those used in our calculation of $J/\psi \rightarrow \eta_c \gamma$ 
described in section~\ref{sec:raddecay}; the $m_ha$ values are arbitrary since they 
correspond to the spectator quark. 
$Z$ is given by the inverse of the fit parameter $V^{nn}_{00}$ given in column 8.  
Column 2 gives the number of configurations used in the calculation and the number 
of time sources for the origin, 0. The $T$ values used are given in column 3. 
Column 7 gives the energy of the NRQCD-$c$ meson obtained from the fit. This 
is not equal to the mass because there is an energy offset in NRQCD. 
 }
\label{tab:zvtable}
\end{table*}

The vector form factor method for determining $Z$ is completely nonperturbative. 
It will therefore be subject to errors coming from lattice 
QCD in the form of discretisation errors. These mean, for example, 
that the $Z$ factor at a particular value of the lattice spacing is 
not completely independent of the mass of the NRQCD spectator quark. 
From our results in Table~\ref{tab:zvtable} we see that changing 
$m_ha$ from 2.8 to 1.5 on coarse set 2 (corresponding to change of 
almost a factor of two)
causes a 2\% change in $Z$. On fine set 4 the 
sensitivity is reduced to a change of 0.2\%, not significant within 
our statistical errors, when $m_ha$ is changed from 2.0 to 
1.5, a change in mass of 30\%. 
Since the change in lattice spacing between the coarse and fine 
sets is a factor of 1.4, pairs of $am_h$ values on 
coarse and fine lattices that correspond
to approximately the same physical mass are (2.8, 2.0) and 
(2.0, 1.5). We will take our central result for $Z_{ff}$ from the 
$Z$ values corresponding to using (2.0, 1.5) and use (2.8, 2.0) 
to check systematic errors coming from $Z$. We see in 
section~\ref{sec:raddecay} that we get the same answer from 
both sets of $Z$ values.  

Table~\ref{tab:zvtable} also shows results for the deliberately 
mistuned $c$ quark mass of 0.66 on set 2 with which we test the 
$c$ quark mass dependence of our $J/\psi \rightarrow \eta_c$ 
form factor. A barely significant change in $Z$ is seen between 
this value and that for the tuned mass of 0.63. 
We also see no significant difference in $Z$ as we change the 
sea light quark masses between set 1 and set 2. 

Discretisation errors also mean that our results for $Z_{ff}$
here do not have to agree exactly with our earlier results 
for $Z_{cc}$. $Z_{cc}$ has a much larger error coming from 
unknown higher order terms in continuum perturbation theory. 
As a result of this, $Z_{ff}$ and $Z_{cc}$ do agree at 
the level of 1$\sigma$. They also agree within errors with the lattice 
QCD perturbation theory for this renormalisation, which has a 
small negative contribution at $\mathcal{O}(\alpha_s)$~\cite{howard}. 

The $Z$ values calculated here form part of a programme to 
normalise nonperturbatively a range of staggered currents 
for a variety of weak semileptonic and electromagnetic 
radiative decay rates~\cite{gordonlat11}. 

\section{Discretisation errors}
\label{appendix:discretisation}

\begin{figure}
\begin{center}
\includegraphics[width=0.9\hsize]{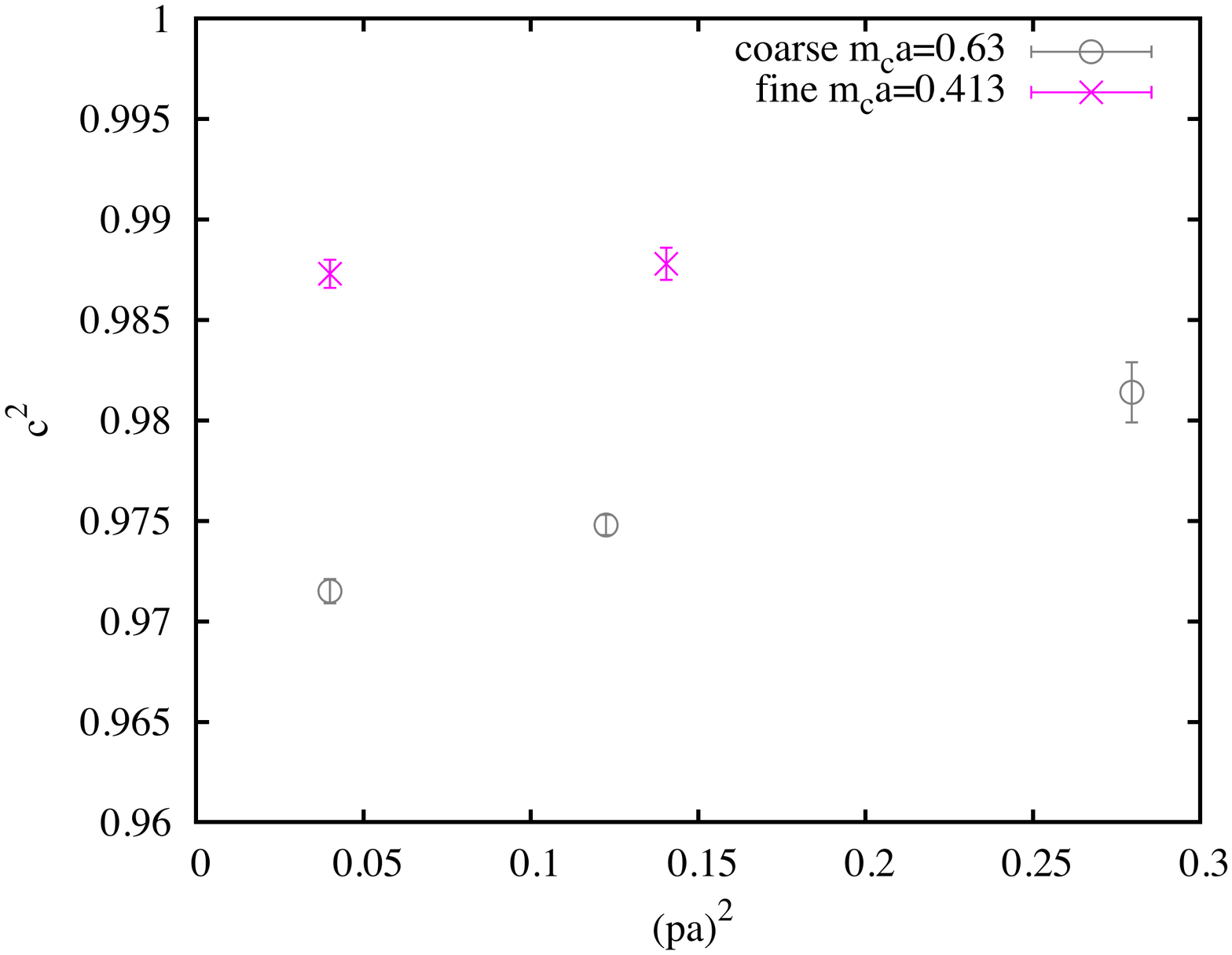}
\includegraphics[width=0.9\hsize]{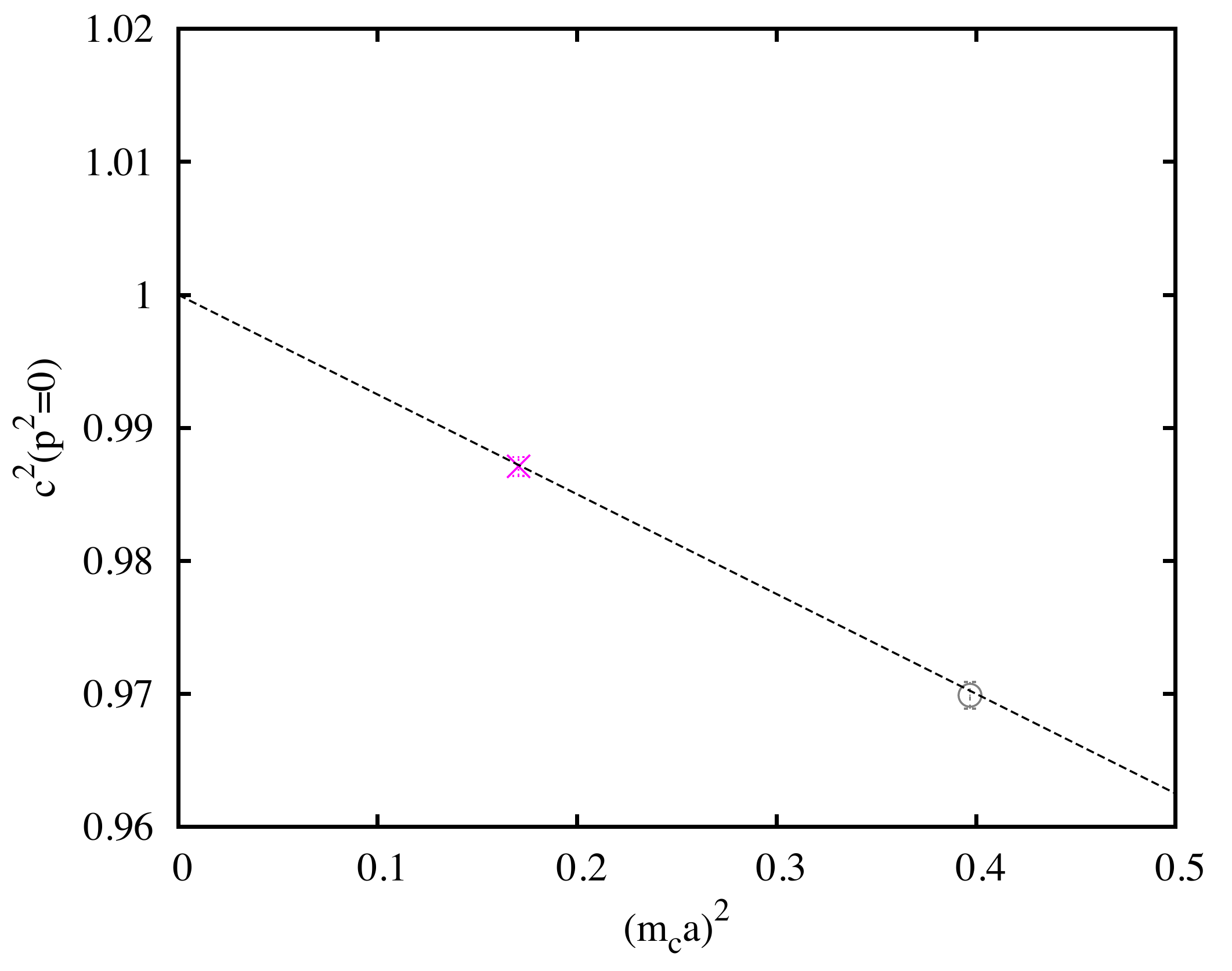}
\end{center}
\caption{ The `speed-of-light', $c^2$, calculated from 
zero and finite momentum $\eta_c$ correlators 
using HISQ $c$ quarks. 
The top figure shows $c^2$ as a function of the 
square of the spatial momentum for coarse lattices, set 
2, where $m_ca=0.63$ (grey open circles) and for 
fine lattices, set 4, where $m_ca=0.413$ (pink crosses). 
The lower figure shows the resulting values of 
$c^2$ as ${\vec{p}}^2a^2 \rightarrow 0$ as a function of 
$(m_ca)^2$. The dashed straight line is drawn to guide 
the eye.  
}
\label{fig:csq}
\end{figure}

\begin{figure}
\begin{center}
\includegraphics[width=0.9\hsize]{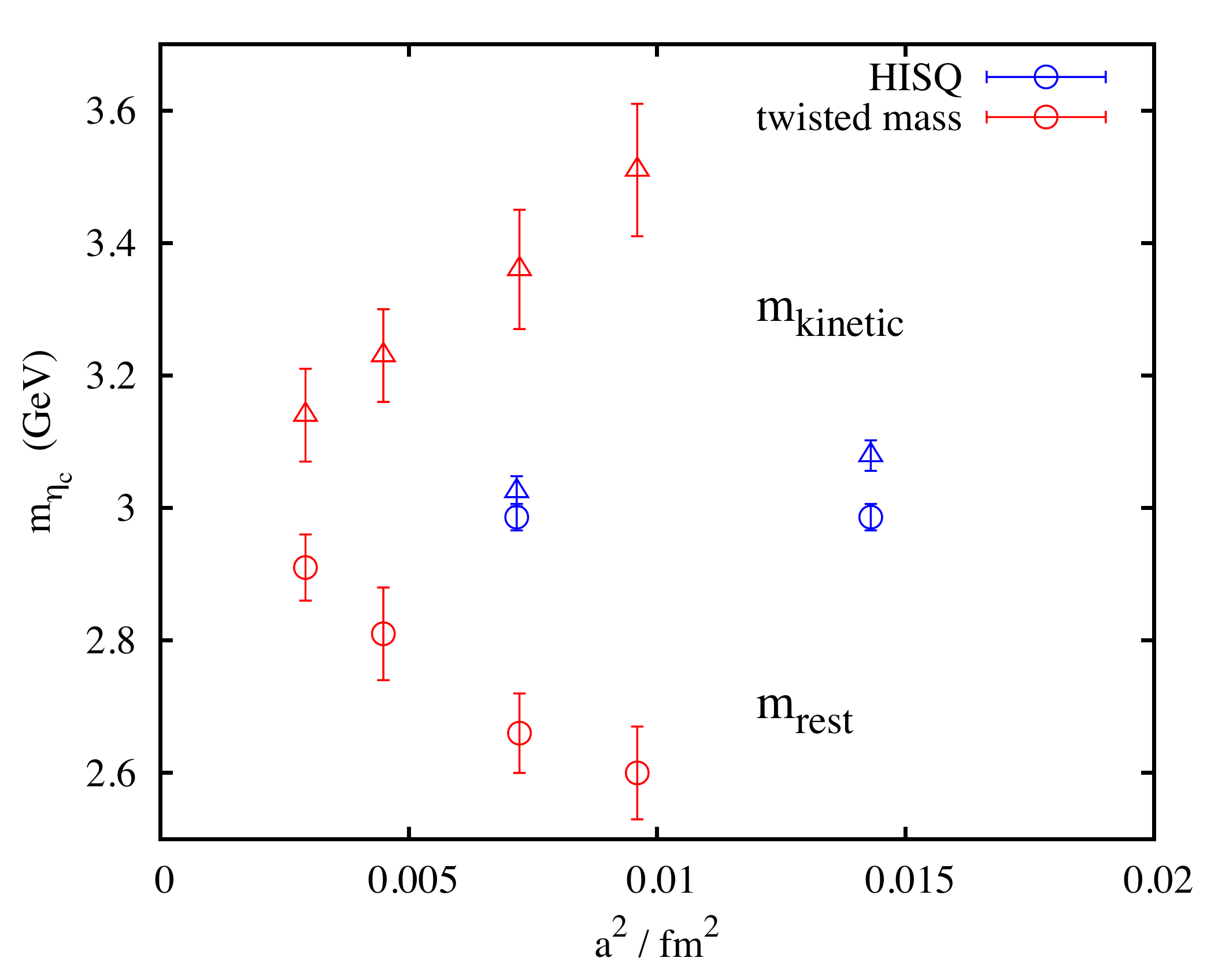}
\end{center}
\caption{ The rest (static) and kinetic 
masses for the $\eta_c$ meson 
compared between the HISQ formalism (this paper) 
and the twisted mass formalism~(Figure 3 from~\cite{etmccharm}),
and plotted against the square of the lattice spacing. 
The rest mass is given by open circles and the kinetic 
mass by open triangles, red for twisted mass and 
blue for HISQ. Errors include the full lattice spacing 
error on each point.
}
\label{fig:mkin}
\end{figure}

\begin{table}
\begin{tabular}{llllll}
\hline
\hline
Set & $m_ca$ & $aM_{\eta_c}$ & $|a\vec{p}|$ & $aE_{\eta_c}$ & $c^2$ \\
\hline
2 & 0.63 & 1.80851(4) & 0.52880 & 1.88286(12) & 0.9814(15)  \\
 &  &  & 0.35000 & 1.84123(5) & 0.9748(5)  \\
 &  &  & 0.20000 & 1.81923(4) & 0.9715(6)  \\
\hline
4 & 0.413  & 1.28042(4) & 0.37486 & 1.33352(6) & 0.9878(8)  \\
 &   &  & 0.20000 & 1.29575(4) & 0.9873(7)  \\
\hline
\hline
\end{tabular}
\caption{Rest masses ($aM_{\eta_c}$) and energies ($aE_{\eta_c}$)
at non-zero momentum $|a\vec{p}|$ for the Goldstone 
$\eta_c$ meson on sets 2 (coarse) and 4 (fine). 
The rest masses differ slightly from those in 
Table~\ref{tab:massres} because they come from 
independent fits; on set 4 we have higher statistics here. 
The zero and non-zero momentum correlator are fitted 
simultaneously and the speed of light, $c^2$, extracted 
using eq.~(\ref{eq:csq}).
}
\label{tab:mkin}
\end{table}

\begin{figure}
\begin{center}
\includegraphics[width=0.9\hsize]{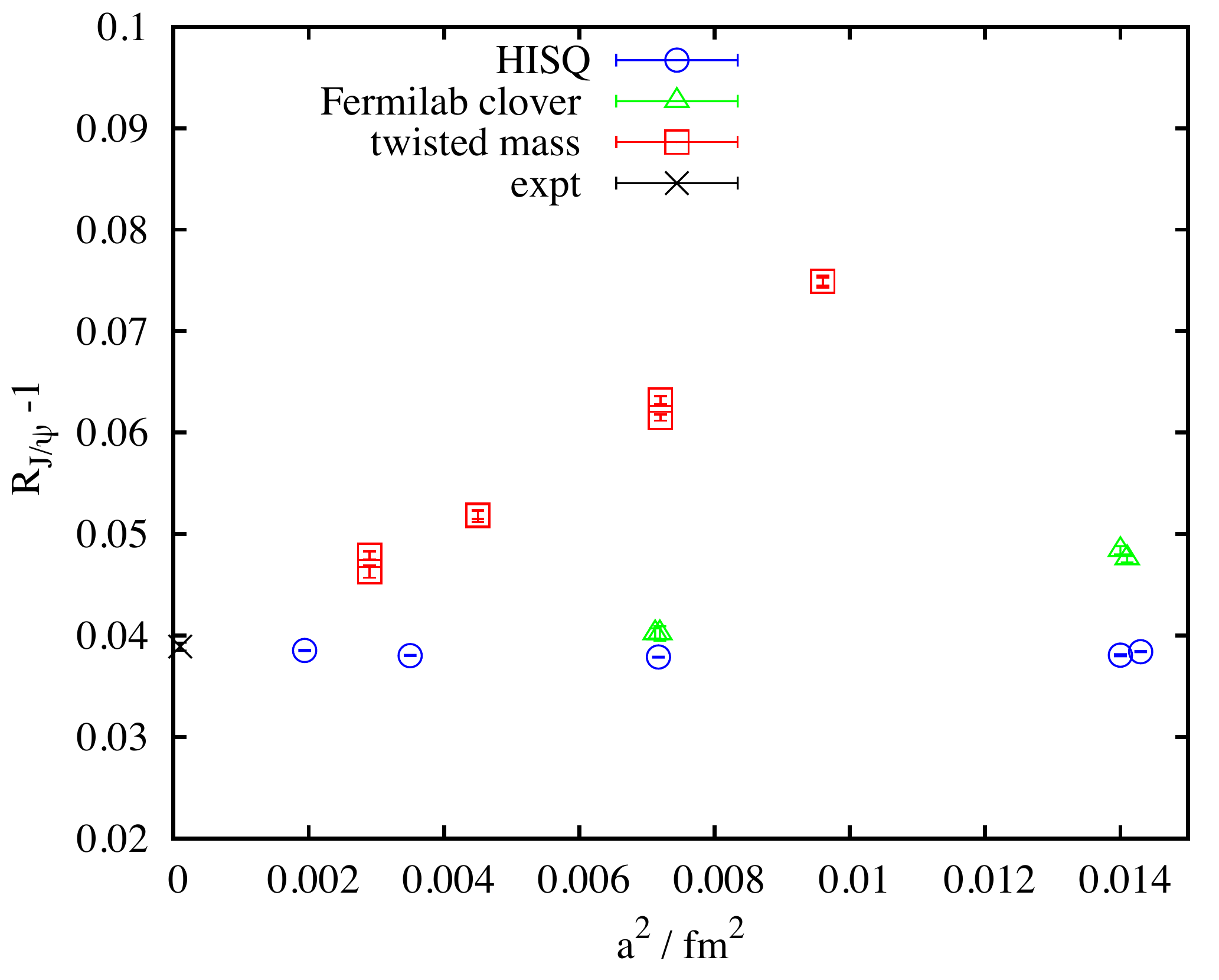}
\end{center}
\caption{ $R_{J/\psi} -1$ plotted against $a^2$ in 
$\mathrm{fm}^2$ for three different quark formalisms 
for the $c$ quark:
HISQ (this paper, blue open circles), twisted mass~(\cite{etmccharm}, red open squares) and Fermilab clover~(\cite{fnalcharm}, green 
open triangles, showing results on the two finest lattices 
only). $R_{J/\psi}$ is $M_{J/\psi}/M_{\eta_c}$ so 
$R_{J/\psi}-1 = \Delta M_{hyp}/M_{\eta_c}$.  
For the twisted mass and Fermilab clover results the heaviest 
and lightest sea quark masses are plotted at each value of 
the lattice spacing. Only statistical errors are shown. 
Additional errors from (twice) the lattice spacing error 
amount to 2\% for HISQ and Fermilab clover and 4-7\% for 
twisted mass. 
The black cross is the experimental average~\cite{pdg}, 
offset slightly from the origin for clarity. 
}
\label{fig:hypcomp}
\end{figure}

\begin{figure}
\begin{center}
\includegraphics[width=0.9\hsize]{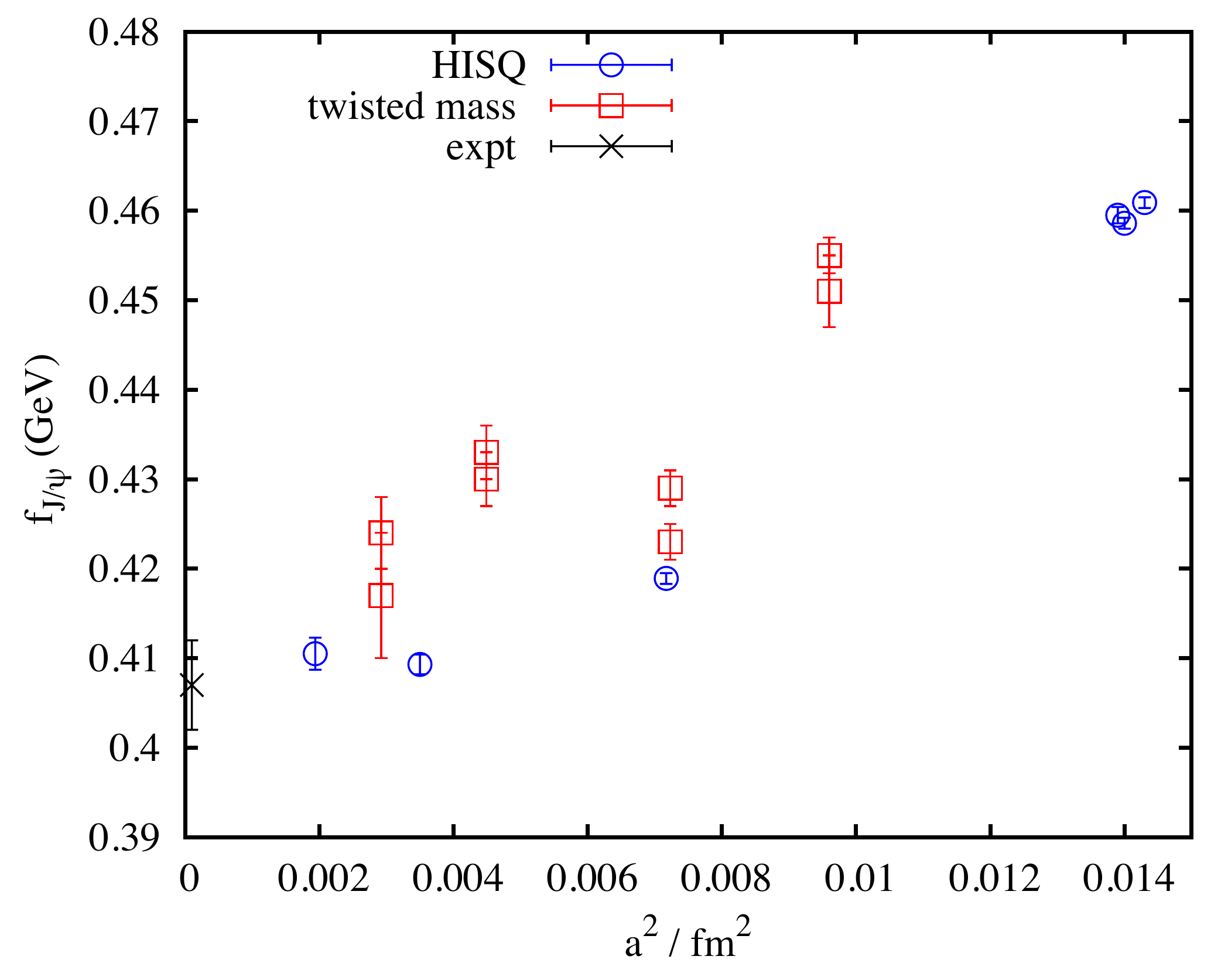}
\end{center}
\caption{ $f_{J/\psi}$ in GeV is plotted against $a^2$ in 
$\mathrm{fm}^2$ for 
the HISQ formalism (this paper, blue open circles) 
and for twisted mass~(\cite{etmccharm}, red open squares). 
For the twisted mass case the heaviest 
and lightest sea quark masses are plotted at each value of 
the lattice spacing. Only statistical errors are shown 
for the twisted mass results. 
For the HISQ results we show the raw data 
from Table~\ref{tab:massres} with statistical and uncorrelated lattice 
spacing errors. There is an additional error of 1.3\% 
from correlated lattice spacing and $Z$ factor uncertainties. 
The black cross is the experimental result from the 
average leptonic width~\cite{pdg}, 
offset slightly from the origin for clarity. 
}
\label{fig:fpsicomp}
\end{figure}

\begin{figure}
\begin{center}
\includegraphics[width=0.9\hsize]{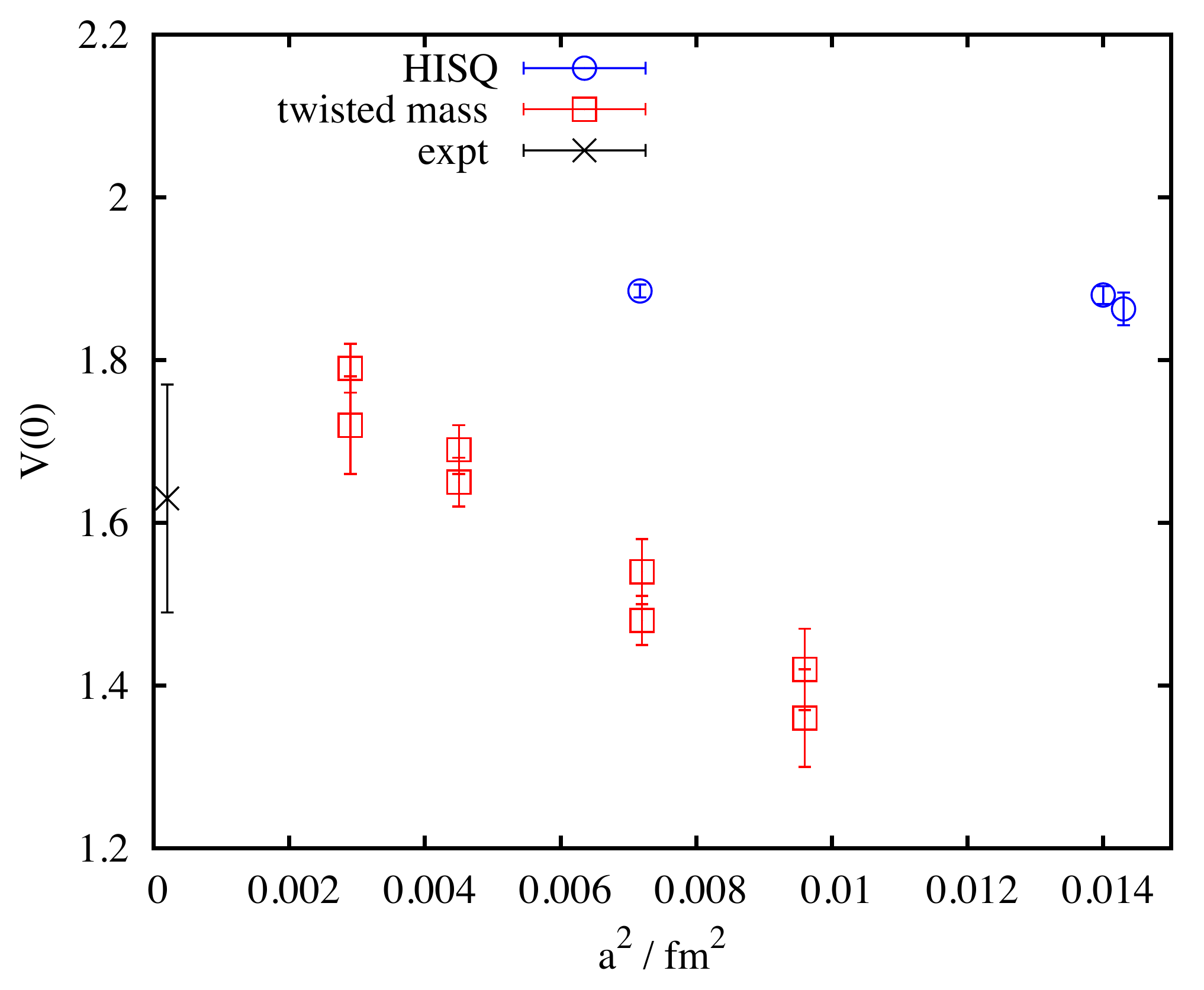}
\end{center}
\caption{ The vector form factor for 
$J/\psi \rightarrow \eta_c$  decay at $q^2=0$, $V(0)$, 
is plotted against $a^2$ in 
$\mathrm{fm}^2$ for 
the HISQ formalism (this paper, jpsigamma0 method, blue open circles) 
and for twisted mass~(\cite{etmccharm}, red open squares). 
For the twisted mass case the heaviest 
and lightest sea quark masses are plotted at each value of 
the lattice spacing. Errors include statistical errors 
and uncertainties in the $Z$ factors. 
The black cross is the experimental result from the 
rate for $J/\psi \rightarrow \eta_c \gamma$ decay~\cite{cleopsieta},
offset slightly from the origin for clarity. 
}
\label{fig:v0comp}
\end{figure}

Finally we discuss discretisation errors. For relativistic 
formalisms the scale of discretisation errors can be set 
by the quark mass when this is larger than $\Lambda_{QCD}$ 
and so they must be monitored closely when working 
with $c$ quarks. 

One simple way to do this is through study of 
the energy of mesons at non-zero spatial momentum. 
Because the HISQ formalism is a relativistic formalism 
we determine meson masses as the result of fitting 
zero-momentum meson correlators as described in 
section~\ref{sec:latt}. For heavy quarks discretisation 
errors mean that this mass, known as the `rest' or 
`static' mass, can differ from the mass that controls 
the momentum-dependence of the energy at non-zero 
spatial momentum. This latter mass is known as the `kinetic mass'. 
An equivalent statement is that the square of the speed-of-light, 
$c^2$, differs from 1, where~\cite{hisqdef} 
\begin{equation}
c^2(\vec{p}) = \frac{E^2(\vec{p}) - m^2}{{\vec{p}}^2}.
\label{eq:csq}
\end{equation}

For $\eta_c$ mesons we are able 
to determine $c^2$ very accurately with HISQ quarks 
when we have $\mathcal{O}(10,000)$ correlators 
as here. We fit zero and non-zero momentum 
simultaneously to the form given in eq.~(\ref{eq:fit1}) 
(although the zero momentum correlators have no 
oscillating component). From the fit we obtain the 
ground state mass, $M_0$, in the zero momentum case 
and the ground state energy, $E_0$, in the 
non-zero momentum case. The simultaneous fit allows us 
to take correlations into account to improve the error 
in $c^2$. Results are given in Table~\ref{tab:mkin}. 
$c^2$ is within 3\% of 1 but 
we can distinguish it from 1 and we can see that it 
depends on the spatial momentum. 
This is shown in the top plot of Figure~\ref{fig:csq} for 
the coarse lattices, set 2, and the fine lattices, set 4. 
All tree-level $a^2$ errors are removed in HISQ but 
$c^2$ can depend linearly on $a^2{\vec{p}}^2$ through 
$\mathcal{O}(\alpha_s)$ corrections. 
We see that the slope
 of $c^2$ with $a^2{\vec{p}}^2$ 
is much smaller on the fine lattices than the 
coarse, as $m_ca$ is reduced. 
This plot can be compared with earlier results 
in Figure 9 of~\cite{hisqscaling}, although note 
that those results show a jump as the lattice 
momentum changes from (1,1,1) to (2,0,0). 
This results from a rotationally-noninvariant discretisation 
error not evident in our results
because of the use of the phased boundary 
condition of eq.~(\ref{eq:twist}) to 
fix the momentum direction and simply change its 
magnitude.  

From the top plot of Figure~\ref{fig:csq} 
we can determine the value of $c^2$ at $a^2\vec{p}^2=0$. 
The results from the coarse 
and fine lattices are then shown in the lower plot 
to lie on a straight line as a 
function of $(m_ca)^2$. The small slope is compatible 
with the $(m_ca)^2$ dependence also being an 
$\mathcal{O}(\alpha_s)$ effect. The straight line 
clearly goes through 1 as $m_ca \rightarrow 0$ as 
it must. 

Another way to look at these discretisation errors 
is to compare the rest and kinetic masses for the 
$\eta_c$. 
The kinetic mass is given by~\cite{gray}: 
\begin{equation}
M_{kin} = \frac{|\vec{p}|^2 - (\Delta E)^2}{2\Delta E}
\label{eq:mkin}
\end{equation}
where $\Delta E = E(\vec{p}) - M_{rest}$. 
We also have 
$c^2 = M_{rest}/M_{kin}$~\cite{hisqdef}. 
In the absence of errors, for a relativistic formalism, 
we should have $M_{rest}=M_{kin}$
(i.e. $c^2=1$). 
In Figure~\ref{fig:mkin} we compare the rest and kinetic 
masses for HISQ $\eta_c$ mesons using $M_{kin}=M_{rest}/(c^2({\vec{p}}^2=0))$ 
determined from Figure~\ref{fig:csq}. 
The rest and kinetic masses differ by 3\% on the 
coarse lattices and 1.3\% on the fine lattices. 

Figure~\ref{fig:mkin} also compares results from the 
twisted mass formalism, from~\cite{etmccharm}. 
For that formalism there is a significantly larger 
difference between rest and kinetic masses for 
the $\eta_c$ meson, amounting to 35\%
on the coarsest lattice spacing. 
The twisted mass formalism has 
tree-level $a^2$ errors, so a larger effect would be
expected. 
The rest and kinetic masses agree in the 
continuum limit, as they must. 

Figure~\ref{fig:hypcomp} 
compares the hyperfine splitting in charmonium in units 
of the $\eta_c$ mass as a function of the square of the 
lattice spacing for the HISQ and twisted mass formalisms. 
The quantity plotted is $R_{J/\psi} -1$ where $R_{J/\psi}$ 
is defined in~\cite{etmccharm} as $M_{J/\psi}/M_{\eta_c}$ 
and results are given on each of their ensembles. The HISQ 
results are taken from Table~\ref{tab:massres}. 
We also show results from the Fermilab clover method~\cite{fnalcharm} 
on their two finest sets of ensembles which correspond 
to the coarse and fine lattices used here. 

All the results tend to the same continuum value which 
agrees with experiment. 
Much 
larger discretisation effects are visible for the twisted 
mass formalism than for HISQ. These are compatible 
with the tree level $a^2$ errors 
expected in that formalism, and with a mass scale of approximately 
2 GeV (i.e. $m_c$). 
These tree level errors are removed in the HISQ formalism, 
but $\alpha_s a^2$ errors remain. These seem to be small for 
this quantity. The Fermilab clover discretisation errors, 
although in principle $\alpha_s a$, are also relatively 
small over this range of $a$. 

Discretisation effects in the hyperfine splitting 
can also enter through tuning of the $c$ quark mass, 
because the hyperfine splitting is very sensitive to this.
As discussed earlier in this section, there can be 
significant differences between rest and kinetic masses 
for mesons made of heavy quarks, and either can be used 
to tune the quark mass. Both relativistic formalisms, HISQ 
and twisted mass, use the rest mass. The Fermilab formalism 
uses the kinetic mass. 

For $f_{J/\psi}$ the difference in discretisation effects 
between the HISQ and twisted mass formalisms is not as large. 
This is shown in Figure~\ref{fig:fpsicomp}. Again, answers 
in agreement are obtained in the continuum limit. 
For the moments of the vector correlator 
our results for $G^V_4$ (Fig~\ref{fig:rn}), which 
show very little dependence on the lattice spacing, can be 
compared to those for the twisted mass formalism 
in~\cite{etmcurrcurr}, where somewhat larger discretisation 
effects are visible. 

Finally in Figure~\ref{fig:v0comp} we compare results for 
the vector form factor for $J/\psi \rightarrow \eta_c$ decay, $V(0)$. 
Large discretisation effects are evident in the 
twisted mass results. Once again discretisation effects 
in the HISQ results are small. Agreement in the continuum limit 
is again clear, however. 

The HISQ results shown here give a further and more testing demonstration 
than that of~\cite{hisqdef, fdsorig} of how small the discretisation 
errors for the HISQ action are, even for a quark as heavy as charm. 

\bibliography{heavy}

\begin{thebibliography}{57}
\expandafter\ifx\csname natexlab\endcsname\relax\def\natexlab#1{#1}\fi
\expandafter\ifx\csname bibnamefont\endcsname\relax
  \def\bibnamefont#1{#1}\fi
\expandafter\ifx\csname bibfnamefont\endcsname\relax
  \def\bibfnamefont#1{#1}\fi
\expandafter\ifx\csname citenamefont\endcsname\relax
  \def\citenamefont#1{#1}\fi
\expandafter\ifx\csname url\endcsname\relax
  \def\url#1{\texttt{#1}}\fi
\expandafter\ifx\csname urlprefix\endcsname\relax\def\urlprefix{URL }\fi
\providecommand{\bibinfo}[2]{#2}
\providecommand{\eprint}[2][]{\url{#2}}

\bibitem[{\citenamefont{Davies et~al.}(2004)}]{ourlatqcd}
\bibinfo{author}{\bibfnamefont{C.}~\bibnamefont{Davies}} \bibnamefont{et~al.}
  (\bibinfo{collaboration}{HPQCD, UKQCD, MILC and Fermilab Lattice
  Collaborations}), \bibinfo{journal}{Phys.Rev.Lett.}
  \textbf{\bibinfo{volume}{92}}, \bibinfo{pages}{022001}
  (\bibinfo{year}{2004}), \eprint{hep-lat/0304004}.

\bibitem[{\citenamefont{Davies et~al.}(2010{\natexlab{a}})\citenamefont{Davies,
  McNeile, Follana, Lepage, Na et~al.}}]{fdsupdate}
\bibinfo{author}{\bibfnamefont{C.}~\bibnamefont{Davies}},
  \bibinfo{author}{\bibfnamefont{C.}~\bibnamefont{McNeile}},
  \bibinfo{author}{\bibfnamefont{E.}~\bibnamefont{Follana}},
  \bibinfo{author}{\bibfnamefont{G.}~\bibnamefont{Lepage}},
  \bibinfo{author}{\bibfnamefont{H.}~\bibnamefont{Na}}, \bibnamefont{et~al.}
  (\bibinfo{collaboration}{HPQCD Collaboration}), \bibinfo{journal}{Phys.Rev.}
  \textbf{\bibinfo{volume}{D82}}, \bibinfo{pages}{114504}
  (\bibinfo{year}{2010}{\natexlab{a}}), \eprint{1008.4018}.

\bibitem[{\citenamefont{Dowdall
  et~al.}(2012{\natexlab{a}})\citenamefont{Dowdall, Davies, Hammant, and
  Horgan}}]{rachelnew}
\bibinfo{author}{\bibfnamefont{R.}~\bibnamefont{Dowdall}},
  \bibinfo{author}{\bibfnamefont{C.}~\bibnamefont{Davies}},
  \bibinfo{author}{\bibfnamefont{T.}~\bibnamefont{Hammant}}, \bibnamefont{and}
  \bibinfo{author}{\bibfnamefont{R.}~\bibnamefont{Horgan}}
  (\bibinfo{collaboration}{HPQCD Collaboration})
  (\bibinfo{year}{2012}{\natexlab{a}}), \eprint{1207.5149}.

\bibitem[{\citenamefont{Davies}(2011)}]{cdlat11}
\bibinfo{author}{\bibfnamefont{C.}~\bibnamefont{Davies}},
  \bibinfo{journal}{PoS} \textbf{\bibinfo{volume}{LATTICE2011}},
  \bibinfo{pages}{019} (\bibinfo{year}{2011}), \eprint{1203.3862}.

\bibitem[{\citenamefont{Laiho et~al.}(2011)\citenamefont{Laiho, Lunghi, and
  Van~de Water}}]{lunghilat11}
\bibinfo{author}{\bibfnamefont{J.}~\bibnamefont{Laiho}},
  \bibinfo{author}{\bibfnamefont{E.}~\bibnamefont{Lunghi}}, \bibnamefont{and}
  \bibinfo{author}{\bibfnamefont{R.}~\bibnamefont{Van~de Water}},
  \bibinfo{journal}{PoS} \textbf{\bibinfo{volume}{LATTICE2011}},
  \bibinfo{pages}{018} (\bibinfo{year}{2011}), \eprint{1204.0791}.

\bibitem[{\citenamefont{Bazavov et~al.}(2010{\natexlab{a}})}]{milc10}
\bibinfo{author}{\bibfnamefont{A.}~\bibnamefont{Bazavov}} \bibnamefont{et~al.}
  (\bibinfo{collaboration}{MILC Collaboration}), \bibinfo{journal}{PoS}
  \textbf{\bibinfo{volume}{LATTICE2010}}, \bibinfo{pages}{074}
  (\bibinfo{year}{2010}{\natexlab{a}}), \eprint{1012.0868}.

\bibitem[{\citenamefont{Beringer et~al.}(2012)}]{pdg}
\bibinfo{author}{\bibfnamefont{J.}~\bibnamefont{Beringer}} \bibnamefont{et~al.}
  (\bibinfo{collaboration}{Particle Data Group}), \bibinfo{journal}{Phys. Rev.}
  \textbf{\bibinfo{volume}{D86}}, \bibinfo{pages}{010001}
  (\bibinfo{year}{2012}).

\bibitem[{\citenamefont{Dudek et~al.}(2006)\citenamefont{Dudek, Edwards, and
  Richards}}]{dudekcharm}
\bibinfo{author}{\bibfnamefont{J.~J.} \bibnamefont{Dudek}},
  \bibinfo{author}{\bibfnamefont{R.~G.} \bibnamefont{Edwards}},
  \bibnamefont{and} \bibinfo{author}{\bibfnamefont{D.~G.}
  \bibnamefont{Richards}}, \bibinfo{journal}{Phys.Rev.}
  \textbf{\bibinfo{volume}{D73}}, \bibinfo{pages}{074507}
  (\bibinfo{year}{2006}), \eprint{hep-ph/0601137}.

\bibitem[{\citenamefont{Chen et~al.}(2011)\citenamefont{Chen, Du, Guo, Li, Liu
  et~al.}}]{clqcdcharm}
\bibinfo{author}{\bibfnamefont{Y.}~\bibnamefont{Chen}},
  \bibinfo{author}{\bibfnamefont{D.-C.} \bibnamefont{Du}},
  \bibinfo{author}{\bibfnamefont{B.-Z.} \bibnamefont{Guo}},
  \bibinfo{author}{\bibfnamefont{N.}~\bibnamefont{Li}},
  \bibinfo{author}{\bibfnamefont{C.}~\bibnamefont{Liu}}, \bibnamefont{et~al.},
  \bibinfo{journal}{Phys.Rev.} \textbf{\bibinfo{volume}{D84}},
  \bibinfo{pages}{034503} (\bibinfo{year}{2011}), \eprint{1104.2655}.

\bibitem[{\citenamefont{Becirevic and Sanfilippo}(2012)}]{etmccharm}
\bibinfo{author}{\bibfnamefont{D.}~\bibnamefont{Becirevic}} \bibnamefont{and}
  \bibinfo{author}{\bibfnamefont{F.}~\bibnamefont{Sanfilippo}}
  (\bibinfo{year}{2012}), \eprint{1206.1445}.

\bibitem[{\citenamefont{Follana et~al.}(2007)}]{hisqdef}
\bibinfo{author}{\bibfnamefont{E.}~\bibnamefont{Follana}} \bibnamefont{et~al.}
  (\bibinfo{collaboration}{HPQCD Collaboration}), \bibinfo{journal}{Phys.Rev.}
  \textbf{\bibinfo{volume}{D75}}, \bibinfo{pages}{054502}
  (\bibinfo{year}{2007}), \eprint{hep-lat/0610092}.

\bibitem[{\citenamefont{Bazavov
  et~al.}(2010{\natexlab{b}})\citenamefont{Bazavov, Toussaint, Bernard, Laiho,
  DeTar et~al.}}]{milcreview}
\bibinfo{author}{\bibfnamefont{A.}~\bibnamefont{Bazavov}},
  \bibinfo{author}{\bibfnamefont{D.}~\bibnamefont{Toussaint}},
  \bibinfo{author}{\bibfnamefont{C.}~\bibnamefont{Bernard}},
  \bibinfo{author}{\bibfnamefont{J.}~\bibnamefont{Laiho}},
  \bibinfo{author}{\bibfnamefont{C.}~\bibnamefont{DeTar}},
  \bibnamefont{et~al.}, \bibinfo{journal}{Rev.Mod.Phys.}
  \textbf{\bibinfo{volume}{82}}, \bibinfo{pages}{1349}
  (\bibinfo{year}{2010}{\natexlab{b}}), \eprint{0903.3598}.

\bibitem[{\citenamefont{Davies et~al.}(2010{\natexlab{b}})\citenamefont{Davies,
  Follana, Kendall, Lepage, and McNeile}}]{oldr1paper}
\bibinfo{author}{\bibfnamefont{C.}~\bibnamefont{Davies}},
  \bibinfo{author}{\bibfnamefont{E.}~\bibnamefont{Follana}},
  \bibinfo{author}{\bibfnamefont{I.}~\bibnamefont{Kendall}},
  \bibinfo{author}{\bibfnamefont{G.~P.} \bibnamefont{Lepage}},
  \bibnamefont{and} \bibinfo{author}{\bibfnamefont{C.}~\bibnamefont{McNeile}}
  (\bibinfo{collaboration}{HPQCD Collaboration}), \bibinfo{journal}{Phys.Rev.}
  \textbf{\bibinfo{volume}{D81}}, \bibinfo{pages}{034506}
  (\bibinfo{year}{2010}{\natexlab{b}}), \eprint{0910.1229}.

\bibitem[{\citenamefont{Gregory et~al.}(2011)\citenamefont{Gregory, Davies,
  Kendall, Koponen, Wong et~al.}}]{gregory}
\bibinfo{author}{\bibfnamefont{E.~B.} \bibnamefont{Gregory}},
  \bibinfo{author}{\bibfnamefont{C.~T.} \bibnamefont{Davies}},
  \bibinfo{author}{\bibfnamefont{I.~D.} \bibnamefont{Kendall}},
  \bibinfo{author}{\bibfnamefont{J.}~\bibnamefont{Koponen}},
  \bibinfo{author}{\bibfnamefont{K.}~\bibnamefont{Wong}}, \bibnamefont{et~al.}
  (\bibinfo{collaboration}{HPQCD Collaboration}), \bibinfo{journal}{Phys.Rev.}
  \textbf{\bibinfo{volume}{D83}}, \bibinfo{pages}{014506}
  (\bibinfo{year}{2011}), \eprint{1010.3848}.

\bibitem[{\citenamefont{Lepage et~al.}(2002)}]{gplbayes}
\bibinfo{author}{\bibfnamefont{G.~P.} \bibnamefont{Lepage}}
  \bibnamefont{et~al.}, \bibinfo{journal}{Nucl. Phys. Proc. Suppl.}
  \textbf{\bibinfo{volume}{106}}, \bibinfo{pages}{12} (\bibinfo{year}{2002}),
  \eprint{hep-lat/0110175}.

\bibitem[{\citenamefont{Davies et~al.}(1998)}]{daviesoldups}
\bibinfo{author}{\bibfnamefont{C.}~\bibnamefont{Davies}} \bibnamefont{et~al.}
  (\bibinfo{collaboration}{UKQCD Collaboration}), \bibinfo{journal}{Phys.Rev.}
  \textbf{\bibinfo{volume}{D58}}, \bibinfo{pages}{054505}
  (\bibinfo{year}{1998}), \eprint{hep-lat/9802024}.

\bibitem[{\citenamefont{Levkova and DeTar}(2011)}]{levkova}
\bibinfo{author}{\bibfnamefont{L.}~\bibnamefont{Levkova}} \bibnamefont{and}
  \bibinfo{author}{\bibfnamefont{C.}~\bibnamefont{DeTar}},
  \bibinfo{journal}{Phys.Rev.} \textbf{\bibinfo{volume}{D83}},
  \bibinfo{pages}{074504} (\bibinfo{year}{2011}), \eprint{1012.1837}.

\bibitem[{\citenamefont{Vinokurova et~al.}(2011)}]{bellehyp}
\bibinfo{author}{\bibfnamefont{A.}~\bibnamefont{Vinokurova}}
  \bibnamefont{et~al.} (\bibinfo{collaboration}{Belle collaboration}),
  \bibinfo{journal}{Phys.Lett.} \textbf{\bibinfo{volume}{B706}},
  \bibinfo{pages}{139} (\bibinfo{year}{2011}), \eprint{1105.0978}.

\bibitem[{\citenamefont{Ablikim et~al.}(2012)}]{bes3etac}
\bibinfo{author}{\bibfnamefont{M.}~\bibnamefont{Ablikim}} \bibnamefont{et~al.}
  (\bibinfo{collaboration}{BESIII Collaboration}),
  \bibinfo{journal}{Phys.Rev.Lett.} \textbf{\bibinfo{volume}{108}},
  \bibinfo{pages}{222002} (\bibinfo{year}{2012}), \eprint{1111.0398}.

\bibitem[{\citenamefont{Erler}(1999)}]{alpha-em}
\bibinfo{author}{\bibfnamefont{J.}~\bibnamefont{Erler}},
  \bibinfo{journal}{Phys.Rev.} \textbf{\bibinfo{volume}{D59}},
  \bibinfo{pages}{054008} (\bibinfo{year}{1999}), \eprint{hep-ph/9803453}.

\bibitem[{\citenamefont{Adams et~al.}(2008)}]{cleo3gamma}
\bibinfo{author}{\bibfnamefont{G.}~\bibnamefont{Adams}} \bibnamefont{et~al.}
  (\bibinfo{collaboration}{CLEO Collaboration}),
  \bibinfo{journal}{Phys.Rev.Lett.} \textbf{\bibinfo{volume}{101}},
  \bibinfo{pages}{101801} (\bibinfo{year}{2008}), \eprint{0806.0671}.

\bibitem[{\citenamefont{Kuhn et~al.}(2007)\citenamefont{Kuhn, Steinhauser, and
  Sturm}}]{kuhnmc07}
\bibinfo{author}{\bibfnamefont{J.~H.} \bibnamefont{Kuhn}},
  \bibinfo{author}{\bibfnamefont{M.}~\bibnamefont{Steinhauser}},
  \bibnamefont{and} \bibinfo{author}{\bibfnamefont{C.}~\bibnamefont{Sturm}},
  \bibinfo{journal}{Nucl.Phys.} \textbf{\bibinfo{volume}{B778}},
  \bibinfo{pages}{192} (\bibinfo{year}{2007}), \eprint{hep-ph/0702103}.

\bibitem[{\citenamefont{Dehnadi et~al.}(2011)\citenamefont{Dehnadi, Hoang,
  Mateu, and Zebarjad}}]{hoangmc11}
\bibinfo{author}{\bibfnamefont{B.}~\bibnamefont{Dehnadi}},
  \bibinfo{author}{\bibfnamefont{A.~H.} \bibnamefont{Hoang}},
  \bibinfo{author}{\bibfnamefont{V.}~\bibnamefont{Mateu}}, \bibnamefont{and}
  \bibinfo{author}{\bibfnamefont{S.~M.} \bibnamefont{Zebarjad}}
  (\bibinfo{year}{2011}), \eprint{1102.2264}.

\bibitem[{\citenamefont{Eichten et~al.}(2008)\citenamefont{Eichten, Godfrey,
  Mahlke, and Rosner}}]{eichtenrosner}
\bibinfo{author}{\bibfnamefont{E.}~\bibnamefont{Eichten}},
  \bibinfo{author}{\bibfnamefont{S.}~\bibnamefont{Godfrey}},
  \bibinfo{author}{\bibfnamefont{H.}~\bibnamefont{Mahlke}}, \bibnamefont{and}
  \bibinfo{author}{\bibfnamefont{J.~L.} \bibnamefont{Rosner}},
  \bibinfo{journal}{Rev.Mod.Phys.} \textbf{\bibinfo{volume}{80}},
  \bibinfo{pages}{1161} (\bibinfo{year}{2008}), \eprint{hep-ph/0701208}.

\bibitem[{\citenamefont{Brambilla et~al.}(2011)\citenamefont{Brambilla,
  Eidelman, Heltsley, Vogt, Bodwin et~al.}}]{qwg10}
\bibinfo{author}{\bibfnamefont{N.}~\bibnamefont{Brambilla}},
  \bibinfo{author}{\bibfnamefont{S.}~\bibnamefont{Eidelman}},
  \bibinfo{author}{\bibfnamefont{B.}~\bibnamefont{Heltsley}},
  \bibinfo{author}{\bibfnamefont{R.}~\bibnamefont{Vogt}},
  \bibinfo{author}{\bibfnamefont{G.}~\bibnamefont{Bodwin}},
  \bibnamefont{et~al.}, \bibinfo{journal}{Eur.Phys.J.}
  \textbf{\bibinfo{volume}{C71}}, \bibinfo{pages}{1534} (\bibinfo{year}{2011}),
  \eprint{1010.5827}.

\bibitem[{\citenamefont{Mitchell et~al.}(2009)}]{cleopsieta}
\bibinfo{author}{\bibfnamefont{R.}~\bibnamefont{Mitchell}} \bibnamefont{et~al.}
  (\bibinfo{collaboration}{CLEO Collaboration}),
  \bibinfo{journal}{Phys.Rev.Lett.} \textbf{\bibinfo{volume}{102}},
  \bibinfo{pages}{011801} (\bibinfo{year}{2009}), \eprint{0805.0252}.

\bibitem[{\citenamefont{de~Divitiis et~al.}(2004)\citenamefont{de~Divitiis,
  Petronzio, and Tantalo}}]{firsttwist}
\bibinfo{author}{\bibfnamefont{G.}~\bibnamefont{de~Divitiis}},
  \bibinfo{author}{\bibfnamefont{R.}~\bibnamefont{Petronzio}},
  \bibnamefont{and} \bibinfo{author}{\bibfnamefont{N.}~\bibnamefont{Tantalo}},
  \bibinfo{journal}{Phys.Lett.} \textbf{\bibinfo{volume}{B595}},
  \bibinfo{pages}{408} (\bibinfo{year}{2004}), \eprint{hep-lat/0405002}.

\bibitem[{\citenamefont{Guadagnoli et~al.}(2006)\citenamefont{Guadagnoli,
  Mescia, and Simula}}]{etmctwist}
\bibinfo{author}{\bibfnamefont{D.}~\bibnamefont{Guadagnoli}},
  \bibinfo{author}{\bibfnamefont{F.}~\bibnamefont{Mescia}}, \bibnamefont{and}
  \bibinfo{author}{\bibfnamefont{S.}~\bibnamefont{Simula}},
  \bibinfo{journal}{Phys.Rev.} \textbf{\bibinfo{volume}{D73}},
  \bibinfo{pages}{114504} (\bibinfo{year}{2006}), \eprint{hep-lat/0512020}.

\bibitem[{\citenamefont{Burch et~al.}(2010)\citenamefont{Burch, DeTar,
  Di~Pierro, El-Khadra, Freeland et~al.}}]{fnalcharm}
\bibinfo{author}{\bibfnamefont{T.}~\bibnamefont{Burch}},
  \bibinfo{author}{\bibfnamefont{C.}~\bibnamefont{DeTar}},
  \bibinfo{author}{\bibfnamefont{M.}~\bibnamefont{Di~Pierro}},
  \bibinfo{author}{\bibfnamefont{A.}~\bibnamefont{El-Khadra}},
  \bibinfo{author}{\bibfnamefont{E.}~\bibnamefont{Freeland}},
  \bibnamefont{et~al.}, \bibinfo{journal}{Phys.Rev.}
  \textbf{\bibinfo{volume}{D81}}, \bibinfo{pages}{034508}
  (\bibinfo{year}{2010}), \eprint{0912.2701}.

\bibitem[{\citenamefont{Liu et~al.}(2012)\citenamefont{Liu, Moir, Peardon,
  Ryan, Thomas et~al.}}]{ryancharm}
\bibinfo{author}{\bibfnamefont{L.}~\bibnamefont{Liu}},
  \bibinfo{author}{\bibfnamefont{G.}~\bibnamefont{Moir}},
  \bibinfo{author}{\bibfnamefont{M.}~\bibnamefont{Peardon}},
  \bibinfo{author}{\bibfnamefont{S.~M.} \bibnamefont{Ryan}},
  \bibinfo{author}{\bibfnamefont{C.~E.} \bibnamefont{Thomas}},
  \bibnamefont{et~al.} (\bibinfo{year}{2012}), \eprint{1204.5425}.

\bibitem[{\citenamefont{Bali et~al.}(2011)\citenamefont{Bali, Collins, and
  Ehmann}}]{saracharm}
\bibinfo{author}{\bibfnamefont{G.~S.} \bibnamefont{Bali}},
  \bibinfo{author}{\bibfnamefont{S.}~\bibnamefont{Collins}}, \bibnamefont{and}
  \bibinfo{author}{\bibfnamefont{C.}~\bibnamefont{Ehmann}},
  \bibinfo{journal}{Phys.Rev.} \textbf{\bibinfo{volume}{D84}},
  \bibinfo{pages}{094506} (\bibinfo{year}{2011}), \eprint{1110.2381}.

\bibitem[{\citenamefont{Namekawa et~al.}(2011)}]{pacscharm}
\bibinfo{author}{\bibfnamefont{Y.}~\bibnamefont{Namekawa}} \bibnamefont{et~al.}
  (\bibinfo{collaboration}{PACS-CS Collaboration}),
  \bibinfo{journal}{Phys.Rev.} \textbf{\bibinfo{volume}{D84}},
  \bibinfo{pages}{074505} (\bibinfo{year}{2011}), \eprint{1104.4600}.

\bibitem[{\citenamefont{Mohler and Woloshyn}(2011)}]{mohlercharm}
\bibinfo{author}{\bibfnamefont{D.}~\bibnamefont{Mohler}} \bibnamefont{and}
  \bibinfo{author}{\bibfnamefont{R.}~\bibnamefont{Woloshyn}},
  \bibinfo{journal}{Phys.Rev.} \textbf{\bibinfo{volume}{D84}},
  \bibinfo{pages}{054505} (\bibinfo{year}{2011}), \eprint{1103.5506}.

\bibitem[{\citenamefont{Asner et~al.}(2009)\citenamefont{Asner, Barnes, Bian,
  Bigi, Brambilla et~al.}}]{besbook}
\bibinfo{author}{\bibfnamefont{D.}~\bibnamefont{Asner}},
  \bibinfo{author}{\bibfnamefont{T.}~\bibnamefont{Barnes}},
  \bibinfo{author}{\bibfnamefont{J.}~\bibnamefont{Bian}},
  \bibinfo{author}{\bibfnamefont{I.}~\bibnamefont{Bigi}},
  \bibinfo{author}{\bibfnamefont{N.}~\bibnamefont{Brambilla}},
  \bibnamefont{et~al.}, \bibinfo{journal}{Int.J.Mod.Phys.}
  \textbf{\bibinfo{volume}{A24}}, \bibinfo{pages}{S1} (\bibinfo{year}{2009}),
  \eprint{0809.1869}.

\bibitem[{\citenamefont{Yuan}(2012)}]{yuannote}
\bibinfo{author}{\bibfnamefont{C.}~\bibnamefont{Yuan}},
  \bibinfo{journal}{private communication}  (\bibinfo{year}{2012}).

\bibitem[{\citenamefont{Follana et~al.}(2008)\citenamefont{Follana, Davies,
  Lepage, and Shigemitsu}}]{fdsorig}
\bibinfo{author}{\bibfnamefont{E.}~\bibnamefont{Follana}},
  \bibinfo{author}{\bibfnamefont{C.}~\bibnamefont{Davies}},
  \bibinfo{author}{\bibfnamefont{G.}~\bibnamefont{Lepage}}, \bibnamefont{and}
  \bibinfo{author}{\bibfnamefont{J.}~\bibnamefont{Shigemitsu}}
  (\bibinfo{collaboration}{HPQCD Collaboration}),
  \bibinfo{journal}{Phys.Rev.Lett.} \textbf{\bibinfo{volume}{100}},
  \bibinfo{pages}{062002} (\bibinfo{year}{2008}), \eprint{0706.1726}.

\bibitem[{\citenamefont{Na et~al.}(2012)\citenamefont{Na, Davies, Follana,
  Lepage, and Shigemitsu}}]{nafd}
\bibinfo{author}{\bibfnamefont{H.}~\bibnamefont{Na}},
  \bibinfo{author}{\bibfnamefont{C.~T.} \bibnamefont{Davies}},
  \bibinfo{author}{\bibfnamefont{E.}~\bibnamefont{Follana}},
  \bibinfo{author}{\bibfnamefont{G.~P.} \bibnamefont{Lepage}},
  \bibnamefont{and}
  \bibinfo{author}{\bibfnamefont{J.}~\bibnamefont{Shigemitsu}}
  (\bibinfo{collaboration}{HPQCD Collaboration}) (\bibinfo{year}{2012}),
  \eprint{1206.4936}.

\bibitem[{\citenamefont{Dowdall et~al.}(2012{\natexlab{b}})}]{dowdallr1}
\bibinfo{author}{\bibfnamefont{R.}~\bibnamefont{Dowdall}} \bibnamefont{et~al.}
  (\bibinfo{collaboration}{HPQCD Collaboration}), \bibinfo{journal}{Phys.Rev.}
  \textbf{\bibinfo{volume}{D85}}, \bibinfo{pages}{054509}
  (\bibinfo{year}{2012}{\natexlab{b}}), \eprint{1110.6887}.

\bibitem[{\citenamefont{Na et~al.}(2010)\citenamefont{Na, Davies, Follana,
  Lepage, and Shigemitsu}}]{na1}
\bibinfo{author}{\bibfnamefont{H.}~\bibnamefont{Na}},
  \bibinfo{author}{\bibfnamefont{C.~T.} \bibnamefont{Davies}},
  \bibinfo{author}{\bibfnamefont{E.}~\bibnamefont{Follana}},
  \bibinfo{author}{\bibfnamefont{G.~P.} \bibnamefont{Lepage}},
  \bibnamefont{and}
  \bibinfo{author}{\bibfnamefont{J.}~\bibnamefont{Shigemitsu}}
  (\bibinfo{collaboration}{HPQCD Collaboration}), \bibinfo{journal}{Phys.Rev.}
  \textbf{\bibinfo{volume}{D82}}, \bibinfo{pages}{114506}
  (\bibinfo{year}{2010}), \eprint{1008.4562}.

\bibitem[{\citenamefont{Na et~al.}(2011)\citenamefont{Na, Davies, Follana,
  Koponen, Lepage et~al.}}]{na2}
\bibinfo{author}{\bibfnamefont{H.}~\bibnamefont{Na}},
  \bibinfo{author}{\bibfnamefont{C.~T.} \bibnamefont{Davies}},
  \bibinfo{author}{\bibfnamefont{E.}~\bibnamefont{Follana}},
  \bibinfo{author}{\bibfnamefont{J.}~\bibnamefont{Koponen}},
  \bibinfo{author}{\bibfnamefont{G.~P.} \bibnamefont{Lepage}},
  \bibnamefont{et~al.} (\bibinfo{collaboration}{HPQCD Collaboration}),
  \bibinfo{journal}{Phys.Rev.} \textbf{\bibinfo{volume}{D84}},
  \bibinfo{pages}{114505} (\bibinfo{year}{2011}), \eprint{1109.1501}.

\bibitem[{\citenamefont{Koponen et~al.}(2011)}]{jonnalat11}
\bibinfo{author}{\bibfnamefont{J.}~\bibnamefont{Koponen}} \bibnamefont{et~al.}
  (\bibinfo{collaboration}{HPQCD Collaboration}), \bibinfo{journal}{PoS}
  \textbf{\bibinfo{volume}{LATTICE2011}}, \bibinfo{pages}{286}
  (\bibinfo{year}{2011}), \eprint{1111.0225}.

\bibitem[{\citenamefont{Donald et~al.}(2011)\citenamefont{Donald, Davies, and
  Koponen}}]{gordonlat11}
\bibinfo{author}{\bibfnamefont{G.}~\bibnamefont{Donald}},
  \bibinfo{author}{\bibfnamefont{C.}~\bibnamefont{Davies}}, \bibnamefont{and}
  \bibinfo{author}{\bibfnamefont{J.}~\bibnamefont{Koponen}}
  (\bibinfo{collaboration}{HPQCD Collaboration}), \bibinfo{journal}{PoS}
  \textbf{\bibinfo{volume}{LATTICE2011}}, \bibinfo{pages}{278}
  (\bibinfo{year}{2011}), \eprint{1111.0254}.

\bibitem[{\citenamefont{Donald et~al.}(2012)\citenamefont{Donald, Koponen
  et~al.}}]{usinprep}
\bibinfo{author}{\bibfnamefont{G.}~\bibnamefont{Donald}},
  \bibinfo{author}{\bibfnamefont{J.}~\bibnamefont{Koponen}},
  \bibnamefont{et~al.} (\bibinfo{collaboration}{HPQCD Collaboration}),
  \bibinfo{journal}{in preparation}  (\bibinfo{year}{2012}).

\bibitem[{\citenamefont{Edwards and Joo}(2005)}]{chroma}
\bibinfo{author}{\bibfnamefont{R.~G.} \bibnamefont{Edwards}} \bibnamefont{and}
  \bibinfo{author}{\bibfnamefont{B.}~\bibnamefont{Joo}}
  (\bibinfo{collaboration}{SciDAC, LHPC and UKQCD Collaborations}),
  \bibinfo{journal}{Nucl.Phys.Proc.Suppl.} \textbf{\bibinfo{volume}{140}},
  \bibinfo{pages}{832} (\bibinfo{year}{2005}), \eprint{hep-lat/0409003}.

\bibitem[{\citenamefont{Chetyrkin et~al.}(2006)\citenamefont{Chetyrkin, Kuhn,
  and Sturm}}]{qcdpt1}
\bibinfo{author}{\bibfnamefont{K.}~\bibnamefont{Chetyrkin}},
  \bibinfo{author}{\bibfnamefont{J.~H.} \bibnamefont{Kuhn}}, \bibnamefont{and}
  \bibinfo{author}{\bibfnamefont{C.}~\bibnamefont{Sturm}},
  \bibinfo{journal}{Eur.Phys.J.} \textbf{\bibinfo{volume}{C48}},
  \bibinfo{pages}{107} (\bibinfo{year}{2006}), \eprint{hep-ph/0604234}.

\bibitem[{\citenamefont{Boughezal et~al.}(2006)\citenamefont{Boughezal, Czakon,
  and Schutzmeier}}]{qcdpt2}
\bibinfo{author}{\bibfnamefont{R.}~\bibnamefont{Boughezal}},
  \bibinfo{author}{\bibfnamefont{M.}~\bibnamefont{Czakon}}, \bibnamefont{and}
  \bibinfo{author}{\bibfnamefont{T.}~\bibnamefont{Schutzmeier}},
  \bibinfo{journal}{Phys.Rev.} \textbf{\bibinfo{volume}{D74}},
  \bibinfo{pages}{074006} (\bibinfo{year}{2006}), \eprint{hep-ph/0605023}.

\bibitem[{\citenamefont{Maier et~al.}(2008)\citenamefont{Maier, Maierhofer, and
  Marqaurd}}]{qcdpt3}
\bibinfo{author}{\bibfnamefont{A.}~\bibnamefont{Maier}},
  \bibinfo{author}{\bibfnamefont{P.}~\bibnamefont{Maierhofer}},
  \bibnamefont{and} \bibinfo{author}{\bibfnamefont{P.}~\bibnamefont{Marqaurd}},
  \bibinfo{journal}{Phys.Lett.} \textbf{\bibinfo{volume}{B669}},
  \bibinfo{pages}{88} (\bibinfo{year}{2008}), \eprint{0806.3405}.

\bibitem[{\citenamefont{Maier et~al.}(2010)\citenamefont{Maier, Maierhofer,
  Marquard, and Smirnov}}]{qcdpt4}
\bibinfo{author}{\bibfnamefont{A.}~\bibnamefont{Maier}},
  \bibinfo{author}{\bibfnamefont{P.}~\bibnamefont{Maierhofer}},
  \bibinfo{author}{\bibfnamefont{P.}~\bibnamefont{Marquard}}, \bibnamefont{and}
  \bibinfo{author}{\bibfnamefont{A.}~\bibnamefont{Smirnov}},
  \bibinfo{journal}{Nucl.Phys.} \textbf{\bibinfo{volume}{B824}},
  \bibinfo{pages}{1} (\bibinfo{year}{2010}), \eprint{0907.2117}.

\bibitem[{\citenamefont{Kiyo et~al.}(2009)\citenamefont{Kiyo, Maier,
  Maierhofer, and Marquard}}]{qcdpt5}
\bibinfo{author}{\bibfnamefont{Y.}~\bibnamefont{Kiyo}},
  \bibinfo{author}{\bibfnamefont{A.}~\bibnamefont{Maier}},
  \bibinfo{author}{\bibfnamefont{P.}~\bibnamefont{Maierhofer}},
  \bibnamefont{and} \bibinfo{author}{\bibfnamefont{P.}~\bibnamefont{Marquard}},
  \bibinfo{journal}{Nucl.Phys.} \textbf{\bibinfo{volume}{B823}},
  \bibinfo{pages}{269} (\bibinfo{year}{2009}), \eprint{0907.2120}.

\bibitem[{\citenamefont{Allison et~al.}(2008)}]{firstcurrcurr}
\bibinfo{author}{\bibfnamefont{I.}~\bibnamefont{Allison}} \bibnamefont{et~al.},
  \bibinfo{journal}{Phys.Rev.} \textbf{\bibinfo{volume}{D78}},
  \bibinfo{pages}{054513} (\bibinfo{year}{2008}), \eprint{0805.2999}.

\bibitem[{\citenamefont{McNeile et~al.}(2010)\citenamefont{McNeile, Davies,
  Follana, Hornbostel, and Lepage}}]{bcmasses}
\bibinfo{author}{\bibfnamefont{C.}~\bibnamefont{McNeile}},
  \bibinfo{author}{\bibfnamefont{C.}~\bibnamefont{Davies}},
  \bibinfo{author}{\bibfnamefont{E.}~\bibnamefont{Follana}},
  \bibinfo{author}{\bibfnamefont{K.}~\bibnamefont{Hornbostel}},
  \bibnamefont{and} \bibinfo{author}{\bibfnamefont{G.}~\bibnamefont{Lepage}}
  (\bibinfo{collaboration}{HPQCD Collaboration}), \bibinfo{journal}{Phys.Rev.}
  \textbf{\bibinfo{volume}{D82}}, \bibinfo{pages}{034512}
  (\bibinfo{year}{2010}), \eprint{1004.4285}.

\bibitem[{\citenamefont{Wingate et~al.}(2003)\citenamefont{Wingate, Shigemitsu,
  Davies, Lepage, and Trottier}}]{mattfirsthl}
\bibinfo{author}{\bibfnamefont{M.}~\bibnamefont{Wingate}},
  \bibinfo{author}{\bibfnamefont{J.}~\bibnamefont{Shigemitsu}},
  \bibinfo{author}{\bibfnamefont{C.~T.} \bibnamefont{Davies}},
  \bibinfo{author}{\bibfnamefont{G.~P.} \bibnamefont{Lepage}},
  \bibnamefont{and} \bibinfo{author}{\bibfnamefont{H.~D.}
  \bibnamefont{Trottier}} (\bibinfo{collaboration}{HPQCD Collaboration}),
  \bibinfo{journal}{Phys.Rev.} \textbf{\bibinfo{volume}{D67}},
  \bibinfo{pages}{054505} (\bibinfo{year}{2003}), \eprint{hep-lat/0211014}.

\bibitem[{\citenamefont{Bazavov et~al.}(2012)}]{fermilabdecay11}
\bibinfo{author}{\bibfnamefont{A.}~\bibnamefont{Bazavov}} \bibnamefont{et~al.}
  (\bibinfo{collaboration}{Fermilab Lattice and MILC Collaborations}),
  \bibinfo{journal}{Phys.Rev.} \textbf{\bibinfo{volume}{D85}},
  \bibinfo{pages}{114506} (\bibinfo{year}{2012}), \eprint{1112.3051}.

\bibitem[{\citenamefont{Trottier}(2008)}]{howard}
\bibinfo{author}{\bibfnamefont{H.}~\bibnamefont{Trottier}},
  \bibinfo{journal}{private communication}  (\bibinfo{year}{2008}).

\bibitem[{\citenamefont{Bazavov et~al.}(2010{\natexlab{c}})}]{hisqscaling}
\bibinfo{author}{\bibfnamefont{A.}~\bibnamefont{Bazavov}} \bibnamefont{et~al.}
  (\bibinfo{collaboration}{MILC collaboration}), \bibinfo{journal}{Phys.Rev.}
  \textbf{\bibinfo{volume}{D82}}, \bibinfo{pages}{074501}
  (\bibinfo{year}{2010}{\natexlab{c}}), \eprint{1004.0342}.

\bibitem[{\citenamefont{Gray et~al.}(2005)\citenamefont{Gray, Allison, Davies,
  Dalgic, Lepage et~al.}}]{gray}
\bibinfo{author}{\bibfnamefont{A.}~\bibnamefont{Gray}},
  \bibinfo{author}{\bibfnamefont{I.}~\bibnamefont{Allison}},
  \bibinfo{author}{\bibfnamefont{C.}~\bibnamefont{Davies}},
  \bibinfo{author}{\bibfnamefont{E.}~\bibnamefont{Dalgic}},
  \bibinfo{author}{\bibfnamefont{G.}~\bibnamefont{Lepage}},
  \bibnamefont{et~al.} (\bibinfo{collaboration}{HPQCD Collaboration}),
  \bibinfo{journal}{Phys.Rev.} \textbf{\bibinfo{volume}{D72}},
  \bibinfo{pages}{094507} (\bibinfo{year}{2005}), \eprint{hep-lat/0507013}.

\bibitem[{\citenamefont{Jansen et~al.}(2011)\citenamefont{Jansen, Petschlies,
  and Urbach}}]{etmcurrcurr}
\bibinfo{author}{\bibfnamefont{K.}~\bibnamefont{Jansen}},
  \bibinfo{author}{\bibfnamefont{M.}~\bibnamefont{Petschlies}},
  \bibnamefont{and} \bibinfo{author}{\bibfnamefont{C.}~\bibnamefont{Urbach}},
  \bibinfo{journal}{PoS} \textbf{\bibinfo{volume}{LATTICE2011}},
  \bibinfo{pages}{234} (\bibinfo{year}{2011}), \eprint{1111.5252}.

\end{thebibliography}

\end{document}